\documentclass[a4paper,12pt,reqno]{amsart}
\usepackage{amssymb}
\usepackage{t1enc}
\usepackage[margin=1in]{geometry}
\usepackage{ifthen}
\usepackage{graphicx}
\usepackage{xcolor}
\usepackage[footnotesize]{caption}
\newcommand{\abs}[1]{\left| #1 \right|}
\newcommand{\expr}[1]{\left( #1 \right)}

\newcommand{\norm}[1]{\left\| #1 \right\|}
\newcommand{\set}[1]{\left\{ #1 \right\}}
\newcommand{\scalar}[1]{\left< #1 \right>}
\newcommand{\tscalar}[1]{\langle #1 \rangle}
\newcommand{\ind}{\mathbf{1}}

\newcommand{\sub}{\subseteq}
\newcommand{\C}{\mathbf{C}}
\newcommand{\R}{\mathbf{R}}

\newcommand{\pr}{\mathbf{P}}
\newcommand{\ex}{\mathbf{E}}

\newcommand{\A}{\mathcal{A}}

\newcommand{\fourier}{\mathcal{F}}
\newcommand{\laplace}{\mathcal{L}}

\newcommand{\domain}{\mathcal{D}}
\newcommand{\eps}{\varepsilon}
\newcommand{\ph}{\varphi}

\newcommand{\thet}{\vartheta}

\newcommand{\ignore}[1]{}

\newcommand{\pvint}{\pv\!\!\int}

\newcommand{\odd}{\mathrm{odd}}
\newcommand{\even}{\mathrm{even}}
\newcommand{\wall}{\mathrm{wall}}
\newcommand{\well}{\mathrm{well}}

\newcommand{\formula}[2][nolabel]
{\ifthenelse{\equal{#1}{nolabel}}
 {\begin{align*} #2 \end{align*}}
 {\ifthenelse{\equal{#1}{}}
  {\begin{align} #2 \end{align}}
  {\begin{align} \label{#1} #2 \end{align}}
 }
}

\DeclareMathOperator{\pv}{pv}

\DeclareMathOperator{\real}{Re}
\DeclareMathOperator{\dist}{dist}

\DeclareMathOperator{\sign}{sign}
\DeclareMathOperator{\supp}{supp}

\theoremstyle{plain}
\newtheorem{theorem}{Theorem}[section]
\newtheorem*{theorem1}{Theorem \ref{th1}'}
\newtheorem*{theorem2}{Theorem \ref{th2}'}

\newtheorem{lemma}[theorem]{Lemma}

\newtheorem{proposition}[theorem]{Proposition}

\theoremstyle{definition}
\newtheorem{definition}[theorem]{Definition}

\newtheorem{remark}[theorem]{Remark}

\theoremstyle{remark}

\title{One-dimensional quasi-relativistic particle in the box}
\author{Kamil Kaleta, Mateusz Kwa{\'s}nicki, Jacek Ma{\l}ecki}
\thanks{The authors were supported by the Polish Ministry of Science and Higher Education grant no.\ N~N201 373136}
\thanks{Mateusz Kwa{\'s}nicki received financial support of the Foundation for Polish Science}
\address{Institute of Mathematics and Computer Science \\ Wroc{\l}aw University of Technology \\ ul. Wybrze{\.z}e Wyspia{\'n}\-skiego 27, 50-370 Wroc{\l}aw, Poland}
\email{kamil.kaleta@pwr.wroc.pl, jacek.malecki@pwr.wroc.pl}
\address{Institute of Mathematics \\ Polish Academy of Sciences \\ ul. {\'S}niadeckich 8, 00-976 Warszawa, Poland}
\email{m.kwasnicki@impan.pl}

\begin{document}

\sloppy

\begin{abstract}
Two-term Weyl-type asymptotic law for the eigenvalues of one-di\-men\-sional quasi-relativistic Hamiltonian $(-\hbar^2 c^2 d^2/dx^2 + m^2 c^4)^{1/2} + V_\well(x)$ (the Klein-Gordon square-root operator with electrostatic potential) with the infinite square well potential $V_\well(x)$ is given: the $n$-th eigenvalue is equal to $(n \pi / 2 - \pi/8) \hbar c / a + O(1/n)$, where $2a$ is the width of the potential well. Simplicity of eigenvalues is proved. Some $L^2$ and $L^\infty$ properties of eigenfunctions are also studied. Eigenvalues represent energies of a `massive particle in the box' quasi-relativistic model.
\end{abstract}

\maketitle

%
%                            ---------- o ----------
%

\section{Introduction and statement of the results}

In quantum physics, the state of a particle, or a system of particles, is represented by the \emph{wave function} $\Psi(t, x)$, which satisfies the Schr{\"o}dinger equation
\formula[eq:sch]{
 i \hbar \, \frac{\partial \Psi}{\partial t} (t, x) & = H \Psi(t, x) .
}
Here $H$ is the \emph{Hamiltonian}, a (typically unbounded) self-adjoint operator, acting in the spatial variable $x$, and $\hbar$ is the reduced Planck constant. When the solution of~\eqref{eq:sch} has a time-harmonic form $\Psi(t, x) = e^{-i E t / \hbar} \Phi(x)$, it is called a \emph{stationary state} with energy $E$. Solutions of this kind are in one-to-one correspondence with eigenfunctions $\Phi(x)$ and eigenvalues $E$ of the Hamiltonian $H$. The function $\Phi(x)$ is often called an \emph{energy eigenstate}, and the corresponding $E$ is said to be an \emph{energy level} of the particle. The number of linearly independent eigenstates $\Phi(x)$ with energy $E$ is the \emph{degeneracy} of the energy level $E$. Given a Hamiltonian $H$, one of the fundamental questions is: What are the energy levels of the system, and what is their degeneracy?

Motion of a single particle subject to an electrostatic field is usually modeled by the Hamiltonian $H = -(\hbar^2 / (2m)) \Delta + V(x)$, where $m$ is the mass of the particle, $V(x)$ is the potential of the electrostatic field, and $\Delta$ is the Laplace operator. This model, however, is inconsistent with special relativity, and leads to erroneous results at high energies (see~\cite{bib:ls10, bib:t92}). A better model for particles at relativistic energies was proposed by Dirac in~\cite{bib:d28}. On the other hand, Dirac's equations are much more complicated, and therefore their mathematical treatment is often problematic (see~\cite{bib:els08}). For this reason, at least in some applications (see~\cite{bib:ls10} and the references therein, and~\cite{bib:gko95, bib:vz99}), one is often satisfied with an intermediate model, with the Hamiltonian $H$ given by
\formula[eq:hamiltonian]{
 H & = H_0 + V(x) = \expr{-\hbar^2 c^2 \Delta + m^2 c^4}^{1/2} + V(x) .
}
Here $c$ is the speed of light, and $H_0 = (-\hbar^2 c^2 \Delta + m^2 c^4)^{1/2}$ is often called the \emph{Klein-Gordon square root operator}, or the \emph{quasi-relativistic Hamiltonian}. We often say that $H$ describes the motion of a \emph{quasi-relativistic particle} in the potential $V(x)$. The purpose of this article is to study the stationary states and their energies for the \emph{one-dimensional} quasi-relativistic particle in the \emph{infinite square potential well}:
\formula[eq:v]{
 V_\well(x) & = \begin{cases} 0 & \text{for $x \in (-a, a)$,} \\ \infty & \text{otherwise.} \end{cases}
}
In the non-relativistic setting, this model is referred to as the (one-dimensional) \emph{particle in the box}, since the potential $V_\well(x)$ constrains the particle not to leave the `box' $(-a, a)$. Therefore, the problem studied in the present article can be described as the \emph{quasi-relativistic particle in the box}. It is well-known that in this case the spectrum of $H$ is purely discrete, and the energy levels form an infinite sequence $E_1 < E_2 \le E_3 \le ... \to \infty$ (see Preliminaries for more details). Our main result can be stated as follows.

\begin{theorem}
\label{th1}
All energy levels are non-degenerate, and
\formula[eq:asymp1]{
 E_n & = \expr{\frac{n \pi}{2} - \frac{\pi}{8}} \frac{\hbar c}{a} + O\expr{\frac{1}{n}} && \text{as $n \to \infty$.}
}
\end{theorem}

It should be emphasized that the operators $H_0$ and $H$ are \emph{non-local}. In the non-relativistic case, one typically solves the equation $H \Phi = E \Phi$ separately in the intervals $(-\infty, -a)$, $(-a, a)$ and $(a, \infty)$, and then matches the boundary conditions at $-a$ and $a$. For the non-local operator $H$, the equation in an interval has no sense unless the values of $\Phi$ in the complement of the interval are known. This means that one cannot apply the non-relativistic method; and indeed, this interval-wise method would yield the erroneous asymptotic formula $E_n = (n \pi / 2) (\hbar c / a) + O(1/n)$, with the \emph{phase shift} $-\pi/8$ missing. Non-locality of this and similar Hamiltonians has caused considerable confusion; see~\cite{bib:jxhs10} for further discussion.

The proof of Theorem~\ref{th1} follows closely the approach of~\cite{bib:kkms10, bib:k11}. The problem studied in~\cite{bib:kkms10} corresponds to a massless particle, described by the Hamiltonian $H = \hbar c (-d^2 / dx^2)^{1/2} + V_\well(x)$, while~\cite{bib:k11} deals with Hamiltonians based on the fractional Laplace operator $(-d^2 / dx^2)^{s}$ for $s \in (0, 1)$. The argument relies heavily on the results of~\cite{bib:k10, bib:kmr11}, where the spectral theory for similar Hamiltonians with a \emph{single infinite potential wall} (that is, the potential $V_\wall(x) = 0$ for $x > 0$, $V_\wall(x) = \infty$ otherwise) was studied. The motivation of the articles cited in this paragraph is purely mathematical. For the discussion of some of their physical implications, see Preliminaries.

We remark that the Weyl-type asymptotic formula $E_n \sim (n \pi / 2) \hbar c / a$ as $n \to \infty$ is known since 1959, due to a general result by R.~M.~Blumenthal and R.~K.~Getoor~\cite{bib:bg59}. Hence, formula~\eqref{eq:asymp1} essentially provides the second term (a phase shift of $-\pi/8$) in the asymptotic expansion. In fact, a stronger result is proved, with detailed estimates of the error term $O(1/n)$. The statement of this theorem requires some additional notation.

\begin{figure}
\begin{minipage}[t]{0.45\textwidth}
\centering
\includegraphics[width=\textwidth]{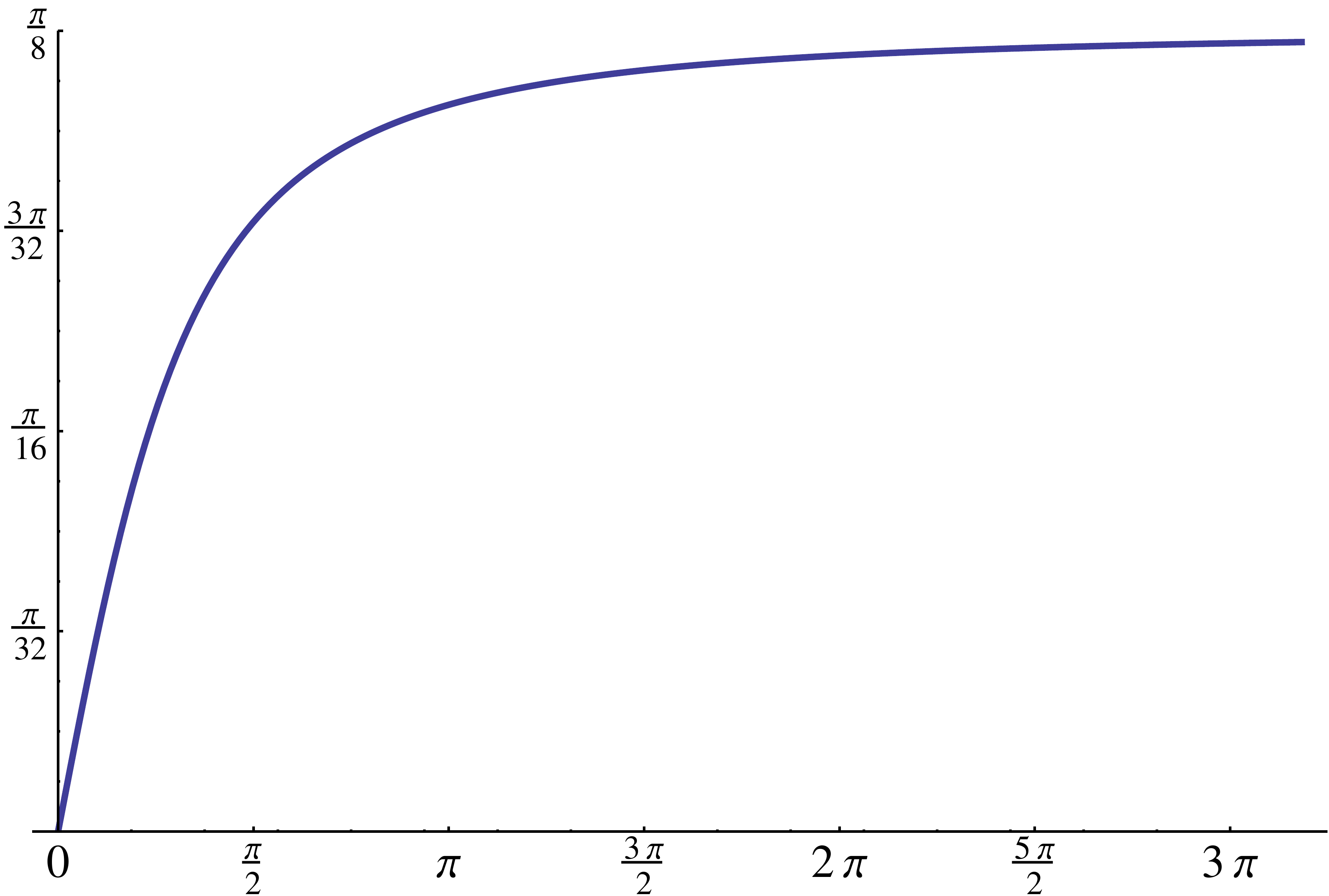}
\caption{Plot of $\thet_\mu$.}
\label{fig:theta}
\end{minipage}
\hspace*{0.02\textwidth}
\begin{minipage}[t]{0.45\textwidth}
\centering
\includegraphics[width=\textwidth]{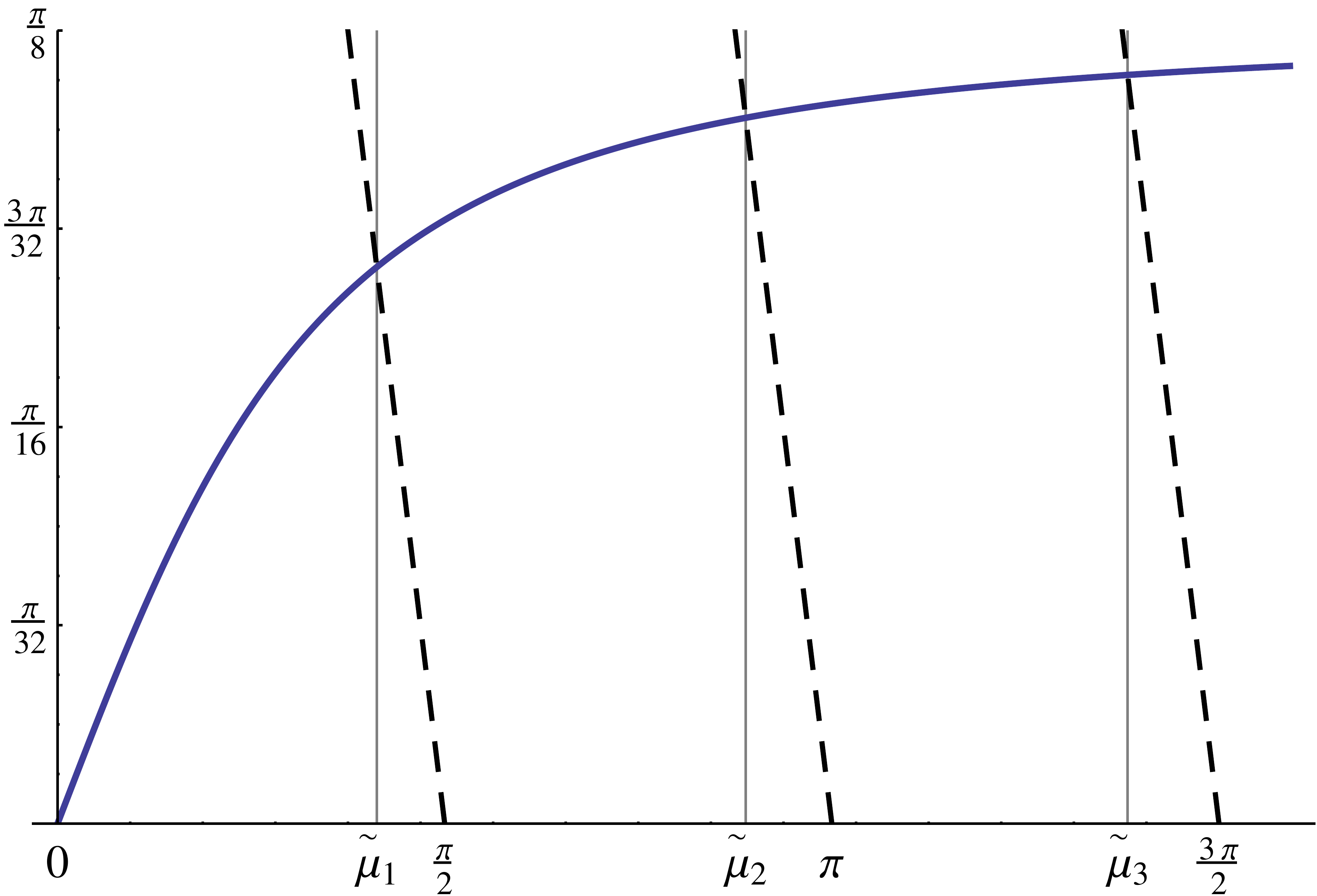}
\caption{Plot of $\thet_\mu$ (solid line) and $n \pi / 2 - (m c a / \hbar) \mu$ for $n = 1, 2, 3$ (dashed line) when $m c a / \hbar = 1$. For the electron ($m = 0.511\,\mathrm{MeV}$), this corresponds to the potential well of width $2a = 0.772\,\mathrm{pm}$. In this case, $\tilde{\mu}_1 \approx 1.295$, $\tilde{\mu}_2 \approx 2.792$, $\tilde{\mu}_3 \approx 4.342$.}
\label{fig:mu}
\end{minipage}
\end{figure}

For $\mu > 0$, we define the phase shift function (see Figure~\ref{fig:theta})
\formula[eq:thetmu]{
 \thet_\mu & = \frac{1}{\pi} \int_0^\infty \frac{\mu}{s^2 - \mu^2} \, \log \expr{\frac{1}{2} + \frac{1}{2} \sqrt{\frac{1 + s^2}{1 + \mu^2}}} ds .
}
Then $\thet_\mu$ is an increasing function, with limits $\thet_{0+} = 0$ and $\thet_\infty = \pi/8$. More precisely,
\formula[]{
\label{eq:thetmuzero}
 \thet_\mu & = \frac{\mu}{\pi} - \frac{7 \mu^3}{18 \pi} + O(\mu^5) && \text{as $\mu \to 0^+$;} \\
\label{eq:thetmuinfinity}
 \thet_\mu & = \frac{\pi}{8} - \frac{1 + 2 \log(2 \mu)}{4 \pi \mu^2} + O\expr{\frac{\log \mu}{\mu^4}} && \text{as $\mu \to \infty$;}
}
these properties are established in Section~\ref{sec:proofs}. We define $\tilde{\mu}_n$ to be the (unique) positive solution of the equation (see Figure~\ref{fig:mu} for a geometric interpretation)
\formula[eq:tildemu0]{
 \thet_{\tilde{\mu}_n} = \frac{n \pi}{2} - \frac{m c a}{\hbar} \, \tilde{\mu}_n .
}

\begin{theorem}
\label{th2}
For all $n = 1, 2, ...$,
\formula[eq:asymp2]{
 \abs{E_n - m c^2 \sqrt{\tilde{\mu}_n^2 + 1}} & < \frac{8}{n} \expr{3 m c^2 + \frac{\hbar c}{a}} \exp\expr{-\frac{m c a}{3 \hbar}} .
}
\end{theorem}

\begin{remark}
\label{rem:limits}
Formula~\eqref{eq:asymp2} provides much more detailed information than~\eqref{eq:asymp1}. Thanks to the explicit estimate of the error term, it enables passing to a limit in various regimes. We discuss three examples of this kind. A formal verification is given in Section~\ref{sec:proofs}.
\begin{enumerate}
\item
By taking $m \to 0^+$ in~\eqref{eq:asymp2}, one recovers the results for $(-d^2/dx^2)^{1/2}$, the quasi-relativistic Hamiltonian of a massless particle in the infinite potential well, obtained earlier in~\cite{bib:kkms10}. In this case we have, as in Theorem~\ref{th1}, $E_n = (n \pi/2 - \pi/8) \hbar c / a + O(1/n)$ as $n \to \infty$.
\item
In the non-relativistic limit $c \to \infty$, we recover the classical expression $(n \pi / (2 a))^2 \hbar / (2 m)$ for the \emph{kinetic} energy in the (non-relativistic) particle in the box model. Indeed, for a fixed $n \ge 1$, we have
\formula[eq:asymp1alt]{
 E_n - m c^2 & = \expr{\frac{n \pi}{2 a}}^2 \frac{\hbar^2}{2 m} + O\expr{\frac{1}{c}} && \text{as $c \to \infty$.}
}
Note that the phase shift $-\pi/8$ is absent in this case, and that in the limit, the energies grow as $n^2$, as opposed to linear growth in~\eqref{eq:asymp1}.
\item
In~\eqref{eq:asymp2}, the error term decays exponentially fast with $c$ and $a$. On the other hand, $\tilde{\mu}_n$ is defined in a rather implicit way. One can try to replace it with a more explicit expression, at the price of a slower decay rate of the error term. For example, we have
\formula[eq:asymp2alt]{
 \abs{E_n - m c^2 \sqrt{\expr{\frac{n \pi}{2} - \thet\expr{\frac{n \pi}{2} \, \frac{\hbar}{m c a}}}^2 \expr{\frac{\hbar}{m c a}}^2 + 1}} & < 12 \, \frac{\hbar c}{n a}
}
(for clarity, we write $\thet(\mu)$ for $\thet_\mu$ here). This estimate can be used for large energy levels of particles in potential wells of large width. If, for example, $\mu > 0$, $n \to \infty$ and $a = n \pi \hbar / (2 m c \mu) + o(n)$, then~\eqref{eq:asymp2alt} and series expansion immediately give
\formula{
 E_n & = m c^2 \sqrt{\mu^2 + 1} \expr{1 - \frac{\mu^2}{\mu^2 + 1} \, \frac{2 \thet_\mu}{n \pi}} + o\expr{\frac{1}{n}} .
}
See~\eqref{eq:asymp2alt2} in Section~\ref{sec:proofs} for a slightly stronger version of~\eqref{eq:asymp2alt}, which is suitable also for non-relativistic limits $c \to \infty$.
\end{enumerate}
\end{remark}

\begin{remark}
By choosing \emph{natural units}, we may assume, without losing generality, that $\hbar = c = m = 1$. This corresponds to \emph{scaling properties} of the Hamiltonian $H$, discussed in detail in Preliminaries.
\end{remark}

Theorem~\ref{th2} can be used to prove the following useful bounds for the energy levels.

\begin{proposition}
\label{prop:simplebounds}
For all $n = 1, 2, ...$, we have
\formula[eq:cs2]{
 m c^2 \sqrt{\expr{\frac{(n - 1) \pi}{2} \, \frac{\hbar}{m c a}}^2 + 1} < E_n & \le m c^2 \sqrt{\expr{\frac{n \pi}{2} \, \frac{\hbar}{m c a}}^2 + 1} \, .
}
\end{proposition}

We remark that the best bounds for $E_n$ known before were
\formula[eq:cs1]{
 \frac{1}{2} \expr{m c^2 \sqrt{\expr{\frac{n \pi}{2} \, \frac{\hbar}{m c a}}^2 + 1} + m c^2} \le E_n & \le m c^2 \sqrt{\expr{\frac{n \pi}{2} \, \frac{\hbar}{m c a}}^2 + 1} ,
}
proved in~\cite{bib:cs05}. Note that~\eqref{eq:cs2} significantly improves the lower bound. For a further discussion, see Preliminaries.

By a general theory, the eigenstates $\Phi_n$ are real-valued and vanish outside $(-a, a)$. The methods of the proofs of main theorems can be used to extract some information about the shape of $\Phi_n$. In principle, the eigenstate $\Phi_n$ (normalized in $L^2(\R)$) corresponding to the energy level $E_n$ is similar to either sine or cosine function (depending on the parity of $n$). Our first result is an $L^2(\R)$ estimate.

\begin{proposition}
\label{prop:l2}
Let
\formula[eq:sincos]{
\begin{aligned}
 f_n(x) & = \frac{1}{\sqrt{a}} \, \cos(\tilde{\mu}_n x) \ind_{(-a, a)}(x) &\qquad& \text{when $n$ is odd,} \\
 f_n(x) & = \frac{1}{\sqrt{a}} \, \sin(\tilde{\mu}_n x) \ind_{(-a, a)}(x) &\qquad& \text{when $n$ is even,} \\
\end{aligned}
}
with $\tilde{\mu}_n$ defined as in Theorem~\ref{th2}. Then for all $n = 1, 2, ...$, the sign of $\Phi_n$ can be chosen so that
\formula[eq:phl2est]{
 \norm{\Phi_n - f_n}_{L^2(\R)} & \le \sqrt{\frac{m c}{\hbar}} \, \min \expr{\frac{10}{\sqrt{n}} , \frac{55 \hbar \sqrt{n}}{m c a}} \, .
}
Furthermore, $\Phi_n$ is an even function when $n$ is odd, and $\Phi_n$ is an odd function when $n$ is even.
\end{proposition}

We remark that the right-hand side of~\eqref{eq:phl2est} is $O(1/\sqrt{n})$ as $n \to \infty$, and $O(1/\sqrt{c})$ as $c \to \infty$. In particular, we recover the classical formula $\Phi_n = f_n$ in the non-relativistic limit $c \to \infty$. The error in~\eqref{eq:phl2est} comes mostly from the wrong \emph{boundary behavior} of the approximation $f_n$. By a suitable modification of $f_n$ near $\pm a$, the decay rate of the error term in~\eqref{eq:phl2est} can be significantly improved; see Proposition~\ref{prop:l2approx} in Section~\ref{sec:proofs} for details.

The next result gives a uniform pointwise bound for the eigenfunctions. Noteworthy, the results for a single potential wall in~\cite{bib:k10, bib:kmr11} suggest that the best admissible constant in~\eqref{eq:phsupest} is likely to be much closer to $1$.

\begin{proposition}
\label{prop:loo}
For all $n = 1, 2, ...$,
\formula[eq:phsupest]{
 \norm{\Phi_n}_{L^\infty(\R)} & \le \frac{8}{\sqrt{a}} \, .
}
\end{proposition}

\medskip

The remaining part of the article is divided into three sections. The article is meant to be accessible both for readers interested in mathematical physics and those working in theoretical mathematics. For this reason, in Preliminaries, we discuss the formal statement of the problem, its various reformulations, and relation to the theory of Markov semigroups. Furthermore, we recall known results used later in the proofs. In Section~\ref{sec:halfline} we present the results of~\cite{bib:k10, bib:kmr11} and discuss their physical interpretation. Finally, all results are proved in Section~\ref{sec:proofs}.

\medskip

The proofs of main theorems are rather technical. For this reason, we conclude the introduction with an intuitive description of the proof of Theorem~\ref{th1}. In~\cite{bib:k10}, the (generalized, or continuum) eigenstates $F_\mu(x)$ were found for the quasi-relativistic particle with an infinite potential wall, $V_\wall(x) = 0$ for $x > 0$ and $V_\wall(x) = \infty$ otherwise; see Section~\ref{sec:halfline} for more details. The case of the potential well, studied in this article, can be thought as a combination of two potential walls, one at $-a$, and the other one at $a$. We use the eigenfunctions $F_\mu(a + x)$ and $F_\mu(a - x)$ corresponding to these single potential walls, to define an \emph{approximate} eigenstate $\tilde{\Phi}_n$ for the case of the potential well. (This approximation is different from the function $f_n$ in Proposition~\ref{prop:l2}.)

More precisely, we define a function $\tilde{\Phi}_n$ in such a way that $\tilde{\Phi}_n(x) = a^{-1/2} F_\mu(a + x)$ for $x$ close to $-a$, and $\tilde{\Phi}_n(x) = (-1)^{n-1} a^{-1/2} F_\mu(a - x)$ for $x$ close to $a$. Inside $(-a, a)$, and away from the boundary, $\tilde{\Phi}_n(x)$ behaves as a sine or a cosine function, in a similar way as in Proposition~\ref{prop:l2}. We prove that for a certain energy $\tilde{E}_n$, we have $H \tilde{\Phi}_n(x) \approx \tilde{E}_n \tilde{\Phi}_n(x)$. Expansion of $\tilde{\Phi}_n(x)$ in the complete orthogonal set of (true, not approximate) eigenstates $\Phi_j(x)$ shows that $\tilde{E}_n$ must be close to some (true) energy level $E_{k(n)}$.

Careful estimates show that for each $n \ge 3$, the corresponding $E_{k(n)}$ are distinct. Furthermore, by an appropriate \emph{trace estimate}, it follows that there can be no other energy levels than $E_1$, $E_2$ and $E_{k(n)}$ for $n \ge 3$, and therefore $k(n) = n$. Theorems~\ref{th1} and~\ref{th2} follow. Other results follow either directly from Theorem~\ref{th2}, or from intermediate steps of its proof.

%
%                            ---------- o ----------
%

\section{Preliminaries}

We begin with some notational remarks. Quantum states are in general complex-valued functions. In our case, however, all states will be real-valued. For this reason, unless explicitly stated otherwise, from now on all functions are assumed to take values in $\R$. By a function we always mean a Borel measurable function, and we extend the definition of any function $f$ defined on a subset $D$ of $\R$ to entire $\R$ by letting $f(x) = 0$ for $x \notin D$.

As usual, $L^p(D)$ ($p \in [1, \infty)$, $D \sub \R$) denotes the space of functions $f$ on $D$ such that $\|f\|_{L^p(D)} = (\int_D |f(x)|^p dx)^{1/p}$ is finite, and $f \in L^\infty(D)$ if and only if the essential supremum of $|f(x)|$ over $x \in D$, denoted by $\|f\|_{L^\infty(D)}$, is finite. By $C_c^\infty(D)$ we denote the space of smooth, compactly supported functions vanishing outside $D$, and $C_0(D)$ is the space of continuous functions in $D$ which vanish continuously at the boundary of $D$ and at $\pm \infty$ (if $D$ is unbounded).

A function $f(x)$ on $(0, \infty)$ is said to be \emph{completely monotone} (on $(0, \infty)$) if $(-1)^n f^{(n)}(x) \ge 0$ for all $x > 0$ and $n = 0, 1, 2, ...$ By Bernstein's theorem (see~\cite[Theorem~1.4]{bib:ssv10}), $f$ is completely monotone if and only if it is the Laplace transform of a (possibly infinite) Radon measure on $[0, \infty)$.

We denote the Fourier transform of $f \in L^2(\R)$ by $\fourier f$; when $f \in L^2(\R) \cap L^1(\R)$, we have $\fourier f(\xi) = \int_{-\infty}^\infty f(x) e^{-i \xi x} dx$. The inverse Fourier transform is denoted by $\fourier^{-1}$. Occasionally, we denote the Laplace transform of a function $f$ by $\laplace f$, $\laplace f(\xi) = \int_0^\infty f(x) e^{-\xi x} dx$. We use mostly $x$, $y$, $z$ to denote spatial variables, and $\xi$ for a `Fourier space' variable.

As usual, we write $f(n) = O(g(n))$ if $\limsup_{n \to \infty} |f(n) / g(n)| < \infty$, and $f(n) = o(g(n))$ when $\lim_{n \to \infty} |f(n) / g(n)| = 0$.

\subsection{Formal definition of the Hamiltonian}

In this subsection we give a formal meaning to the definition~\eqref{eq:hamiltonian} of the Hamiltonian $H$. Further explanation of the notation used in formula~\eqref{eq:hamiltonian} is given in Subsection~\ref{subsec:fk} using a Feynman-Kac path integral approach. 

The \emph{free} one-dimensional quasi-relativistic Hamiltonian, or the Klein-Gordon square-root operator, is denoted by $H_0$:
\formula[eq:hamiltonian0]{
 H_0 & = \expr{-\hbar^2 c^2 \, \frac{d^2}{d x^2} + m^2 c^4}^{1/2} .
}
The operator $H_0$ is an unbounded, non-local, self-adjoint operator, acting on its domain $\domain(H_0) \sub L^2(\R)$, and it is defined via the Fourier transform: $f \in \domain(H_0)$ if and only if $(1 + \xi^2)^{1/2} \fourier f(\xi)$ is square integrable, and
\formula[eq:hamiltonian0f]{
 \fourier H_0 f(\xi) & = \sqrt{\hbar^2 c^2 \xi^2 + m^2 c^4} \, \fourier f(\xi) .
}
In other words, $H_0$ is a \emph{Fourier multiplier} with symbol $(\hbar^2 c^2 \xi^2 + m^2 c^4)^{1/2}$. Note that $\domain(H_0)$ coincides with the Sobolev space $H^1(\R)$. By Plancherel's theorem, $H_0$ is positive-definite, and the quadratic form associated to $H_0$ satisfies
\formula[eq:hpositive]{
 \tscalar{f, H_0 f}_{L^2(\R)} = \frac{1}{2 \pi} \, \tscalar{\fourier f(\xi), \sqrt{\hbar^2 c^2 \xi^2 + m^2 c^4} \, \fourier f(\xi)}_{L^2(\R)} & \ge m c^2 \norm{f}_{L^2(\R)}
}
for all $f \in \domain(H_0)$. Formula~\eqref{eq:hpositive} states that the total energy, measured by $H_0$, is greater than the rest energy $m c^2$.

Let $V_\well(x)$ denote the infinite square potential well of width $2a$, as in~\eqref{eq:v}. For brevity, we denote $D = (-a, a)$; hence $V_\well(x) = 0$ for $x \in D$ and $V_\well(x) = \infty$ otherwise. Since $V_\well(x)$ is highly singular, the definition of $H = H_0 + V_\well(x)$ is problematic: multiplication by $V_\well(x)$ is not well-defined on $\domain(H_0)$. Clearly, $H f$ can only be defined when $f$ vanishes almost everywhere outside $D$. Furthermore, when $f \in \domain(H_0)$ and $f(x) = 0$ for almost all $x \notin D$, then we expect that $H f(x) = H_0 f(x)$ for $x \in D$. This motivates the following definition.

\begin{definition}
\label{def:h}
Let $\tilde{H}$ be the unbounded, positive-definite symmetric operator with domain $\domain(\tilde{H}) = C_c^\infty(D)$, $D = (-a, a)$, given by the formula $\tilde{H} f(x) = H_0 f(x) \ind_D(x)$. We define $H$ to be the \emph{Friedrichs extension} (or the \emph{minimal self-adjoint extension}) of $\tilde{H}$. This means that the quadratic form of $H$ is the closure of the quadratic form of $\tilde{H}$. More precisely, a function $f \in L^2(D)$ belongs to the domain $\domain(H^{1/2})$ of the quadratic form of $H$ if and only if:
\begin{enumerate}
\item[(a)] there is a sequence $f_k \in C_c^\infty(D)$ convergent to $f$ in $L^2(D)$, and such that $\tscalar{\mbox{$f_k - f_l$}, \mbox{$\tilde{H}(f_k - f_l)$}}_{L^2(D)} \to 0$ as $k, l \to \infty$.
\end{enumerate}
The operator $H$ is associated to the form defined above; that is, $f \in L^2(D)$ is in the domain $\domain(H)$ of $H$ if and only if it satisfies~(a), and furthermore:
\begin{enumerate}
\item[(b)] there exists $H f \in L^2(D)$ such that $\tscalar{\tilde{H} g, f}_{L^2(D)} = \tscalar{g, H f}_{L^2(D)}$ for $g \in C_c^\infty(D)$.
\end{enumerate}
Condition~(b) automatically defines the action of $H$ on $\domain(H)$. Note that in condition~(a), we have $\tscalar{f_k - f_l, \tilde{H}(f_k - f_l)}_{L^2(D)} = \| H_0^{1/2} (f_k - f_l) \|_{L^2(\R)} = \| H^{1/2} (f_k - f_l) \|_{L^2(D)}$.

More generally, suppose that $D$ is an open subset of $\R$, $V(x) = 0$ for $x \in D$ and $V(x) = \infty$ for $x \notin D$. Then by $H_0 + V(x)$ we mean the Friedrichs extension in $L^2(D)$ of the operator with domain $C_c^\infty(D)$, which maps a function $f \in C_c^\infty(D)$ to $H_0 f(x) \ind_D(x)$.
\end{definition}

The above definition mimics the well-recognized non-relativistic case. There the Hamiltonian corresponding to the particle in the box model is based on the \emph{Dirichlet Laplace operator} in $D$, which is the Friedrichs extension (in $L^2(D)$) of the Laplace operator restricted to $C_c^\infty(D)$. Noteworthy, the self-adjoint extension of the Laplace operator restricted to $C_c^\infty(D)$ is not unique (an example of a different extension is the Neumann Laplace operator in $D$), hence the choice of the Friedrichs extension might seem arbitrary. On the other hand, Proposition~\ref{prop:esa} below states that the operator $\tilde{H}$ in Definition~\ref{def:h} is essentially self-adjoint, and hence $H$ is the \emph{unique} self-adjoint extension of $\tilde{H}$. That is, the phrase `Friedrichs extension of $\tilde{H}$' in Definition~\ref{def:h} can be replaced with `the closure of $\tilde{H}$'.

The following characterization of the domain of the quadratic form of $H$ seems to be well-known to specialists.

\begin{proposition}
\label{prop:domain}
Suppose that $f \in L^2(D)$, $D = (-a, a)$. Then the condition~(a) of Definition~\ref{def:h} is equivalent to each of the following:
\begin{enumerate}
\item[(a1)] $f \in \domain(H^{1/2})$;
\item[(a2)] $f \in \domain(H_0^{1/2})$;
\item[(a3)] $(\hbar^2 c^2 \xi^2 + m^2 c^4)^{1/2} |\fourier f(\xi)|^2$ is integrable on $\R$.
\end{enumerate}
In other words, $\domain(H^{1/2})$ is the set of functions in $\domain(H_0^{1/2})$ vanishing outside $D$.
\end{proposition}

Note that an analogous statement is true for $D = (0, \infty)$ (with a very similar proof), but not for arbitrary open sets $D \sub \R$ (a typical example of a domain $D$ for which the equivalence of~(a1) and~(a2) fails is the complement of the Cantor set).

\begin{proof}
Equivalence of~(a2) and~(a3) follows right from the fact that $H_0$ takes a diagonal form in the Fourier space. Also, (a) is equivalent to~(a1) by Definition~\ref{def:h} and the definition of the square root of an operator. Finally, (a) clearly implies~(a2). Hence, it suffices to prove that~(a3) implies~(a).

For an arbitrary function $f \in L^2(D)$, denote $M_\eps f(x) = f((1 + \eps)^{-1} x)$. Suppose that (a3) is satisfied. Fix $h \in C_c^\infty(\R)$ satisfying $h(x) = 0$ for $x \notin (-a, a)$, $h(x) \ge 0$ for $x \in (-a, a)$, and $\int h(x) dx = 1$. Let $h_\eps$ be the approximation to the identity, given by $h_\eps(x) = \eps^{-1} h(\eps^{-1} x)$. Note that $h_\eps$ is supported in $(-\eps a, \eps a)$. We define $g_\eps = h_\eps * f$ and $f_\eps = M_\eps g_\eps$. Since $g_\eps \in C_c^\infty(\R)$ is supported in a compact subset of $(-(1 + \eps) a, (1 + \eps) a)$, we have $f_\eps \in C_c^\infty(D)$. By Plancherel's theorem,
\formula[eq:domain:limit]{
 \norm{f_\eps - f}_{L^2(\R)}^2 + \norm{H_0^{1/2} (f_\eps - f)}_{L^2(\R)}^2 & = \frac{1}{2 \pi} \int_{-\infty}^\infty v(\xi) |\fourier f_\eps(\xi) - \fourier f(\xi)|^2 d\xi ,
}
where $v(\xi) = 1 + (\hbar^2 c^2 \xi^2 + m^2 c^4)^{1/2}$. We claim that the right-hand side converges to $0$ as $\eps \to 0^+$. Once this is proved, the sequence $f_{1/k}$ converges to $f$ in $L^2(D)$ and $\tscalar{f_{1/k} - f_{1/l}, \tilde{H} (f_{1/k} - f_{1/l})}_{L^2(D)} = \|H_0^{1/2} (f_{1/k} - f_{1/l})\|_{L^2(\R)}^2 \to 0$ as $k, l \to \infty$, that is, $f_{1/k}$ satisfies condition~(a). Hence, it remains to prove the claim.

Observe that
\formula{
 \fourier f_\eps(\xi) & = (1 + \eps) \fourier g_\eps((1 + \eps) \xi) = (1 + \eps) \fourier h(\eps (1 + \eps) \xi) \fourier f((1 + \eps) \xi) ,
}
and that $\fourier f$ is a continuous function. Hence $\fourier f_\eps(\xi)$ converges pointwise to $\fourier f(\xi)$. Furthermore, we have $0 \le \fourier h(\xi) \le 1$ and $0 \le v(\xi) \le v((1 + \eps) \xi)$, so that
\formula{
  v(\xi) |\fourier f_\eps(\xi) - \fourier f(\xi)|^2 & = v(\xi) \bigl| (1 + \eps) \fourier h(\eps (1 + \eps) \xi) \fourier f((1 + \eps) \xi) - \fourier f(\xi) \bigr|^2 d\xi \\
 & \le 2 v(\xi) \bigl| (1 + \eps) \fourier h(\eps (1 + \eps) \xi) \fourier f((1 + \eps) \xi) \bigr|^2 + 2 v(\xi) |\fourier f(\xi)|^2 \\
 & \le 2 (1 + \eps)^2 v((1 + \eps) \xi) |\fourier f((1 + \eps) \xi)|^2 + 2 v(\xi) |\fourier f(\xi)|^2 .
}
By the assumption~(a3), $v(\xi) |\fourier f(\xi)|^2$ is integrable. Therefore, the family of functions $v(\xi) |\fourier f_\eps(\xi) - \fourier f(\xi)|^2$ ($\eps \in (0, 1)$) is tight and uniformly integrable. This proves our claim.
\end{proof}

The properties of $H$ for multi-dimensional domains $D$ have been studied in~\cite{bib:cks12, bib:r02}, which adopt a probabilistic definition of $H$; see Subsection~\ref{subsec:fk} for further discussion. 

\subsection{Scaling}

A significant simplification of the spectral problem for $H$ is possible thanks to the well-known scaling properties of the operators $H_0$ and $H$. Below we show that the number of parameters can be reduced to one, and we may assume that, for example, $c = \hbar = m = 1$, with the only free parameter $a$, half of the width of the potential well~\eqref{eq:v}. This procedure is standard, see, for example, \cite{bib:cks12, bib:r02}, and in physical terms, this is simply the choice of natural units.

Recall that $D = (-a, a)$. In this paragraph only, we denote by $\bar{H}$, $\bar{H}_0$, $\bar{\Phi}_n$, $\bar{E}_n$ and $\bar{D}$ the corresponding objects with $\hbar, c, m, a$ replaced by $1, 1, 1, \bar{a} = \hbar^{-1} m c a$, respectively. By~\eqref{eq:hamiltonian0} (or~\eqref{eq:hamiltonian0f}), if $f(x) = \bar{f}(\hbar^{-1} m c x)$, then
\formula{
 H_0 f(x) & = m c^2 \bar{H}_0 \bar{f}(\hbar^{-1} m c x) .
}
In particular, this means that $f \in \domain(H_0)$ if and only if $\bar{f} \in \domain(\bar{H}_0)$. By Definition~\ref{def:h},
\formula{
 H f(x) & = m c^2 \bar{H} \bar{f}(\hbar^{-1} m c x) .
}
It follows that
\formula[eq:scaling]{
 \Phi_n(x) & = \sqrt{\frac{m c}{\hbar}} \, \bar{\Phi}_n\expr{\frac{m c x}{\hbar}} && \text{and} & E_n & = m c^2 \bar{E}_n .
}
We have thus proved that it suffices to consider only the one-parameter family of Hamiltonians, corresponding to $c = \hbar = m = 1$ and $a > 0$.

It is easy to check that the above scaling is consistent with the results given in Theorems~\ref{th1} and~\ref{th2}, and Propositions~\ref{prop:simplebounds}--\ref{prop:loo}. Therefore, it is sufficient to prove them only when $c = \hbar = m = 1$. For easier reference, we state the simplified versions of main theorems.

\begin{theorem1}
Let $E_n$ be the sequence of energy levels of the Hamiltonian
\formula[eq:h]{
 H & = \expr{-\frac{d^2}{d x^2} + 1}^{1/2} + V_\well(x) ,
}
where $V_\well(x) = 0$ for $(-a, a)$ and $V_\well(x) = \infty$ otherwise. Then $E_n$ are non-degenerate, and
\formula[eq:asymp1prime]{
 E_n & = \frac{n \pi}{2 a} - \frac{\pi}{8 a} + O\expr{\frac{1}{n}} && \text{as $n \to \infty$.}
}
\end{theorem1}

\begin{theorem2}
For $n \ge 1$, let $\tilde{\mu}_n$ be the unique solution of
\formula[eq:tildemu0prime]{
 \thet_{\tilde{\mu}_n} = \frac{n \pi}{2} - a \tilde{\mu}_n ,
}
where
\formula{
 \thet_\mu & = \frac{1}{\pi} \int_0^\infty \frac{\mu}{s^2 - \mu^2} \, \log \expr{\frac{1}{2} + \frac{1}{2} \sqrt{\frac{1 + s^2}{1 + \mu^2}}} ds .
}
Then
\formula[eq:asymp2prime]{
 \abs{E_n - \sqrt{\tilde{\mu}_n^2 + 1}} & < \frac{8}{n} \expr{3 + \frac{1}{a}} \exp\expr{-\frac{a}{3}} .
}
\end{theorem2}

To simplify the notation, whenever possible, we assume that $c = \hbar = m = 1$; for example, we prove Theorems~\ref{th1}' and~\ref{th2}' rather than their original versions. The general, non-scaled case only needs to be considered in the verification of Remark~\ref{rem:limits}.

\subsection{Free Hamiltonian and its Markov semigroup}
\label{subsec:free}

The evolution of a quantum state is described by the group of unitary operators $\exp(i t H)$ generated by the Hamiltonian $H$. However, when studying the spectral theory of $H$, it is often more convenient to work with the semigroup of self-adjoint contractions $\exp(-t H)$, $t \ge 0$ (note that all of the operators $H$, $\exp(i t H)$ and $\exp(-t H)$ have the same eigenfunctions). It turns out that, as in the non-relativistic case, this semigroup is a sub-Markov semigroup, related to a certain stochastic process. Using this semigroup approach, one can relatively easily obtain some simple properties of $H$ and its eigenvalues. Below we only state the results, and we refer the reader to~\cite[Chapter~5]{bib:bbkrsv09} and to~\cite{bib:cms90} for the full exposition.

In this subsection, we focus on the free Hamiltonian (in natural units $c = \hbar = m = 1$)
\formula[eq:h0]{
 H_0 & = \expr{-\frac{d^2}{d x^2} + 1}^{1/2} ,
}
and the closely related operator
\formula{
 A_0 & = H_0 - 1 = \expr{-\frac{d^2}{d x^2} + 1}^{1/2} - 1 = w\expr{-\frac{d^2}{dx^2}} ,
}
where for brevity we let
\formula{
 w(\xi) & = \sqrt{\xi + 1} - 1 .
}
Note that $A_0$ measures the total energy (given by the Hamiltonian $H_0$) minus the rest energy $m c^2 = 1$. For this reason, it is often called the \emph{kinetic energy} operator. Both $H_0$ and $A_0$ have the same domain, and are nonnegative-definite on $L^2(\R)$ (cf.~\eqref{eq:hpositive}). Their spectra are $\sigma(H_0) = [1, \infty)$ and $\sigma(A_0) = [0, \infty)$, and both operators are Fourier multipliers with symbols, respectively, $(\xi^2 + 1)^{1/2}$ and $w(\xi^2) = (\xi^2 + 1)^{1/2} - 1$. 

The operator $-A_0$ generates a strongly continuous semigroup of self-adjoint contractions
\formula{
 T_0(t) & = \exp(-t A_0) ,
}
with $t \ge 0$; $T_0(0)$ is the identity operator, $T_0(t)$ is the Fourier multiplier with symbol $\exp(-t w(\xi^2))$, and
\formula[eq:a0gen]{
\begin{aligned}
 \domain(A_0) & = \set{f \in L^2(\R) : \lim_{t \to 0^+} \frac{T_0(t) f - f}{t} \text{ exists in } L^2(\R)} , \\ -A_0 f & = \lim_{t \to 0^+} \frac{T_0(t) f - f}{t} \, .
\end{aligned}
}
For $t > 0$, the operator $T_0(t)$ is a convolution operator, and it has an everywhere positive, continuous, symmetric kernel function $T_0(t; x - y)$: for all $f \in L^2(\R)$,
\formula[eq:t0]{
 T_0(t) f(x) & = \int_{-\infty}^\infty T_0(t; x - y) f(y) dy .
}
The Fourier transform (in the $x$ variable) of $T_0(t; x)$ is $\exp(-t w(\xi^2))$, and $T_0(t; x)$ is integrable function, with total mass $1$. Hence, $T_0(t)$ is a \emph{Markov operator}, and formula~\eqref{eq:t0} defines a contraction on every $L^p(\R)$ ($p \in [1, \infty]$), and also on $C_0(\R)$, the space of continuous functions on $\R$ convergent to $0$ at $\pm \infty$. In each of these Banach spaces, we can define the generator of the semigroup $T_0(t)$ in a similar way as in~\eqref{eq:a0gen}; for example, 
\formula[eq:a0gen2]{
\begin{aligned}
 \domain(A_0; C_0(\R)) & = \set{f \in C_0(\R) : \lim_{t \to 0^+} \frac{T_0(t) f - f}{t} \text{ exists in } C_0(\R)} , \\
 -A_0 f & = \lim_{t \to 0^+} \frac{T_0(t) f - f}{t} \, .
\end{aligned}
}
We abuse the notation and use the same symbol $-A_0$ for the generator of the semigroup $T_0(t)$ in any of the spaces $L^p(\R)$ or $C_0(\R)$. This causes no confusion: if $f \in \domain(A_0; L^p(\R))$ for different values of $p$, then $A_0 f$ does not depend on the choice of $p$. Also, if $f \in \domain(A_0; L^p(\R)) \cap \domain(A_0; C_0(\R))$, then both definitions of $A_0 f$ coincide. Finally, $C_c^\infty(\R)$ is contained in $\domain(A_0, L^p(\R))$ ($p \in [1, \infty]$) and in $\domain(A_0; C_0(\R))$, and it is the core of $A_0$ in each of these Banach spaces except $L^\infty(\R)$. Whenever we write $\domain(A_0)$, we mean $\domain(A_0; L^2(\R))$.

For $f \in C_c^\infty(\R)$ we have, by Fourier inversion formula,
\formula[eq:a0pw]{
 A_0 f(x) & = \pvint_{-\infty}^\infty (f(x) - f(x + z)) \nu(z) dz ,
}
where
\formula[eq:nu]{
 \nu(z) & = \frac{1}{\pi} \, \frac{K_1(|z|)}{|z|} \, ,
}
$K_1$ is the modified Bessel function of the second kind, and `$\pvint$' denotes the Cauchy principal value integral:
\formula{
 \pvint_{-\infty}^\infty (f(x) - f(x + z)) \nu(z) dz & = \lim_{\eps \to 0^+} \int_{\R \setminus (-\eps, \eps)} (f(x) - f(x + z)) \nu(z) dz .
}
In particular, $A_0$ is a non-local operator; see~\cite[Lemma~2]{bib:r02} for a proof of~\eqref{eq:a0pw} and~\eqref{eq:nu}.

We choose, following~\cite{bib:cms90} and others, to work with the semigroup generated by $-A_0$ rather than $-H_0$, because the constant $1$ is an invariant function for $T_0(t) = \exp(-t A_0)$. This choice is not essential, since the relation $\exp(-t H_0) = e^{-t} T_0(t)$ enables one to switch freely between the two semigroups.

\subsection{Relativistic L{\'e}vy process}
\label{subsec:fk}

Recall that the Hamiltonian for the quasi-relativistic particle in the box model is given by $H = H_0 + V_\well(x)$, formally defined in Definition~\ref{def:h}. We define in a similar manner $A = A_0 + V_\well(x)$, and we let $T(t) = \exp(-t A)$. It is easy to see that $A = H - 1$ and $T(t) = e^t \exp(-t H)$. We denote, as usual, $D = (-a, a)$.

In this subsection, we give an alternative construction of $H$, $A$ and $T(t)$, involving some theory of Markov processes. Probabilistic concepts described here are never used outside this subsection. Hence, the reader unfamiliar with Markov processes may freely skip this part, noting only that the operators $T(t)$ admit a kernel function $T(t; x, y)$ such that $0 \le T(t; x, y) \le T_0(t; x - y)$ and $T(t; x, y) = 0$ whenever $x$ or $y$ is outside $D$. Note that here $D = (-a, a)$, but the extension to more general open sets (in particular, to $D = (0, \infty)$, corresponding to a single potential wall $V_\wall(x)$) is straightforward.

The operator $-A_0$ (acting on the $C_0(\R)$ space) is the Feller generator of a stochastic process $X(t)$, called the \emph{relativistic L{\'e}vy process}, or simply \emph{relativistic process}. It is a L{\'e}vy process: it has independent and stationary increments, and c{\`a}dl{\`a}g (right-continuous, with left-limits) paths. In fact, $X(t)$ is a subordinate Brownian motion; see~\cite[Chapter~5]{bib:bbkrsv09}. We write $\pr_x$ for the probability law for the process starting at $x \in \R$, and $\ex_x$ for the corresponding expectation. Hence, $\pr_x(X(t) \in dy) = T_0(t; x - y) dy$; that is, $T_0(t; x - y)$ is the \emph{transition density} of $X(t)$. The paths of $X(t)$ are discontinuous: $X(t)$ is a jump-type process (the intensity of jumps, or the \emph{L{\'e}vy measure}, is given by $\nu(y) dy$, see~\eqref{eq:nu}). Let $D = (-a, a)$, and define the \emph{time of first exit from $D$}
\formula{
 \tau_D & = \inf \set{t \ge 0 : X(t) \notin D} .
}
Let $X^D(t)$ be the \emph{part of $X(t)$ in $D$}, or \emph{$X(t)$ killed upon leaving $D$}: the (sub-Markovian) stochastic process with lifetime $\tau_D$, equal to $X(t)$ for $t \in [0, \tau_D)$. Then for $t > 0$ and $x, y \in \R$,  the transition density of $X_t^D$ is given by \emph{Hunt's formula},
\formula{
 T(t; x, y) & = T_0(t; x - y) - \ex_x(T_0(t - \tau_D; X(\tau_D) - y); t \ge \tau_D) .
}
It is known that $T(t; x, y) = T(t; y, x)$, $T(t; x, y) = 0$ whenever $x \notin D$ or $y \notin D$, and $0 \le T(t; x, y) \le T_0(t; x, y)$. The corresponding semigroup of sub-Markov operators is
\formula{
 T(t) f(x) & = \int_{-\infty}^\infty T(t; x, y) f(y) dy = \ex_x(f(X(t)) \ind_{t < \tau_D})
}
for $f \in L^p(\R)$ ($p \in [1, \infty]$) or $f \in C_0(\R)$. Since $T(t) f = 0$ for any $f$ supported outside $D$, $T(t)$ is not strongly continuous in $L^2(\R)$. It is, however, strongly continuous in $L^2(D)$ (with the usual identification of $L^2(D)$ as a subspace of $L^2(\R)$). On its $L^2(D)$ domain, the generator $-A$ of $T(t)$ coincides with the operator $-(H - 1)$, where $H$ is defined analytically in Definition~\ref{def:h}. Again, the definition of $A$ can be extended to $L^p(D)$ ($p \in [1, \infty]$) and $C_0(D)$, with appropriate domains. We remark that $T(t)$ maps $C_0(D)$ to $C_0(D)$ also for $D = (0, \infty)$, but not for all open sets $D \sub \R$ ($D = \R \setminus \{0\}$ being a simple counterexample).

Finally, the semigroup $T(t)$ can be thought as a Feynman-Kac transform of $T_0(t)$, corresponding to the potential $V_\well(x)$. Indeed, since $\{t : X(t) \notin D\} \cap (\tau_D, \tau_D + \eps)$ almost surely has positive Lebesgue measure, the functional $\int_0^t V_\well(X(s)) ds$ is almost surely equal to zero when $t \le \tau_D$, and infinity otherwise. (The last sentence remains true for $D = (0, \infty)$, which corresponds to the single potential wall $V_\wall(x)$; however, it is not true for general open sets $D$.) Hence,
\formula[eq:fk]{
 T(t) f(x) & = \ex_x \expr{\exp\expr{-\int_0^t V_\well(X(s)) ds} f(X(t))} ,
}
with $\exp(-\infty)$ understood to be $0$. If $V(x)$ is bounded, $A_0 + V(x)$ can be understood literally (because multiplication by $V(x)$ is well-defined on $\domain(A_0)$), and a formula similar to~\eqref{eq:fk} defines a semigroup generated by $A_0 + V(x)$ (see~\cite{bib:cms90, bib:ks06}). In the case of the potential well, $A_0 + V_\well(x)$ is formally not well-defined, but~\eqref{eq:fk} still makes sense, and it is natural to understand $A_0 + V_\well(x)$ as the generator of the semigroup~\eqref{eq:fk}. This explains the notation used in~\eqref{eq:hamiltonian}, \eqref{eq:h} and Definition~\ref{def:h}.

We turn back to the analytical setting, and refer the interested reader to~\cite{bib:cks12, bib:ks06, bib:r02} for recent developments on the relativistic process $X(t)$, or to~\cite{bib:ksv11} and the references therein for research related to processes $X(t)$ corresponding to more general Bernstein functions $w(\xi)$. For various estimates of $T(t; x, y)$ and the Green function for $A$ in the quasi-relativistic setting (for multi-dimensional domains $D$), see~\cite{bib:r02, bib:cks12}. Similar estimates for some other Hamiltonians $H_0$ can be found in~\cite{bib:b00, bib:bgr10, bib:cks10, bib:ksv10, bib:k97}.

\subsection{Eigenfunctions and eigenvalues}
\label{subsec:green}

Some spectral properties of $A$ and $H$ follow immediately from the general theory of strongly continuous semigroups of compact self-adjoint Markov operators. There is a complete orthonormal set of eigenfunctions $\Phi_n \in L^2(D)$ of $T(t)$, and the corresponding eigenvalues have the form $e^{-t \lambda_n}$, where the sequence $\lambda_n$ is nondecreasing and unbounded. Furthermore, $\Phi_n \in L^\infty(D)$ (because the kernel $T(t; x, y)$ is bounded for each $t > 0$), and even $\Phi_n \in C_0(D)$ (by an argument used for example in~\cite{bib:cz95} for the non-relativistic case). By the symmetry of $D = (-a, a)$, we have $T(t; x, y) = T(t; -x, -y)$, and hence the spaces of odd and even $L^2(D)$ functions are invariant under the action of $T(t)$. Therefore, we may assume that every $\Phi_n$ is either an odd or even function. The \emph{ground state eigenvalue} $\lambda_1$ is positive and simple, and the corresponding \emph{ground state eigenfunction} has constant sign in $D$; we choose it to be positive in $D$. The functions $\Phi_n$ are also the eigenfunctions of $A$, the generator of the semigroup $T(t)$, and $H = A + 1$; the corresponding eigenvalues are $\lambda_n$ and $E_n = \lambda_n + 1$, respectively. Therefore, we have
\formula{
 & 0 < \lambda_1 < \lambda_2 \le \lambda_3 \le ... \to \infty , && \text{or, equivalently,} && 1 < E_1 < E_2 \le E_3 \le ... \to \infty .
}
Relatively little is known about $\lambda_n$, $E_n$ and $\Phi_n$, in particular, no closed-form formulae are available. The asymptotic result $\lambda_n \sim n \pi / (2 a)$, already mentioned in Introduction, follows directly from a very general result of~\cite{bib:bg59} (see Theorem~2.3 therein). Best known estimates of $\lambda_n$ are found in~\cite[Theorem~4.4]{bib:cs05}, where it is proved, also in a much more general setting, that:
\formula[eq:cs]{
 \frac{1}{2} \expr{\sqrt{\expr{\frac{n \pi}{2 a} }^2 + 1} - 1} \le \lambda_n & \le \sqrt{\expr{\frac{n \pi}{2 a}}^2 + 1} - 1 .
}
In the non-natural units, the above estimate takes the form stated in Introduction in~\eqref{eq:cs1}. The upper bound in~\eqref{eq:cs} is a rather simple consequence of \emph{operator monotonicity} of the function $w(\xi) = (\xi + 1)^{1/2} - 1$, while the lower bound requires much more effort. In the proof of Theorems~\ref{th1} and~\ref{th2} below we only use the upper bound of~\eqref{eq:cs} for $n \le 2$ to show simplicity of $\lambda_2$. We also apply~\eqref{eq:cs} for small $n$ in the proof of Proposition~\ref{prop:simplebounds}. It should be pointed out that for large $n$ or large $a$, the estimates in Theorem~\ref{th2} are much better than both~\eqref{eq:cs2} and~\eqref{eq:cs1}. They also improve bounds for \emph{sums} of eigenvalues $E_n$, obtained in~\cite{bib:hy09}.

\subsection{Essential self-adjointness}
\label{subsec:esa}

The following interesting result seems to be new. It provides an important property of the quasi-relativistic particle in the box model, which contrasts sharply with the non-relativistic setting. For this reason we decided to include it, despite the fact that it is not required for the main results of the article (it is referred to only in the remark following Definition~\ref{def:h}).

\begin{proposition}
\label{prop:esa}
The operator $\tilde{H}$ given in Definition~\ref{def:h} is essentially self-adjoint on $L^2(D)$.
\end{proposition}

\begin{proof}
Denote by $J_0 = (-d^2 / dx^2)^{1/2}$ the one-dimensional fractional Laplace operator, and let $\tilde{J}$ and $J$ be defined as in the first part of Definition~\ref{def:h}, but with $H_0$ replaced by $J_0$. On $C_c^\infty(\R)$, the operator $J_0 - H_0$ is a Fourier multiplier with bounded symbol $|\xi| - (\xi^2 + 1)^{1/2}$, and hence it extends to a bounded operator on $L^2(\R)$. Therefore, also $\tilde{J} - \tilde{H}$ extends to a bounded operator on $L^2(D)$. Hence $\tilde{J}$ is essentially self-adjoint if and only if $\tilde{H}$ is essentially self-adjoint. Below we prove that $\tilde{J}$ is essentially self-adjoint. (With some effort, the following argument could be modified to work directly with $H$; however, the theory for the fractional Laplace operator $J_0$ is much better developed than the corresponding one for $H_0$, so references to literature are much easier in the case of $J_0$.)

Clearly, $\tilde{J}$ is a symmetric unbounded operator on $L^2(D)$, with domain $\domain(\tilde{J}) = C_c^\infty(D)$. Furthermore, there is $\eps > 0$ such that $\|\tilde{J} f\|_{L^2(D)} \ge \eps \|f\|_{L^2(D)}$ for $f \in \domain(\tilde{J})$ (because the ground state eigenvalue of $J$ is positive; see~\cite{bib:bk04, bib:bk09, bib:k98, bib:cs97}). Hence, $\tilde{J}^{-1}$ is a bounded operator defined on the range of $\tilde{J}$. We claim that the range of $\tilde{J}$ is dense in $L^2(D)$. Once this is proved, it follows that the closure of $\tilde{J}^{-1}$ is a bounded self-adjoint operator on entire $L^2(D)$. This means that the closure of $\tilde{J}$ is the inverse of a bounded self-adjoint operator, and the desired result follows.

Suppose, contrary to our claim, that there is a nonzero $g \in L^2(D)$ orthogonal to the range of $\tilde{J}$. Then for any test function $f \in C_c^\infty(D)$,
\formula{
 0 & = \tscalar{g, \tilde{J} f}_{L^2(D)} = \tscalar{g, J_0 f}_{L^2(\R)} .
}
By Plancherel's theorem and Fourier inversion formula, the right term is equal to \mbox{$J_0 (\check{f} * g)(0)$}, where $\check{f}(y) = f(-y)$ (note that $\check{f} * g \in C_c^\infty(\R) \sub \domain(J_0)$). Replacing $f$ by $\check{f}$, we obtain that for all $f \in C_c^\infty(D)$,
\formula{
 J_0 (f * g)(0) = 0 .
}
Let $h_\eps \in C_c^\infty(D)$ ($\eps \in (0, a/2)$) be an approximation to the identity: $\supp h_\eps \sub [-\eps, \eps]$, $h_\eps \ge 0$ and $\int h_\eps(x) dx = 1$. Furthermore, we assume that $h_\eps \le 1 / \eps$ and $h_\eps(-y) = h_\eps(y)$. For $x \in (-a + \eps, a - \eps)$ the function $f(y) = h_\eps(x + y)$ is in $C_c^\infty(D)$. Hence,
\formula{
 0 & = J_0 (f * g)(0) = J_0 (h_\eps * g)(x) .
}
That is, $J_0 (h_\eps * g)(x) = 0$ for $x \in (-a + \eps, a - \eps)$.

In the remaining part of the proof we use the notion of harmonicity with respect to $J_0$ (or with respect to the corresponding stochastic process, the symmetric $1$-stable L{\'e}vy process), explained in detail, for example, in~\cite{bib:bb99} (see~\cite{bib:cks12, bib:r02} for the closely related notion of harmonicity with respect to $H_0$). Since in this article the concept of harmonicity is used only in this proof, we are very brief here. A function $h$ is said to be \emph{$J_0$-harmonic} in a domain $U \sub \R$ if for all intervals $[x_0 - r, x_0 + r] \sub U$ and all $x \in (x_0 - r, x_0 + r)$ it satisfies a mean-value property
\formula{
 h(x) & = \int_{\R \setminus [x_0 - r, x_0 + r]} h(y) P_{x_0, r}(x, y) dy ,
}
where $P_{x_0, r}(x, y)$ is the \emph{Poisson kernel for $J_0$}. A closed-form formula for $P_{x_0, r}(x, y)$ is known due to M.~Riesz (\cite{bib:r38}), we have
\formula{
 P_{x_0, r}(x, y) & = \frac{1}{\pi} \, \sqrt{\frac{r^2 - |x - x_0|^2}{|y - x_0|^2 - r^2}} \, \frac{1}{|x - y|} \, .
}
Here $|x - x_0| < r < |y - x_0|$; if this condition is not satisfied, we take $P_{x_0, r}(x, y) = 0$.

It is proved in~\cite[Theorem~3.12]{bib:bb99}, that \emph{weak harmonicity} of $h_\eps * g$ (that is, the condition $J_0 (h_\eps * g)(x) = 0$ for $x \in (-a + \eps, a - \eps)$) implies (strong) harmonicity. Hence, for all $x \in (-a + 2 \eps, a - 2 \eps)$,
\formula{
 h_\eps * g(x) & = \int_{\R \setminus (-a + 2 \eps, a - 2 \eps)} P_{0, a - 2 \eps}(x, y) h_\eps * g(y) dy \\
 & = \int_{\R \setminus (-a + 2 \eps, a - 2 \eps)} \int_{(-a, a)} P_{0, a - 2 \eps}(x, y) h_\eps(y - z) g(z) dz dy .
}
By the explicit formula for the Poisson kernel,
\formula{
 P_{0, a - 2 \eps}(x, y) & \le \frac{1}{\pi} \, \sqrt{\frac{2 a (a - 2 \eps - |x|)}{a (|y| - (a - 2 \eps)}} \, \frac{1}{a - 2 \eps - |x|} = \frac{\sqrt{2}}{\pi} \, \frac{1}{\sqrt{a - 2 \eps - |x|}} \, \frac{1}{\sqrt{|y| - (a - 2 \eps)}} \, .
}
The properties of $h_\eps$ and a short calculation yield that for some constant $C$,
\formula{
 \int_{\R \setminus (-a + 2 \eps, a - 2 \eps)} P_{0, a - 2 \eps}(x, y) h_\eps(y - z) dy & \le \frac{C}{\sqrt{a - 2 \eps - |x|}} \, \frac{1}{\sqrt{\eps}} \, \ind_{\R \setminus (-a + 3 \eps, a - 3 \eps)}(z) ;
}
see~\cite{bib:b97, bib:bb99, bib:bkk08} for a similar \emph{regularization of the Poisson kernel}. We conclude that
\formula{
 |h_\eps * g(x)| & \le \int_{(-a, a)} \expr{\int_{\R \setminus (-a + 2 \eps, a - 2 \eps)} P_{0, a - 2 \eps}(x, y) h_\eps(y - z) dy} |g(z)| dz \\
 & \le \frac{C}{\sqrt{a - 2 \eps - |x|}} \, \frac{1}{\sqrt{\eps}} \int_{(-a, a)} |g(z)| \ind_{\R \setminus (-a + 3 \eps, a - 3 \eps)}(z) dz .
}
By Schwarz inequality,
\formula{
 |h_\eps * g(x)| & \le \frac{C}{\sqrt{a - 2 \eps - |x|}} \, \frac{1}{\sqrt{\eps}} \int_{(-a, a) \setminus (-a + 3 \eps, a - 3 \eps)} |g(z)| dz \\
 & \le \frac{C}{\sqrt{a - 2 \eps - |x|}} \, \sqrt{6} \expr{\int_{(-a, a) \setminus (-a + 3 \eps, a - 3 \eps)} |g(z)|^2 dz}^{1/2} .
}
Let $\eps = 1/n$ and $n \to \infty$. The left-hand side converges to $|g(x)|$ (for almost all $x \in D$), and the right-hand side tends to $0$. It follows that $g(x) = 0$ for almost all $x \in D$, contrary to our assumption that $g$ is nonzero. The proposition is proved.
\end{proof}

Noteworthy, the above argument works also for the fractional Laplace operator $(-d^2 / dx^2)^{s}$ and the related operator $(-d^2 / dx^2 + 1)^{s}$, but only when $s \in (0, 1/2]$. For $s \in (1/2, 1]$, Proposition~\ref{prop:esa} fails: the corresponding operators are not essentially self-adjoint on $L^2(D)$.

%
%                            ---------- o ----------
%

\section{Quasi-relativistic particle in the presence of a single infinite potential wall}
\label{sec:halfline}

The main subject of this article is the quasi-relativistic particle in an infinite potential well. As described in Introduction, a potential well can be viewed as a superposition of two infinite potential walls, and the case of a single infinite potential wall was recently solved in~\cite{bib:k10, bib:kmr11}. The framework of these papers is purely mathematical. In this section we state some of the results of~\cite{bib:k10, bib:kmr11} in the context of quasi-relativistic particles.

\subsection{Non-relativistic case}

It seems instructive to begin with the well-understood classical, non-relativistic model, where the free Hamiltonian is given by $H_0 = -d^2 / dx^2$ (for simplicity, we omit the multiplicative constant $\hbar / (2 m)$). The eigenfunctions of $H_0$ are the free wave solutions $e^{i \mu x}$ ($\mu \in \R$), corresponding to energies $E = \mu^2$. These eigenfunctions are not $L^2(\R)$ functions. Nevertheless, any (complex-valued) $L^2(\R)$ function $\Phi$ can be written as a `mixture' of free waves (or a `wave packet') by means of the Fourier transform, at least when $\fourier \Phi$ is integrable:
\formula{
 \Phi(x) & = \frac{1}{2 \pi} \int_{-\infty}^\infty \fourier \Phi(\mu) e^{i \mu x} d\mu .
}
On its domain, $H_0$ acts under the integral sign,
\formula{
 H_0 \Phi(x) & = \frac{1}{2 \pi} \int_{-\infty}^\infty \mu^2 \fourier \Phi(\mu) e^{i \mu x} d\mu ,
}
at least when $\mu^2 \fourier \Phi(\mu)$ is integrable. In other words, the Fourier transform diagonalizes the action of $H_0$: for any $\Phi \in \domain(H_0)$,
\formula{
 \fourier (H_0 \Phi)(\mu) & = \mu^2 \fourier \Phi(\mu) .
}
For real-valued functions, one can replace complex-valued free waves $e^{i \mu x}$ ($\mu \in \R$) by real-valued ones, $\sin(\mu x)$ and $\cos(\mu x)$ ($\mu > 0$).

Consider now a non-relativistic model with an infinite potential wall $V_\wall(x) = 0$ for $x > 0$, $V_\wall(x) = \infty$ otherwise. The corresponding Hamiltonian $H_\wall = H_0 + V_\wall(x)$ is defined as in the second part of Definition~\ref{def:h} (with non-relativistic $H_0$ and $D = (0, \infty)$), and it is the Dirichlet Laplace operator in $(0, \infty)$. The eigenfunctions of $H_\wall$ must vanish in $(-\infty, 0]$ and they are eigenfunctions of $H_0$ on $(0, \infty)$ (here we use the locality of the operator $H_0$). Therefore, $F_\mu(x) = \sin(\mu x) \ind_{(0, \infty)}(x)$ ($\mu > 0$) is the eigenfunction of $H_\wall$ corresponding to the energy $E = \mu^2$. Again, $F_\mu \notin L^2((0, \infty))$, but the Fourier sine transform
\formula{
 \fourier_{\sin} f(\mu) & = \int_0^\infty f(x) \sin(\mu x) dx = \int_0^\infty f(x) F_\mu(x) dx
}
diagonalizes the action of $H_\wall$. More precisely, $(2 / \pi)^{1/2} \fourier_{\sin}$ extends to a unitary operator on $L^2((0, \infty))$, and for all $\Phi \in \domain(H)$,
\formula{
 \fourier_{\sin} (H_\wall \Phi)(\mu) & = \mu^2 \fourier_{\sin} \Phi(\mu) .
}

The free wave solutions both for $H_0$ and for $H_\wall$ are not square-integrable, so that, strictly speaking, they are not eigenstates of $H_0$ and $H_\wall$. Instead, they are \emph{generalized eigenfunctions}, or \emph{continuum eigenfunctions}, of $H_0$ and $H_\wall$, and the integral decomposition of any state into the free waves is called \emph{generalized eigenfunction expansion}. The primary goal of~\cite{bib:k10, bib:kmr11} was to provide similar generalized eigenfunction expansion for a class of Hamiltonians $H_\wall = H_0 + V_\wall(x)$ with a single infinite potential wall, based on non-local free Hamiltonians $H_0$.

\subsection{Quasi-relativistic case}

We return to the quasi-relativistic setting. We assume that $c = \hbar = m = 1$ (scaling), so that $H_0 = (-d^2 / dx^2 + 1)^{1/2}$ is a Fourier multiplier with symbol $(\xi^2 + 1)^{1/2}$. As in the non-relativistic case, the free wave solutions $e^{i \mu x}$ for $\mu \in \R$ (or $\cos(\mu x)$ and $\sin(\mu x)$ for $\mu > 0$, when one prefers real-valued functions), are the generalized eigenfunctions of $H_0$. The only difference here is in the corresponding energies, which are now given by $E = (\mu^2 + 1)^{1/2}$ instead of the classical expression $E = \mu^2$.

With the presence of the infinite potential wall $V_\wall(x) = 0$ for $x > 0$, $V_\wall(x) = \infty$ otherwise, the situation is more complicated due to non-locality of $H_0$. Nevertheless, it can be expected that the effect of non-locality decays as the distance to the potential wall increases, that is, the generalized eigenfunction corresponding to the energy $E = (\mu^2 + 1)^{1/2}$ should be asymptotically equal to $\sin(\mu x + \thet_\mu)$ as $x \to \infty$, for some $\thet_\mu \in [0, 2 \pi)$. This is indeed the case, and the phase shift $\thet_\mu$ is given by~\eqref{eq:thetmu}.

The operator $H_\wall = H_0 + V_\wall(x)$ is defined in the second part of Definition~\ref{def:h}, with $D = (0, \infty)$. The results of Subsection~\ref{subsec:fk} extend to the case of a single potential wall; in particular, $H_\wall$ acts (as an unbounded operator) on any of the spaces $L^p((0, \infty))$ ($p \in [1, \infty]$) or $C_0((0, \infty))$, with the corresponding domain denoted by $\domain(H_\wall; L^p((0, \infty)))$ etc. We remark that also Proposition~\ref{prop:esa} holds true for the single potential wall, but the proof requires some modifications (roughly, one needs to consider $J_0 + 1$ instead of $J_0$).

As in Preliminaries, we define the kinetic energy operators $A_0 = H_0 - 1$ and $A_\wall = A_0 + V_\wall(x) = H_\wall - 1$. The following definition and theorem is a compilation of~\cite[Theorems~1.1, 1.3 and Example~6.2]{bib:k10} and~\cite[Theorem~1.10]{bib:kmr11}, applied to the operator monotone function $w(\xi) = (\xi + 1)^{1/2} - 1$. Note that in~\cite{bib:k10, bib:kmr11}, the function $w$ is denoted by the symbol $\psi$.

\begin{figure}
\begin{minipage}[b]{0.45\textwidth}
\centering
\includegraphics[width=\textwidth]{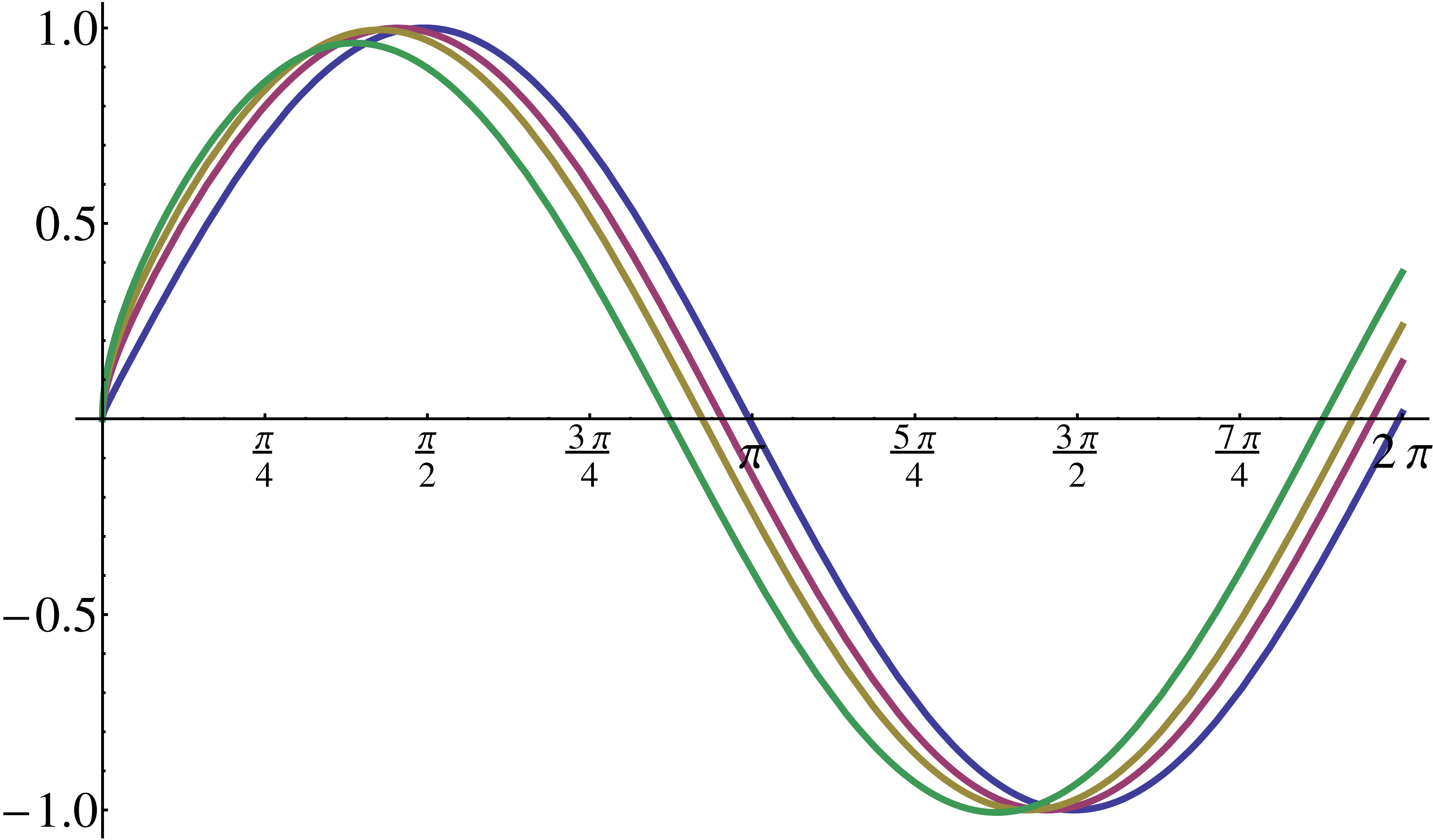}
\caption{Plot of $F_\mu(x / \mu)$ for $\mu = 1/20, 1/2, 1, 10$ (blue, magenta, yellow, green).}
\label{fig:f}
\end{minipage}
\hspace*{0.02\textwidth}
\begin{minipage}[b]{0.45\textwidth}
\includegraphics[width=\textwidth]{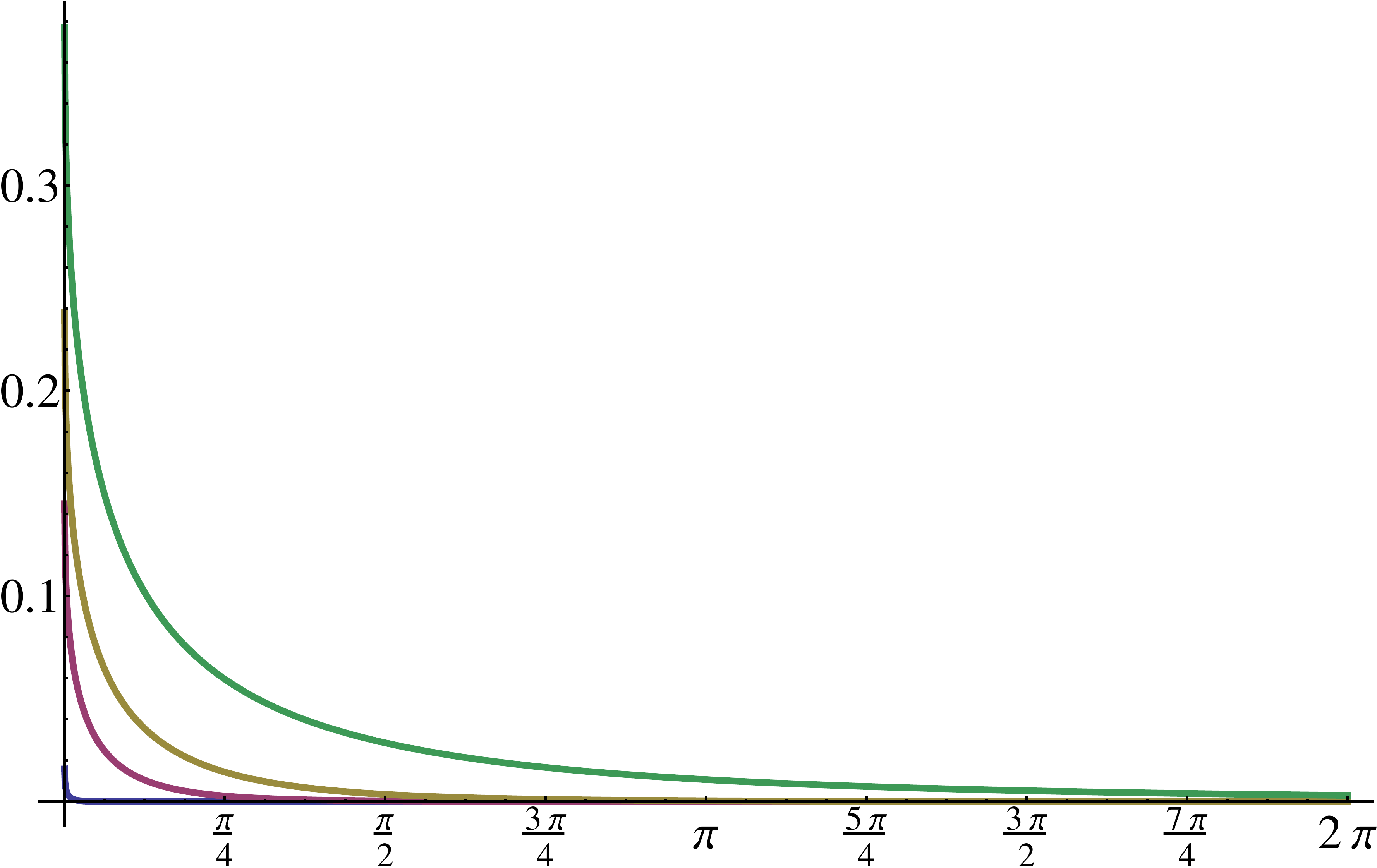}
\caption{Plot of $G_\mu(x / \mu)$ for $\mu = 1/20, 1/2, 1, 10$ (blue, magenta, yellow, green).}
\label{fig:g}
\end{minipage}
\end{figure}

\begin{definition}
\label{def:halfline}
For $\mu > 0$, let
\formula[eq:hf:f]{
 F_\mu(x) & = \sin(\mu x + \thet_\mu) \ind_{(0, \infty)}(x) - G_\mu(x) ,
}
where $\thet_\mu$ is given by~\eqref{eq:thetmu} and $G_\mu$ is a completely monotone function on $(0, \infty)$, given by the formula
\formula{
 G_\mu(x) & = \expr{\int_1^\infty e^{-x r} \gamma_\mu(r) dr} \ind_{(0, \infty)}(x) ,
}
with
\formula[eq:gamma]{
 \gamma_\mu(r) & = \frac{\sqrt{2}}{2 \pi} \, \frac{\mu}{\sqrt{1 + \mu^2}} \, \frac{\sqrt{r^2 - 1}}{\mu^2 + r^2} \, \exp \expr{-\frac{1}{\pi} \int_0^\infty \frac{r \log \bigl(1 + \sqrt{\frac{1 + s^2}{1 + \mu^2}} \bigr)}{r^2 + s^2} \, ds} \ind_{(1, \infty)}(r) .
}
\end{definition}

\begin{theorem}
\label{th:halfline}
The functions $F_\mu$ are generalized eigenfunctions of the Hamiltonian $H_\wall$ and the kinetic energy operator $A_\wall$ for the quasi-relativistic particle in the presence of a single infinite potential wall; the corresponding energies are $E = (\mu^2 + 1)^{1/2}$ (total energy) and $\lambda = (\mu^2 + 1)^{1/2} - 1$ (kinetic energy). The operators $H_\wall$ and $A_\wall$ admit the generalized eigenfunction expansion. More precisely, let
\formula{
 \Pi f(\mu) & = \int_0^\infty f(x) F_\mu(x) dx
}
for $f \in L^1(\R) \cap L^2(\R)$. Then $(2 / \pi)^{1/2} \Pi$ extends to a unitary mapping on $L^2((0, \infty))$. Furthermore, for $\Phi \in L^2((0, \infty))$,
\formula{
 \Phi \in \domain(H_\wall) & \iff \sqrt{\mu^2 + 1} \, \Pi \Phi(\mu) \in L^2((0, \infty)) ,
}
and if $\Phi \in \domain(H_\wall)$, then
\formula{
 \Pi (H_\wall \Phi)(\mu) & = \sqrt{\mu^2 + 1} \, \Pi \Phi(\mu) , & \Pi (A_\wall \Phi) & = \expr{\sqrt{\mu^2 + 1} - 1} \Pi \Phi(\mu) .
}
Furthermore, $F_\mu \in \domain(H_\wall, L^\infty((0, \infty)))$ for all $\mu > 0$, and $H_\wall F_\mu = (\mu^2 + 1)^{1/2} F_\mu$, $A_\wall F_\mu = ((\mu^2 + 1)^{1/2} - 1) F_\mu$.
\end{theorem}

The derivation of the formula for $F_\mu$ relies on a Wiener-Hopf method, and the proof of Theorem~\ref{th:halfline} involves complex variable methods; see~\cite{bib:k10, bib:kmr11} for details.

As a part of~\cite[Theorem~1.1]{bib:k10} we have $\thet_\mu \in [0, \pi/2)$. By~\cite[Proposition~4.17]{bib:k10}, $\thet_\mu$ is continuous and differentiable in $\mu > 0$. The formula for the Laplace transform of $F_\mu(x)$ is also given in \cite[Theorem~1.1]{bib:k10}, for $\xi \in \C$ with $\real \xi > 0$,
\formula[eq:laplace:gen]{
 \laplace F_\mu(\xi) & = \int_0^\infty F_\mu(x) e^{-\xi x} dx = \frac{\sqrt{2}}{2} \, \frac{\mu}{\mu^2 + \xi^2} \, \exp\expr{\frac{1}{\pi} \int_0^\infty \frac{\xi \log \expr{1 + \sqrt{\frac{1 + s^2}{1 + \mu^2}}}}{\xi^2 + s^2} \, ds} .
}
We need an estimate of $\laplace F_\mu(\xi)$. For real $\xi$ this is proved in~\cite[Corollary~5.1]{bib:kmr11}, which in our case gives
\formula[eq:laplaceest:gen]{
 \laplace F_\mu(\xi) & \le C \, \frac{\mu}{\mu^2 + \xi^2} \, \sqrt{1 + \sqrt{\frac{1 + \xi^2}{1 + \mu^2}}}
}
for $\xi > 0$ (here we can take $C = \sqrt{2}$; a reverse inequality also holds true with a different constant, but we only need the upper bound.) For complex $\xi$ with $\real \xi > 0$, by combining~\cite[Proposition~2.21(c) and Remark~4.12]{bib:k10}, we obtain
\formula{
 |\mu^2 + \xi^2| |\laplace F_\mu(\xi)| & \le C (\mu^2 + |\xi|^2) \laplace F_\mu(|\xi|)
}
(with constant $C = \sqrt{2}$). Here we use the fact that $(\mu^2 + \xi^2) \laplace F_\mu(\xi)$ is a complete Bernstein function, see~\cite[Proposition~2.19 and Lemma~3.8]{bib:k10}. This and~\eqref{eq:laplaceest:gen} give
\formula[eq:laplaceest2:gen]{
 |\laplace F_\mu(\xi)| & \le C \abs{\frac{\mu}{\mu^2 + \xi^2}} \, \sqrt{1 + \sqrt{\frac{1 + |\xi|^2}{1 + \mu^2}}}
}
when $\xi \in \C$, $\real \xi > 0$ (we can take $C = 2$ here).

%
%                            ---------- o ----------
%

\section{Proofs}
\label{sec:proofs}

This section is modeled after~\cite[Appendix~C and Sections~8--10]{bib:kkms10}. Some of the results of this part do not depend on detailed properties of $H_0$, and can be applied to a wider class of models based on the free Hamiltonians $H_0 = w(-d^2 / dx^2) + C$, where $w$ is an \emph{operator monotone function} (also called \emph{complete Bernstein function}) and $C \in \R$. This more general case will be studied in a separate article.

\subsection{Estimates for $\mathcal{A}_0$}

Recall that $A_0 = H_0 - 1 = (-d^2 / dx^2 + 1)^{1/2}$, and for $f \in C_c^\infty(\R)$, by~\eqref{eq:a0pw} we have
\formula[eq:gen2]{
\begin{aligned}
 A_0 f(x) & = \pvint_{-\infty}^\infty (f(x) - f(y)) \nu(x - y) dy \\
 & = \int_0^\infty (2 f(x) - f(x + z) - f(x - z)) \nu(z) dz ,
\end{aligned}
}
where $\nu(z)$ is a symmetric, unimodal, positive kernel function given by~\eqref{eq:nu}. We denote the right-hand side by $\A_0 f(x)$ (with a calligraphic letter $\A$) whenever the integral converges.

\begin{proposition}
\label{prop:genest1}
Let $x \in \R$, $b > 0$, and let $g$ have an absolutely continuous derivative in $(x - b, x + b)$. Then
\formula{
 \abs{\A_0 g(x)} & \le \expr{\sup_{y \in (x - b, x + b)} |g''(y)|} \int_0^b z^2 \nu(z) dz + \int_{\R \setminus (x - b, x + b)} (|g(x)| + |g(y)|) \nu(y - x) dy .
}
\end{proposition}

\begin{proof}
The result follows directly from the definition of $\A_0$ (see~\eqref{eq:gen2}) and Taylor's expansion for $g$,
\formula{
 |2 g(x) - g(x + z) - g(x - z)| & \le z^2 \sup_{y \in (x - b, x + b)} |g''(y)|
}
when $z \in (-b, b)$.
\end{proof}

For $b > 0$, we define, as in~\cite[Appendix~C]{bib:kkms10}, an auxiliary function:
\formula[eq:q]{
 q(x) & = \begin{cases}
 0 & \text{for } x \in (-\infty, -b] , \\
 (1/2) (x/b + 1)^2 & \text{for } x \in [-b, 0] , \\
 1 - (1/2) (x/b - 1)^2 & \text{for } x \in [0, b] , \\
 1 & \text{for } x \in [b, \infty) .
 \end{cases}
}
Note that $q$ is $C^1$, $q'$ is absolutely continuous, $0 \le q''(x) \le 1 / b^2$ ($x \in \R \setminus \{-b, 0, b\}$), and $q(x) + q(-x) = 1$.

\begin{proposition}
\label{prop:genest}
Let $b > 0$, let $f \in L^1(\R)$, and suppose that $f''(x)$ exists and is continuous for $x \in [-b, b]$. Define
\formula{
 M_{-1} & = \int_0^\infty |f(x)| dx , & M_0 & = \sup_{x \in [-b, b]} |f(x)| , & M_1 & = \sup_{x \in [-b, b]} |f'(x)| , & M_2 & = \sup_{x \in [-b, b]} |f''(x)| ,
}
Let $q(x)$ be given by~\eqref{eq:q}, and define $g(x) = q(x) f(x)$. For $x \in (-\infty, 0)$, we have
\formula{
 \abs{\A_0 g(x)} & \le C(b, \nu) (M_{-1} + M_0 + M_1 + M_2) .
}
More precisely, for $x \in (-\infty, -b]$ we have
\formula{
 \abs{\A_0 g(x)} & \le \frac{M_0}{2 b^2} \int_0^{2b} z^2 \nu(z) dz + \nu(2b) M_{-1}
}
and for $x \in (-b, 0)$,
\formula{
 \abs{\A_0 g(x)} & \le \expr{\frac{M_0}{b^2} + \frac{2 M_1}{b} + M_2} \int_0^b z^2 \nu(z) dz + 2 M_0 \int_b^\infty \nu(z) dz + \nu(b) M_{-1} .
}
\end{proposition}

\begin{proof}
We have $g(y) = q(y) f(y) = 0$ for $y \le -b$. Furthermore, $q(y)$, and hence also $g(y)$, have absolutely continuous derivative in $[-b, b]$. For $y \in (-\infty, b) \setminus \{-b, 0\}$ we have $|q(y)| \le 1$, $|q'(y)| \le 1/b$, $|q''(y)| \le 1/b^2$, and $g''(y) = f(y) q''(y) + 2 f'(y) q'(y) + f''(y) q(y)$. It follows that for $y \in (-\infty, b) \setminus \{-b, 0\}$,
\formula{
 |g''(y)| & \le \frac{M_0}{b^2} + \frac{2 M_1}{b} + M_2 .
}
Let $x \in (-b, 0)$. By Proposition~\ref{prop:genest1}, we have
\formula{
 |\A_0 g(x)| & \le \expr{\frac{M_0}{b^2} + \frac{2 M_1}{b} + M_2} \int_0^b z^2 \nu(z) dz + \int_{\R \setminus (x - b, x + b)} (M_0 + |g(y)|) \nu(y - x) dy
}
Since $x - b \le -b$, we have $g(y) = 0$ for $y \in (-\infty, x - b]$. Furthermore, for $y \in [x + b, \infty)$ we have $|g(y)| \le |f(y)|$ and $\nu(y - x) \le \nu(b)$. It follows that
\formula{
 |\A_0 g(x)| & \le \expr{\frac{M_0}{b^2} + \frac{2 M_1}{b} + M_2} \int_0^b z^2 \nu(z) dz + 2 M_0 \int_b^\infty \nu(z) dz + \nu(b) M_{-1} .
}
When $x \in (-\infty, -b)$, then $g(x) = 0$ and $g''(x) = 0$, and we simply have (again using monotonicity of $\nu$)
\formula{
 |\A_0 g(x)| & \le \int_{-b}^\infty |g(y)| \nu(y - x) dy \le M_0 \int_{-b}^b q(y) \nu(y - x) dy + \nu(2 b) \int_b^\infty |f(y)| dy .
}
Using the inequalities $q(y) \le (b + y)^2 / (2 b^2)$ and $\nu(y - x) \le \nu(b + y)$ for the first integral on the right-hand side, we obtain
\formula{
 |\A_0 g(x)| & \le \frac{M_0}{2 b^2} \int_0^{2b} z^2 \nu(z) dz + \nu(2b) M_{-1} .
\qedhere
}
\end{proof}

\subsection{Approximate eigenfunctions}

Definition~\ref{def:halfline} and Theorem~\ref{th:halfline} yield generalized eigenfunctions $F_\mu(x)$ for the operators $H_\wall = H_0 + V_\wall(x)$ and $A_\wall = A_0 + V_\wall(x) = H_\wall - 1$, corresponding to the eigenvalues $(\mu^2 + 1)^{1/2}$ and $(\mu^2 +1)^{1/2} - 1$, respectively. Below we construct approximations to eigenvalues and eigenfunctions of $H = H_0 + V_\well(x)$ and $A = A_0 + V_\well(x) = H - 1$.

Recall that $D = (-a, a)$ for some $a > 0$, the half of the width of the potential well~\eqref{eq:v}. We fix an auxiliary number $b \in (0, a)$; later we choose $b = a / 3$ to optimize constants. For $n \ge 1$ let $\tilde{\mu}_n$ be a solution of
\formula[eq:tildemu]{
 a \tilde{\mu}_n + \thet_{\tilde{\mu}_n} = \frac{n \pi}{2}
}
(this is a scaled version of~\eqref{eq:tildemu0}), and let
\formula[eq:tildelambda]{
 \tilde{E}_n & = \sqrt{\tilde{\mu}_n^2 + 1} \, , & \tilde{\lambda}_n & = \sqrt{\tilde{\mu}_n^2 + 1} - 1 .
}
Here $\thet_\mu$ is defined by~\eqref{eq:thetmu}. Since $a \mu + \thet_\mu$ is a continuous function of $\mu$, it takes all values between $\thet_{0+} = 0$ and $\infty$. In particular, a number $\tilde{\mu}_n$ satisfying~\eqref{eq:tildemu} always exists. Furthermore, since $\thet_\mu$ is nondecreasing in $\mu > 0$ (see the next section), the solution $\tilde{\mu}_n$ of~\eqref{eq:tildemu} is unique.

\begin{figure}
\centering
\begin{tabular}{cc}
\includegraphics[width=0.35\textwidth]{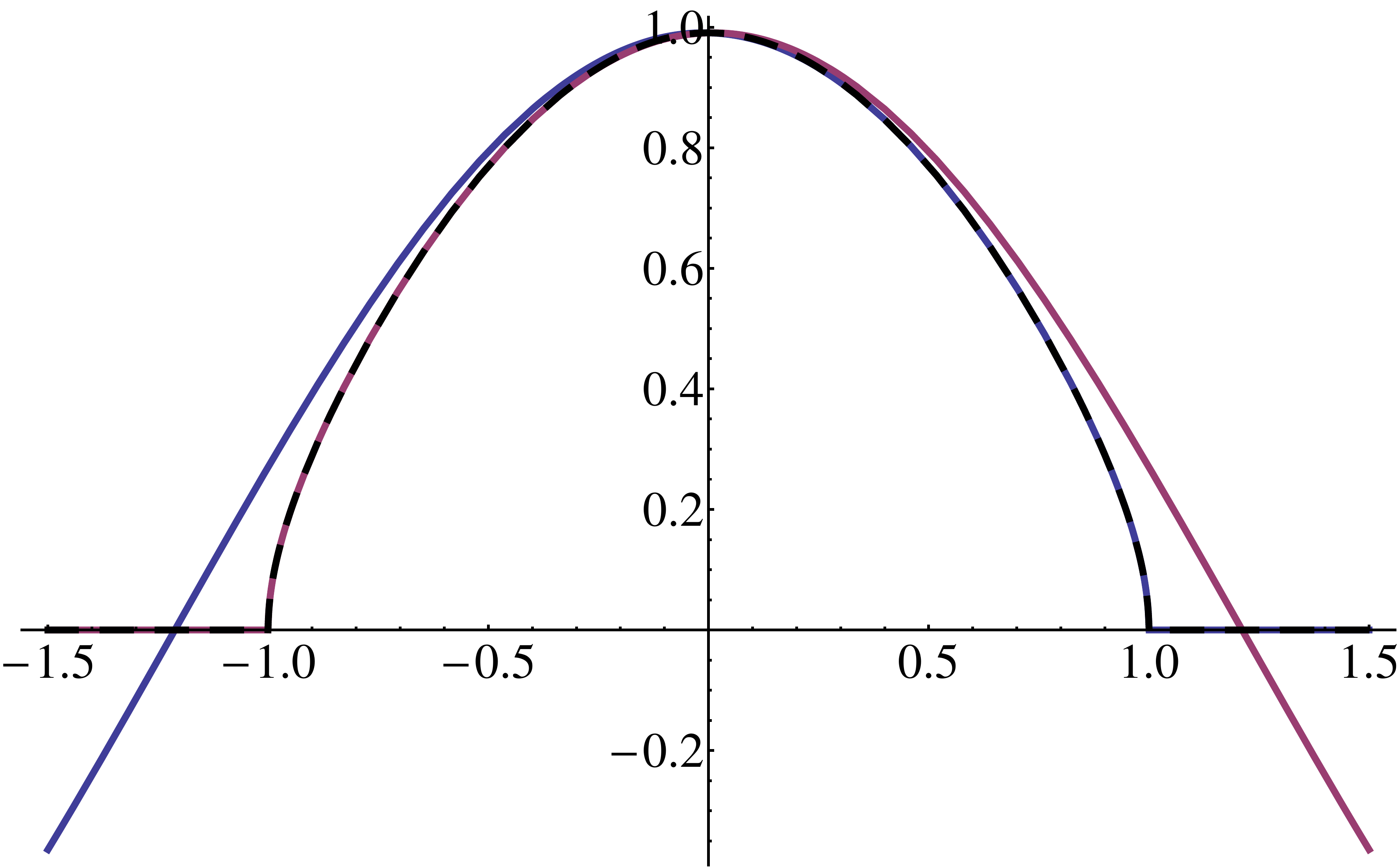} & \includegraphics[width=0.35\textwidth]{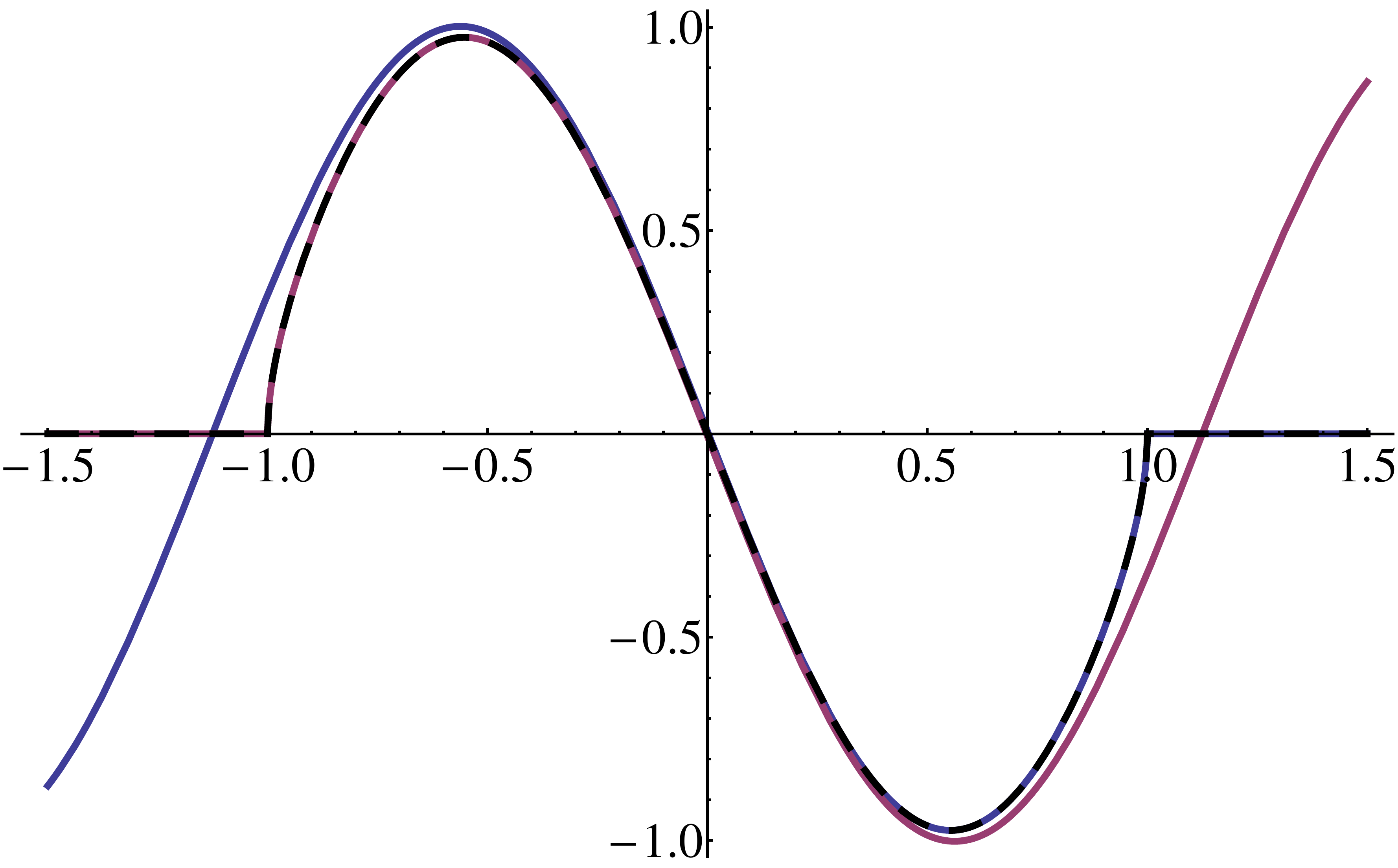} \\
{\footnotesize (a)} & {\footnotesize (b)} \\[1em]
\includegraphics[width=0.35\textwidth]{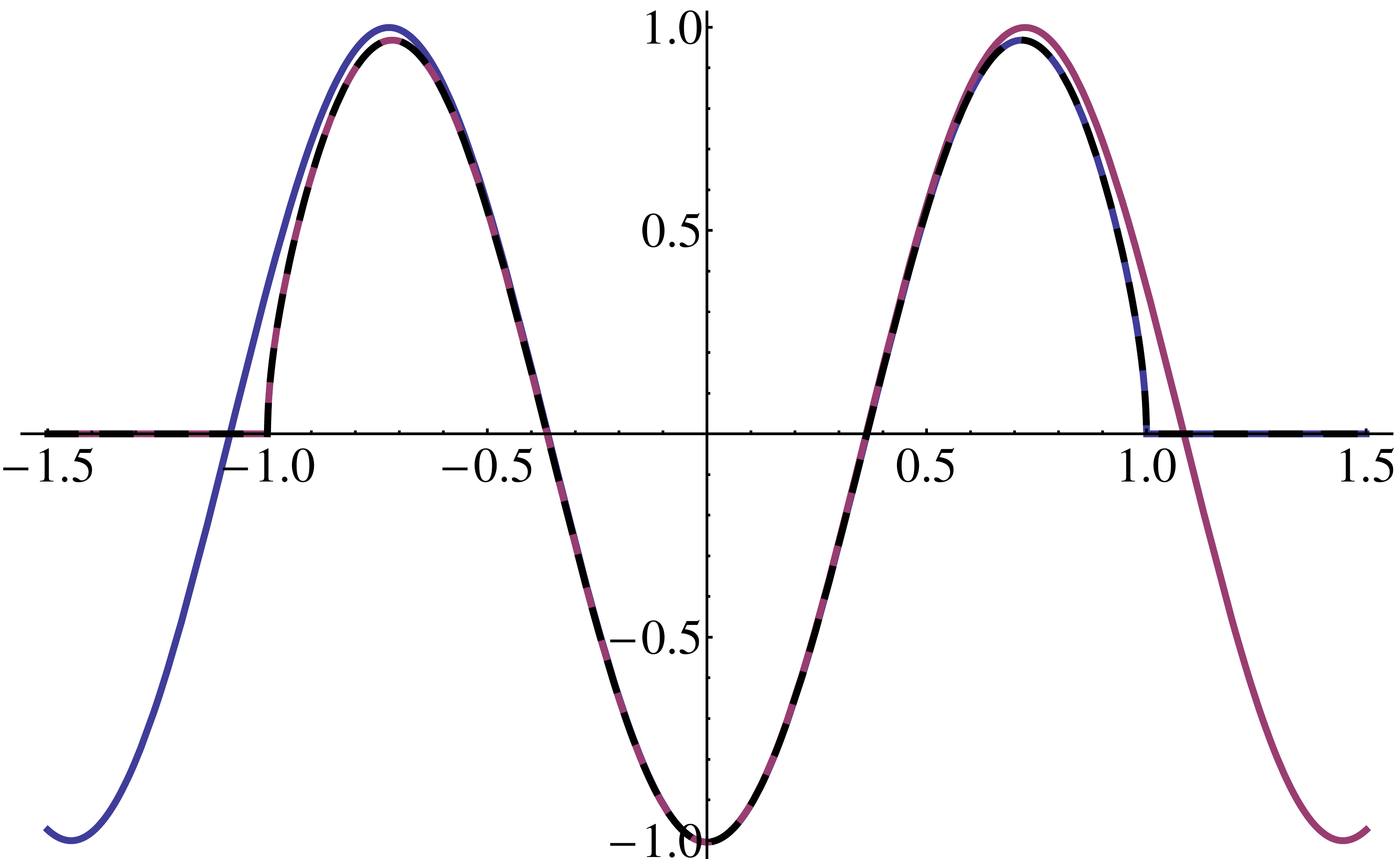} & \includegraphics[width=0.35\textwidth]{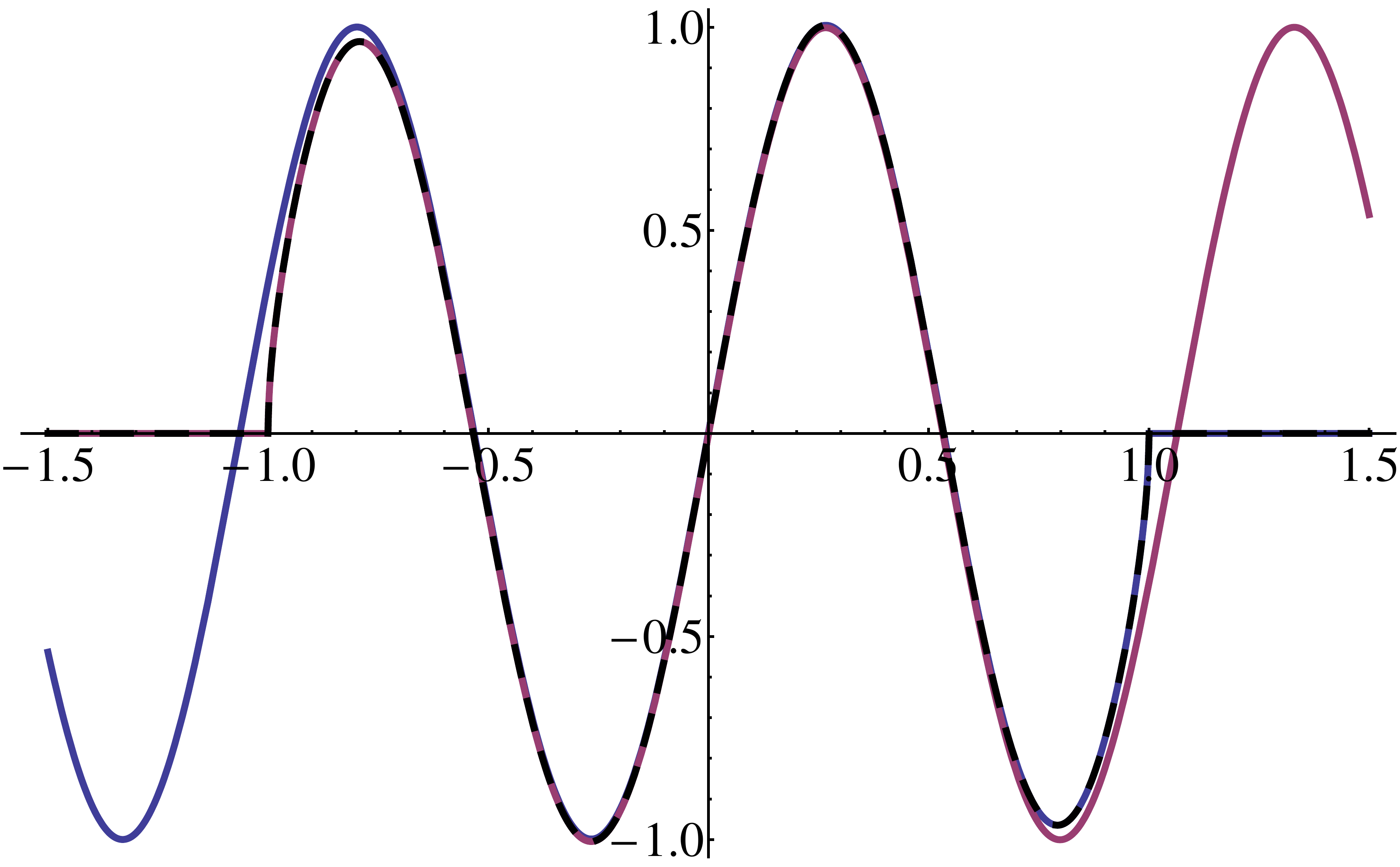} \\
{\footnotesize (c)} & {\footnotesize (d)}
\end{tabular}
\caption{Plot of $F_{\tilde{\mu}_n}(a + x)$ (magenta), $-(-1)^n F_{\tilde{\mu}_n}(a - x)$ (blue) and $\tilde{\ph}_n(x)$ (black dashed) for $a = 1$ and (a) $n = 1$; (b) $n = 2$; (c) $n = 3$; (d) $n = 4$.}
\label{fig:phitilde}
\end{figure}

Our goal is to show that $\tilde{E}_n$ is close to some eigenvalue of $H$, or, equivalently, that $\tilde{\lambda}_n$ is close to some eigenvalue of $A$. To this end, we define an approximate eigenfunction $\tilde{\ph}_n$ of $H$ and $A$, using the eigenfunctions $F_{\tilde{\mu}_n}(a + x)$, $F_{\tilde{\mu}_n}(a - x)$ for single potential walls at $-a$ and $a$. We let (see Figure~\ref{fig:phitilde})
\formula[eq:phitilde]{
 \tilde{\ph}_n(x) & = q(-x) F_{\tilde{\mu}_n}(a + x) - (-1)^n q(x) F_{\tilde{\mu}_n}(a - x) ,
}
with the auxiliary function $q$ defined by~\eqref{eq:q}. Here $x \in \R$, but we have $\tilde{\ph}_n(x) = 0$ for $x \notin D$, so that $\tilde{\ph}_n \in L^2(D)$. The notation introduced above is kept throughout this section. Note that $\tilde{\lambda}_n$ depends on $a$ and $n$, while $\tilde{\ph}_n(x)$ depends also on $b$. Note also that $\tilde{\ph}_n$ is not normed in $L^2(\R)$, its norm is approximately equal to $\sqrt{a}$ (see Lemma~\ref{lem:norm0}); this is why we denote it with a lowercase letter, and reserve $\tilde{\Phi}_n(x)$ for $\tilde{\ph}_n(x) / \|\tilde{\ph}_n\|_{L^2(D)}$.

\begin{lemma}
\label{lem:domain}
We have $\tilde{\ph}_n \in \domain(A)$ and $A \tilde{\ph}_n(x) = \A_0 \tilde{\ph}_n(x)$ for almost all $x \in D$.
\end{lemma}

\begin{proof}
For brevity, in this proof we write $\tilde{\mu} = \tilde{\mu}_n$ and $\tilde{\ph} = \tilde{\ph}_n$. The operator $A = A_0 + V_\well(x)$ is defined in Definition~\ref{def:h}. Hence, we need to check two conditions: (a) there are $f_k \in C_c^\infty(D)$ such that $f_k \to \tilde{\ph}$ in $L^2(D)$ and $\tscalar{f_k - f_l, A_0 (f_k - f_l)}_{L^2(D)} \to 0$ as $k, l \to \infty$; and (b) $\tscalar{\tilde{\ph}, A_0 g}_{L^2(D)} = \tscalar{\A_0 \tilde{\ph}, g}_{L^2(D)}$ for $g \in C_c^\infty(D)$. We first verify~(b).

Note that $\A_0 \tilde{\ph}(x)$ is well-defined for all $x \in D \setminus \{-b, b\}$, since $\tilde{\ph}$ is smooth in $D \setminus \{-b, b\}$ and bounded on $\R$. Furthermore, by the definition of $\A_0$ (see~\eqref{eq:gen2}), for any $g \in C_c^\infty(D)$,
\formula{
 \int_{-a}^a \A_0 \tilde{\ph}(x) g(x) dx - \int_{-a}^a \tilde{\ph}(x) \A_0 g(x) dx & = \\
 & \hspace*{-15em} \int_{-a}^a \int_0^\infty (g(x + z) \tilde{\ph}(x) + g(x - z) \tilde{\ph}(x) - g(x) \tilde{\ph}(x + z) - g(x) \tilde{\ph}(x - z)) \nu(z) dz dx ,
}
which is $0$ by Fubini, provided that the double integral converges. Denote the integrand by $I(x, z) \nu(z)$, and let $\eps = \dist(\supp g, \R \setminus D) / 3$, so that $\supp g \sub (-a + 3 \eps, a - 3 \eps)$. When $z \ge \eps$, then $|I(x, z)| \le C_1(\tilde{\ph}, g)$. Suppose that $z \in (0, \eps)$. If $x \notin (-a + 2 \eps, a - 2 \eps)$, then $I(x, z) = 0$. Otherwise, by first-order Taylor's expansion of $I(x, z)$ around $z = 0$ (note that $I(x, 0) = (\partial / \partial z) I(x, 0) = 0$) with the remainder in the integral form, we obtain that $|I(x, z)| \le C_2(\tilde{\ph}, g) z^2$. We conclude that $|I(x, z) \nu(z)| \le C_3(\tilde{\ph}, g) \min(1, z^2) \nu(z)$, which implies joint integrability of $I(x, z) \nu(z)$. Hence, condition~(b) is satisfied.

By Proposition~\ref{prop:domain}, in order to verify~(a), one only needs to show that $(1 + \xi^2)^{1/2} |\fourier \tilde{\ph}(\xi)|^2$ is integrable on $\R$. Let $f(x) = q(a - x) F_{\tilde{\mu}}(x)$, so that $\tilde{\ph}(x) = f(a + x) - (-1)^n f(a - x)$ (see~\eqref{eq:phitilde}). Hence, it suffices to prove integrability of $(1 + \xi^2)^{1/2} |\fourier f(\xi)|^2$.

Fix $\eps > 0$ and let $\tilde{q}(x) = q(a - x) e^{\eps x}$. It is easy to check that the distributional derivatives $q$, $q'$ and $q''$ are integrable functions, and the third distributional derivative of $q(x)$ is a finite signed measure on $\R$. Hence, $\tilde{q}(x)$ has the same property. Therefore, $\fourier q(\xi)$ and $\fourier q'''(\xi) = -i \xi^3 \fourier q(\xi)$ are bounded functions, and so $|\fourier \tilde{q}(\xi)| \le C_4(\eps, a, b) / (1 + |\xi|)^3$. 

The Fourier transform of $e^{-\eps x} F_{\tilde{\mu}}(x)$ is $\laplace F_{\tilde{\mu}}(\eps + i \xi)$, and the Fourier transform of $f(x) = q(a - x) F_{\tilde{\mu}}(x) = \tilde{q}(x) e^{-\eps x} F_{\tilde{\mu}}(x)$ is given by the convolution
\formula{
 \fourier f(\xi) & = \frac{1}{2 \pi} \int_{-\infty}^\infty \fourier \tilde{q}(\xi - s) \laplace F_{\tilde{\mu}}(\eps + i s) ds .
}
Suppose that $\xi > 0$. To estimate $|\fourier f(\xi)|$, we write
\formula[eq:fourierdecomposition]{
 \fourier f(\xi) & = \frac{1}{2 \pi} \int_{\xi/2}^\infty \fourier \tilde{q}(\xi - s) \laplace F_{\tilde{\mu}}(\eps + i s) ds + \frac{1}{2 \pi} \int_{\xi/2}^\infty \fourier \tilde{q}(s) \laplace F_{\tilde{\mu}}(\eps + i (\xi - s)) ds .
}
By~\eqref{eq:laplaceest2:gen}, we have
\formula{
 |\laplace F_{\tilde{\mu}}(\eps + i s)| & \le C_5(\eps, \tilde{\mu}) \, \frac{1}{1 + s^2} \, \sqrt{1 + \sqrt{1 + s^2}} \le C_6(\eps, \tilde{\mu}) \, \frac{1}{(1 + |s|)^{3/2}} \, .
}
Hence,
\formula{
 \hspace*{3em} & \hspace*{-3em} \abs{\int_{\xi/2}^\infty \fourier \tilde{q}(\xi - s) \laplace F_{\tilde{\mu}}(\eps + i s) ds} \le \frac{C_6(\eps, \tilde{\mu})}{(1 + \xi/2)^{3/2}} \, \int_{\xi/2}^\infty |\fourier \tilde{q}(\xi - s)| ds \\
 & \le \frac{C_6(\eps, \tilde{\mu}) C_4(\eps, a, b)}{(1 + \xi/2)^{3/2}} \int_{\xi/2}^\infty \frac{1}{(1 + (\xi - s))^3} \, ds \le \frac{C_6(\eps, \tilde{\mu}) C_4(\eps, a, b)}{(1 + \xi/2)^{3/2}} \, .
}
For the other integral in~\eqref{eq:fourierdecomposition}, we use $|\laplace F_{\tilde{\mu}}(\eps + i s)| \le C_6(\eps, \tilde{\mu})$:
\formula{
 \abs{\int_{\xi/2}^\infty \fourier \tilde{q}(s) \laplace F_{\tilde{\mu}}(\eps + i (\xi - s)) ds} & \le C_6(\eps, \tilde{\mu}) \int_{\xi/2}^\infty |\fourier \tilde{q}(s)| ds \le \frac{C_6(\eps, \tilde{\mu}) C_4(\eps, a, b)}{2 (1 + \xi/2)^2} \, .
}
Therefore, for $\xi > 0$,
\formula{
 |\fourier f(\xi)| & \le C_7(\eps, a, b, \tilde{\mu}) \, \frac{1}{(1 + |\xi/2|)^{3/2}} \, .
}
Since $\fourier f(-\xi) = \overline{\fourier f(\xi)}$, the above estimate holds for all $\xi \in \R$. We conclude that for all $\xi \in \R$,
\formula{
 (1 + \xi^2)^{1/2} |\fourier f(\xi)|^2 & \le 4 (C_7(\eps, a, b, \tilde{\mu}))^2 \, \frac{1}{(1 + |\xi/2|)^2} \, ,
}
and the right-hand side is integrable. The proof is complete.
\end{proof}

For brevity, in the remaining part of the article we use the following notation:
\formula[eq:constants]{
\begin{aligned}
 \nu_0(x) & = \int_0^x z^2 \nu(z) dz , \qquad & \nu_\infty(x) & = \int_x^\infty \nu(z) dz , \\
 I_\mu & = \int_0^\infty G_\mu(x) dx , \qquad & G_{\mu,b}(x) & = G_\mu(x - b) + G_\mu(x + b) .
\end{aligned}
}
The following results are stated for $A$ and $\tilde{\lambda}_n$; their reformulation for $H$ and $\tilde{E}_n$ is straightforward.

\begin{lemma}
\label{lem:approx0}
We have
\formula{
 \| A \tilde{\ph}_n - \tilde{\lambda}_n \tilde{\ph}_n \|_{L^2(D)}^2 & \le C(a, b) (G_{\tilde{\mu}_n,b}(a) - G_{\tilde{\mu}_n,b}'(a) + G_{\tilde{\mu}_n,b}''(a) + I_{\tilde{\mu}_n} + 1 / \tilde{\mu}_n) .
}
More precisely, we have
\formula[eq:approx0]{
\begin{aligned}
 \| A \tilde{\ph}_n - \tilde{\lambda}_n \tilde{\ph}_n \|_{L^2(D)}^2 & \le 2 (a - b) \expr{\frac{G_{\tilde{\mu}_n,b}(a) \nu_0(2b)}{2 b^2} + \nu(2b) I_{\tilde{\mu}_n} + \frac{2 \nu(a)}{\tilde{\mu}_n}}^2 \\
 & \hspace*{-6em} + 2 b \Biggl( \frac{(G_{\tilde{\mu}_n,b}(a) - 2 b G_{\tilde{\mu}_n,b}'(a) + b^2 G_{\tilde{\mu}_n,b}''(a)) \nu_0(b)}{b^2} \\
 & \hspace*{1em} + 2 G_{\tilde{\mu}_n,b}(a) \nu_\infty(b) + \nu(b) I_{\tilde{\mu}_n} + \frac{\tilde{\lambda}_n G_{\tilde{\mu}_n,b}(a)}{2} + \frac{2 \nu(a)}{\tilde{\mu}_n} \Biggr)^2 .
\end{aligned}
}
\end{lemma}

\begin{proof}
The argument below follows exactly the method developed in~\cite[Lemma~1]{bib:kkms10}, and applied also in~\cite[Lemma~1]{bib:k11}. Since in this proof we consider $n$ fixed, we simplify the notation and write $\tilde{\mu} = \tilde{\mu}_n$, $\tilde{\lambda} = \tilde{\lambda}_n = (\tilde{\mu}^2 + 1)^{1/2} - 1$ and $\tilde{\ph} = \tilde{\ph}_n$.

Recall that $F_{\tilde{\mu}}(x) = \sin(\tilde{\mu} x + \thet_{\tilde{\mu}}) \ind_{(0, \infty)(x)} - G_{\tilde{\mu}}(x)$, and $G_{\tilde{\mu}}(x) = 0$ for $x \le 0$. Note that by~\eqref{eq:tildemu},
\formula{
 \sin(\tilde{\mu} (a + x) + \thet_{\tilde{\mu}}) & = \sin(\tfrac{n \pi}{2} + \tilde{\mu} x) \\
 & = -(-1)^n \sin(\tfrac{n \pi}{2} - \tilde{\mu} x) = -(-1)^n \sin(\tilde{\mu} (a - x) + \thet_{\tilde{\mu}}) .
}
Hence, by~\eqref{eq:phitilde} and~\eqref{eq:hf:f}, for all $x \in \R$,
\formula[eq:tildephapprox]{
\begin{aligned}
 \tilde{\ph}(x) & = q(-x) F_{\tilde{\mu}}(a + x) - (-1)^n q(x) F_{\tilde{\mu}}(a - x) \\
 & = \sin(\tfrac{n \pi}{2} + \tilde{\mu} x) \ind_D(x) - q(-x) G_{\tilde{\mu}}(a + x) + (-1)^n q(x) G_{\tilde{\mu}}(a - x) .
\end{aligned}
}
In a similar way, for all $x \in \R$,
\formula{
 F_{\tilde{\mu}}(a + x) + (-1)^n F_{\tilde{\mu}}(a - x) & = \sin(\tfrac{n \pi}{2} + \tilde{\mu} x) \ind_{\R \setminus D}(x) \sign x \\
 & \hspace*{5em} - (G_{\tilde{\mu}}(a + x) + (-1)^n G_{\tilde{\mu}}(a - x)) ,
}
so that, again by~\eqref{eq:phitilde}, for all $x \in \R$,
and
\formula{
 \tilde{\ph}(x) - F_{\tilde{\mu}}(a + x) & = (q(-x) - 1) F_{\tilde{\mu}}(a + x) - (-1)^n q(x) F_{\tilde{\mu}}(a - x) \\
 & = -q(x) (F_{\tilde{\mu}}(a + x) + (-1)^n F_{\tilde{\mu}}(a - x)) \\
 & = q(x) (G_{\tilde{\mu}}(a + x) + (-1)^n G_{\tilde{\mu}}(a - x)) - \sin(\tfrac{n \pi}{2} + \tilde{\mu} x) \ind_{[a, \infty)}(x) .
}
Denote
\formula{
 f(x) & = G_{\tilde{\mu}}(a + x) + (-1)^n G_{\tilde{\mu}}(a - x) , \\
 g(x) & = q(x) f(x) , \\	
 h(x) & = \sin(\tfrac{n \pi}{2} + \tilde{\mu} x) \ind_{[a, \infty)}(x) .
}
It follows that $\tilde{\ph}(x) = F_{\tilde{\mu}}(a + x) + g(x) - h(x)$. For $x \in (-a, 0)$, we have $\A_0 F_{\tilde{\mu}}(a + x) - \tilde{\lambda} F_{\tilde{\mu}}(a + x) = 0$ and $h(x) = 0$. Hence, for $x \in (-a, 0)$,
\formula[eq:decomposition]{
 |\A_0 \tilde{\ph}(x) - \tilde{\lambda} \tilde{\ph}(x)| & \le |\A_0 g(x)| + |\A_0 h(x)| + |\tilde{\lambda} g(x)| .
}
We estimate each of the summands on the right-hand side separately.

First, we apply Proposition~\ref{prop:genest} for the function $f$. Since $G_{\tilde{\mu}}(x)$ is a completely monotone function, $G_{\tilde{\mu}}(x)$, $-G_{\tilde{\mu}}'(x)$ and $G_{\tilde{\mu}}''(x)$ are nonnegative convex functions of $x \in (0, \infty)$. Hence,
\formula{
 M_0 & = \sup_{x \in (-b, b)} |f(x)| \le G_{\tilde{\mu}}(a - b) + G_{\tilde{\mu}}(a + b) = G_{\tilde{\mu},b}(a), \\
 M_1 & = \sup_{x \in (-b, b)} |f'(x)| \le -G_{\tilde{\mu}}'(a - b) - G_{\tilde{\mu}}'(a + b) = -G_{\tilde{\mu},b}'(a) , \\
 M_2 & = \sup_{x \in (-b, b)} |f''(x)| \le G_{\tilde{\mu}}''(a - b) + G_{\tilde{\mu}}''(a + b) = G_{\tilde{\mu},b}''(a) .
}
Furthermore,
\formula{
 M_{-1} & = \int_0^\infty |f(x)| dx \le \int_0^\infty G_{\tilde{\mu}}(a + x) dx + \int_0^a G_{\tilde{\mu}}(a - x) dx = \int_0^\infty G_{\tilde{\mu}}(y) dy = I_{\tilde{\mu}} .
}
By Proposition~\ref{prop:genest}, for $x \in (-a, -b)$,
\formula{
 |\A_0 g(x)| & \le \frac{G_{\tilde{\mu},b}(a) \nu_0(2b)}{2 b^2} + \nu(2b) I_{\tilde{\mu}} ,
}
and for $x \in (-b, 0)$,
\formula{
 |\A_0 g(x)| & \le \frac{(G_{\tilde{\mu},b}(a) - 2 b G_{\tilde{\mu},b}'(a) + b^2 G_{\tilde{\mu},b}''(a)) \nu_0(b)}{b^2} + 2 G_{\tilde{\mu},b}(a) \nu_\infty(b) + \nu(b) I_{\tilde{\mu}} .
}
Note that $|g(x)| = 0$ for $x \in (-a, -b)$, and for $x \in (-b, 0)$ we have
\formula{
 |\tilde{\lambda} g(x)| & \le \frac{\tilde{\lambda} |f(x)|}{2} \le \frac{\tilde{\lambda} M_0}{2} \le \frac{\tilde{\lambda} G_{\tilde{\mu},b}(a)}{2} \, .
}
Finally, $\nu(y)$ decreases for $y > 0$. Hence, integrating by parts, we obtain that for $x < 0$,
\formula{
 |\A_0 h(x)| & = \abs{\int_a^\infty \sin(\tfrac{n \pi}{2} + \tilde{\mu} x) \nu(y - x) dy} \le \frac{2 \nu(a)}{\tilde{\mu}} \, .
}
By combining the above estimates and~\eqref{eq:decomposition}, we conclude that for $x \in (-a, -b)$,
\formula{
 |\A_0 \tilde{\ph}(x) - \tilde{\lambda} \tilde{\ph}_n(x)| & \le \frac{G_{\tilde{\mu},b}(a) \nu_0(2b)}{2 b^2} + \nu(2b) I_{\tilde{\mu}} + \frac{2 \nu(a)}{\tilde{\mu}} \, ,
}
and for $x \in (-b, 0)$,
\formula{
 |\A_0 \tilde{\ph}(x) - \tilde{\lambda} \tilde{\ph}_n(x)| & \le \frac{(G_{\tilde{\mu},b}(a) - 2 b G_{\tilde{\mu},b}'(a) + b^2 G_{\tilde{\mu},b}''(a)) \nu_0(b)}{b^2} \\
 & \hspace*{7em} + 2 G_{\tilde{\mu}, b}(a) \nu_\infty(b) + \nu(b) I_{\tilde{\mu}} + \frac{2 \nu(a)}{\tilde{\mu}} \, .
}
By symmetry, similar estimates hold also for $x \in (0, a)$. Integration in $x$ gives formula~\eqref{eq:approx0} with $A \tilde{\ph}(x)$ in the left-hand side replaced by $\ind_D(x) \A_0 \tilde{\ph}(x)$. The result follows by Lemma~\ref{lem:domain}.
\end{proof}

\begin{lemma}
\label{lem:norm0}
We have
\formula{
 \bigl| \|\tilde{\ph}_n\|_{L^2(D)}^2 - a \bigr| & \le C(a) (I_{\tilde{\mu}_n} + 1 / \tilde{\mu}_n) .
}
More precisely,
\formula[eq:norm0]{
 a - \frac{\sin(\thet_{\tilde{\mu}_n})}{\tilde{\mu}_n} - 4 I_{\tilde{\mu}_n} & \le \|\tilde{\ph}_n\|_{L^2(D)}^2 \le a + \frac{\sin(\thet_{\tilde{\mu}_n})}{\tilde{\mu}_n} + 4 I_{\tilde{\mu}_n} (1 + \sin \thet_{\tilde{\mu}_n}) .
}
\end{lemma}

\begin{proof}
As in the previous proof, we write $\tilde{\mu} = \tilde{\mu}_n$, $\tilde{\lambda} = \tilde{\lambda}_n$ and $\tilde{\ph} = \tilde{\ph}_n$. By~\eqref{eq:tildephapprox},
\formula{
 \tilde{\ph}(x) & = \sin(\tfrac{n \pi}{2} + \tilde{\mu} x) \ind_D(x) - q(-x) G_{\tilde{\mu}}(a + x) + (-1)^n q(x) G_{\tilde{\mu}}(a - x) ,
}
It follows that
\formula{
 \|\tilde{\ph}\|_{L^2(D)}^2 - a & = \int_{-a}^a \expr{\expr{\sin(\tfrac{n \pi}{2} + \tilde{\mu} x)}^2 - \tfrac{1}{2}} dx \\
 & \hspace*{3em} + \int_{-a}^a \expr{q(-x) G_{\tilde{\mu}}(a + x) - (-1)^n q(x) G_{\tilde{\mu}}(a - x)}^2 dx \\
 & \hspace*{3em} - 2 \int_{-a}^a \sin(\tfrac{n \pi}{2} + \tilde{\mu} x) \expr{q(-x) G_{\tilde{\mu}}(a + x) - (-1)^n q(x) G_{\tilde{\mu}}(a - x)} dx .
}
By direct integration and~\eqref{eq:tildemu},
\formula{
 \hspace*{3em} & \hspace*{-3em} \abs{\int_{-a}^a \expr{\expr{\sin(\tfrac{n \pi}{2} + \tilde{\mu} x)}^2 - \tfrac{1}{2}} dx} = \abs{\frac{1}{2} \int_{-a}^a \cos(n \pi + 2 \tilde{\mu} x) dx} \\
 & = \frac{1}{4 \tilde{\mu}} \abs{\sin(n \pi - 2 \tilde{\mu} a) - \sin(n \pi + 2 \tilde{\mu} a)} = \frac{\sin (2 \thet_{\tilde{\mu}})}{2 \tilde{\mu}} \le \frac{\sin(\thet_{\tilde{\mu}})}{\tilde{\mu}} \, .
}
Furthermore,
\formula{
 & \abs{\int_{-a}^a \sin(\tfrac{n \pi}{2} + \tilde{\mu} x) \expr{q(-x) G_{\tilde{\mu}}(a + x) - (-1)^n q(x) G_{\tilde{\mu}}(a - x)} dx} \\
 & \hspace*{10em} \le \int_{-a}^a \expr{G_{\tilde{\mu}}(a + x) + G_{\tilde{\mu}}(a - x)} dx \le 2 I_{\tilde{\mu}} .
}
Finally, using $G_{\tilde{\mu}}(x) \le G_{\tilde{\mu}}(0+) = \sin(\thet_{\tilde{\mu}})$,
\formula{
 0 & \le \int_{-a}^a \expr{q(-x) G_{\tilde{\mu}}(a + x) - (-1)^n q(x) G_{\tilde{\mu}}(a - x)}^2 dx \\
 & \le 2 \expr{\int_{-a}^a (G_{\tilde{\mu}}(a + x))^2 dx + \int_{-a}^a (G_{\tilde{\mu}}(a - x))^2 dx} \\
 & \le 2 \sin \thet_{\tilde{\mu}} \expr{\int_{-a}^a G_{\tilde{\mu}}(a + x) dx + \int_{-a}^a G_{\tilde{\mu}}(a - x) dx} \le 4 I_{\tilde{\mu}} \sin \thet_{\tilde{\mu}} ,
}
and the lemma is proved.
\end{proof}

Let $\sigma(A)$ denote the spectrum of $A$. Recall that the spectra of $A$ and $H$ are purely discrete (see Subsection~\ref{subsec:green}), and the eigenvalues of $A$ and $H$ (repeated according to multiplicity) are denoted by $\lambda_n$ and $E_n = \lambda_n + 1$, respectively.

\begin{proposition}
\label{prop:dist0}
We have
\formula[eq:dist0]{
 \dist(\tilde{\lambda}_n, \sigma(A)) & \le \frac{\| A \tilde{\ph}_n - \tilde{\lambda}_n \tilde{\ph}_n \|_{L^2(D)}}{\|\tilde{\ph}_n\|_{L^2(D)}} \, .
}
More precisely, let $A^{\even}$ and $A^{\odd}$ be the restrictions of $A$ to the (invariant) subspaces of $L^2(D)$ consisting of even and odd functions, respectively. Then~\eqref{eq:dist0} holds with $\sigma(A)$ replaced by $\sigma(A^{\even})$ when $n$ is odd, and by $\sigma(A^{\odd})$ when $n$ is even.
\end{proposition}

\begin{proof}
Let $E(d\lambda)$ be the spectral measure of $A$. Since $\tilde{\ph}_n \in L^2(D)$, we have
\formula{
 \| A \tilde{\ph}_n - \tilde{\lambda}_n \tilde{\ph}_n \|_{L^2(D)}^2 & = \int_{\sigma(A)} |\lambda - \tilde{\lambda}_n|^2 \scalar{\tilde{\ph}_n, E(d\lambda) \tilde{\ph}_n} \\
 & \hspace*{-8em} \ge (\dist(\tilde{\lambda}_n, \sigma(A)))^2 \int_{\sigma(A)} \scalar{\tilde{\ph}_n, E(d\lambda) \tilde{\ph}_n} = (\dist(\tilde{\lambda}_n, \sigma(A)))^2 \|\tilde{\ph}_n\|_{L^2(D)}^2 ,
}
as desired. Furthermore, since $\tilde{\ph}_n$ is either an even or odd function, we can replace $A$ by $A^{\even}$ or $A^{\odd}$, respectively, depending on the parity of $n$.
\end{proof}

\subsection{Estimates of $\vartheta_\mu$ and $G_\nu$}

Before we apply the results of the previous subsection, we need some detailed estimates. Recall that the (singular) kernel $\nu$ of $A$ is given by~\eqref{eq:nu}, $\nu(z) = K_1(|z|) / (\pi |z|)$, where $K_1$ is the modified Bessel function of the second kind.

\begin{proposition}
For $x \in \R \setminus \{0\}$, we have
\formula[eq:nuest]{
 \frac{e^{-|x|}}{\pi x^2} \le \nu(x) & \le \frac{(1 + |x|) e^{-|x|}}{\pi x^2} \, .
}
\end{proposition}

\begin{proof}
Let $x > 0$. By the definition of the Bessel function and a substitution $\sinh t = s$, we have
\formula{
 e^x K_1(x) & = \int_0^\infty e^{-x (\cosh t - 1)} \cosh t \, dt = \int_0^\infty e^{-x (\sqrt{1 + s^2} - 1)} ds .
}
Since $\sqrt{1 + s^2} - 1 \ge \max(0, s - 1)$,
\formula{
 e^x K_1(x) & \le \int_0^1 1 \, ds + \int_1^\infty e^{-x (s - 1)} ds = 1 + \frac{1}{x} \, .
}
We conclude that $e^x \nu(x) = e^x K_1(x) / (\pi x) \le (x + 1) / (\pi x^2)$, which is the desired upper bound. In a similar manner, $\sqrt{1 + s^2} - 1 \le s$, and therefore $e^x K_1(x) \ge 1$. The lower bound follows.
\end{proof}

Since $(1 + x) e^{-x} \le 1$, we have, in particular, $\nu(x) \le 1 / (\pi x^2)$. This simple bound is often used below. With the notation of~\eqref{eq:constants}, for $x > 0$ we therefore have
\formula[eq:nu0est]{
 \nu_0(x) & = \int_0^x z^2 \nu(z) dz \le \frac{1}{\pi} \, \min \expr{x, \int_0^\infty (1 + z) e^{-z} dz} = \frac{1}{\pi} \, \min(x, 2) .
}
Furthermore,
\formula[eq:nuinftyest]{
 \nu_\infty(x) & = \int_x^\infty \nu(z) dz \le \int_x^\infty \frac{1}{\pi z^2} \, dz = \frac{1}{\pi x}
}
for $x > 0$.

\begin{proposition}
The function $\thet_\mu$ is increasing in $\mu$, converges to $0$ as $\mu \to 0^+$, and to $\pi/8$ as $\mu \to \infty$. Furthermore, for $\mu > 0$ we have
\formula[eq:dthetaest]{
 0 < \frac{d \thet_\mu}{d \mu} & < \frac{2}{\pi} \, \frac{1}{1 + \mu^2} \, .
}
In particular,
\formula[eq:thetaest]{
 \max \expr{0, \frac{\pi}{8} - \frac{2}{\pi \mu}} \le \thet_\mu & \le \min \expr{\frac{2 \mu}{\pi}, \frac{\pi}{8}} . 
}
\end{proposition}

\begin{proof}
By~\eqref{eq:thetmu} and dominated convergence, for $\mu > 0$ we have (see~\cite[Proposition~4.17]{bib:k10})
\formula[eq:dtheta]{
\begin{aligned}
 \frac{d \thet_\mu}{d \mu} & = \frac{1}{\pi} \, \frac{1}{1 + \mu^2} \int_0^\infty \frac{1}{\sqrt{1 + r^2} \, (\sqrt{1 + \mu^2} + \sqrt{1 + r^2})} \, dr \\
 & = \frac{1}{\pi} \, \frac{1}{\mu (1 + \mu^2)} \log \frac{1 + \mu + \sqrt{1 + \mu^2}}{1 - \mu + \sqrt{1 + \mu^2}} \, ;
\end{aligned}
}
the integral above is evaluated using the first Euler substitution. In particular, $\thet_\mu$ increases with $\mu$. Furthermore,~\eqref{eq:thetmu} can be rewritten as (see~\cite[Proposition~4.16]{bib:k10})
\formula{
 \thet_\mu & = \frac{1}{\pi} \int_0^1 \frac{1}{1 - s^2} \, \log \frac{1 + \sqrt{\frac{1 + \mu^2 / s^2}{1 + \mu^2}}}{1 + \sqrt{\frac{1 + \mu^2 s^2}{1 + \mu^2}}} \, ds .
}
Hence, by dominated convergence, $\thet_\mu \to 0$ as $\mu \to 0$ and
\formula{
 \lim_{\mu \to \infty} \thet_\mu & = \frac{1}{\pi} \int_0^1 \frac{1}{1 - s^2} \, \log \frac{1 + 1/s}{1 + s} \, ds = \frac{1}{\pi} \int_0^1 \frac{1}{1 - s^2} \, \log \frac{1}{s} \, ds = \frac{\pi}{8} \, ;
}
for the last equality, see~\cite[Proposition~4.15]{bib:k10}. These properties of $\thet_\mu$ were already noticed in~\cite[Example~6.2]{bib:k10}. By~\eqref{eq:dtheta} and the inequality $\log(1 + s) < s$ ($s > 0$), we also obtain that
\formula{
 \frac{d \thet_\mu}{d \mu} & < \frac{1}{\pi} \, \frac{1}{\mu (1 + \mu^2)} \, \frac{2 \mu}{1 - \mu + \sqrt{1 + \mu^2}} < \frac{2}{\pi} \, \frac{1}{1 + \mu^2} \, , && \mu > 0 ,
}
as desired.
\end{proof}

By series expansion of $d \thet_\mu / d \mu$ (using~\eqref{eq:dtheta}) near $0$ and $\infty$, one easily obtains~\eqref{eq:thetmuzero} and~\eqref{eq:thetmuinfinity}.

For any $a > 0$, $a \mu + \thet_\mu$ is increasing in $\mu \in (0, \infty)$. Hence, as it was observed in the previous section (see the remark following~\eqref{eq:tildemu}), the solution $\tilde{\mu}_n$ of the equation $a \mu + \thet_\mu = n \pi / 2$ ($a > 0$, $n \ge 1$ fixed) is unique.

\begin{proposition}
For $a > 0$ and $n \ge 1$, we have
\formula[eq:muest2]{
 \frac{1}{a} \expr{\frac{n \pi}{2} - \thet_{n \pi / (2 a)}} & < \tilde{\mu}_n < \frac{1}{a} \expr{\frac{n \pi}{2} - \thet_{n \pi / (2 a)}} + \frac{\min(8 n / (\pi a), 1)}{4 a^2 + \pi^2 (n - 1/4)^2} \/ .
}
In particular,
\formula[eq:muest1]{
 \frac{n \pi}{2 a} - \frac{\pi}{8 a} & < \tilde{\mu}_n < \frac{n \pi}{2 a} \/ .
}
\end{proposition}

\begin{proof}
Since $\thet_\mu$ is increasing, we have
\formula[eq:muest1a]{
 \frac{1}{a} \expr{\frac{n \pi}{2} - \thet_{n \pi / (2 a)}} & < \tilde{\mu}_n < \frac{n \pi}{2 a} \/ .
}
This gives~\eqref{eq:muest1} and the lower bound of~\eqref{eq:muest2}. Furthermore, by the mean value theorem, \eqref{eq:dthetaest} and~\eqref{eq:muest1},
\formula{
 \tilde{\mu}_n - \frac{1}{a} \expr{\frac{n \pi}{2} - \thet_{n \pi / (2 a)}} & = \frac{1}{a} \expr{\thet_{n \pi / (2 a)} - \thet_{\tilde{\mu}_n}} < \frac{2}{\pi a} \, \frac{1}{1 + \tilde{\mu}_n^2} \expr{\frac{n \pi}{2 a} - \tilde{\mu}_n} < \frac{2}{\pi a^2} \frac{\thet_{n \pi / (2 a)}}{1 + \tilde{\mu}_n^2} .
}
This, combined with the inequality $\tilde{\mu}_n^2 > ((n - 1/4) \pi / (2 a))^2$ (which follows from~\eqref{eq:muest1a}) and~\eqref{eq:thetaest}, gives the upper bound of~\eqref{eq:muest2}.
\end{proof}

We now turn to estimates of $G_\mu$.

\begin{proposition}
For $\mu, x > 0$,
\formula[eq:gest1]{
 G_\mu(x) & \le \min \expr{\frac{2 \mu}{\pi} , 0.383} ,
}
and
\formula[eq:gest2]{
 \int_0^\infty G_\mu(x) dx & \le \min \expr{\frac{\mu}{8} , \frac{0.217}{\mu}} .
}
\end{proposition}

\begin{proof}
By~\cite[Lemma~4.21]{bib:k10}, $G_\mu(x) \le G_\mu(0+) = \sin(\thet_\mu)$. This and~\eqref{eq:thetaest} give~\eqref{eq:gest1}:
\formula{
 G_\mu(x) & \le \sin(\thet_\mu) \le \min\expr{\sin \frac{2 \mu}{\pi} , \sin \frac{\pi}{8}} \le \min \expr{\frac{2 \mu}{\pi} , 0.383} .
}
Furthermore, again by~\cite[Lemma~4.21]{bib:k10}, with $w(\xi) = (\xi + 1)^{1/2} - 1$ we have
\formula[eq:gest0]{
\begin{aligned}
 \int_0^\infty G_\mu(x) dx & = \frac{\cos(\thet_\mu)}{\mu} - \sqrt{\frac{w'(\mu^2)}{w(\mu^2)}} \\
 & = \frac{1}{\mu} \expr{\cos \thet_\mu - \sqrt{\frac{\mu^2}{2 \sqrt{1 + \mu^2} \, (\sqrt{1 + \mu^2} - 1)}}} \\
 & = \frac{1}{\mu} \expr{\cos \thet_\mu - \sqrt{\frac{1 + \sqrt{1 + \mu^2}}{2 \sqrt{1 + \mu^2}}}} .
\end{aligned}
}
We have $1 / \sqrt{1 + s} \ge 1 - s/2 + s^2/32$ for $s \le 8$. Hence, whenever $\mu^2 \le 8$,
\formula{
 \sqrt{\frac{1 + \sqrt{1 + \mu^2}}{2 \sqrt{1 + \mu^2}}} & = \sqrt{\frac{1}{2} + \frac{1}{2 \sqrt{1 + \mu^2}}} \ge \sqrt{1 - \frac{\mu^2}{4} + \frac{\mu^4}{64}} = 1 - \frac{\mu^2}{8} \, .
}
When $\mu^2 > 8$, the right-hand side is negative, so the inequality between left and right-hand sides holds for all $\mu$. It follows that $\int_0^\infty G_\mu(x) dx \le \mu^{-1} (\cos(\thet_\mu) - 1 + \mu^2 / 8) \le \mu / 8$. This proves one part of~\eqref{eq:gest2}. Furthermore, by~\eqref{eq:thetaest} and concavity of $\cos$ on $[0, \pi/8]$,
\formula{
 \cos(\thet_\mu) & \le \cos \expr{\max \expr{\frac{\pi}{8} - \frac{2}{\pi \mu}, 0}} \le \min \expr{\cos(\pi/8) + \frac{2 \sin (\pi/8)}{\pi \mu}, 1}
}
for all $\mu > 0$. Hence, by~\eqref{eq:gest0},
\formula[eq:gest00]{
 \mu \int_0^\infty G_\mu(x) dx & \le \min \expr{\cos(\pi/8) + \frac{2 \sin (\pi/8)}{\pi \mu}, 1} - \sqrt{\frac{1 + \sqrt{1 + \mu^2}}{2 \sqrt{1 + \mu^2}}} \, .
}
When $\mu \le \pi / (2 \tan (\pi/8))$, the minimum in~\eqref{eq:gest00} is equal to $1$, and hence the right-hand side of~\eqref{eq:gest00} is clearly an increasing function of $\mu \in (0, \pi / (2 \tan (\pi/8))]$. We claim that it is increasing also when $\mu \ge \pi / (2 \tan (\pi/8))$. Indeed,
\formula{
 & \mu^2 \frac{d}{d\mu} \expr{\cos(\pi/8) + \frac{2 \sin (\pi/8)}{\pi \mu} - \sqrt{\frac{1 + \sqrt{1 + \mu^2}}{2 \sqrt{1 + \mu^2}}}} \\
 & \hspace*{5em} = \frac{1}{2 \sqrt{2} (1 + \mu^{-2})^{5/4} \sqrt{\mu^{-1} + \sqrt{1 + \mu^{-2}}}} - \frac{2 \sin(\pi/8)}{\pi} \, ;
}
the right-hand side is increasing in $\mu \ge \pi / (2 \tan(\pi/8))$ and positive for $\mu = \pi / (2 \tan(\pi/8))$. Hence, the derivative of the right-hand side of~\eqref{eq:gest00} is positive for $\mu \ge \pi / (2 \tan (\pi/8))$, and our claim is proved. The right-hand side of~\eqref{eq:gest00} converges to $\cos(\pi / 8) - 1 / \sqrt{2}$ as $\mu \to \infty$. We conclude that
\formula{
 \int_0^\infty G_\mu(x) dx & \le \frac{1}{\mu} \expr{\cos(\pi/8) - \frac{1}{\sqrt{2}}} < \frac{0.217}{\mu} \, ,
}
proving the other part of~\eqref{eq:gest2}.
\end{proof}

\begin{proposition}
Let $u(x) = (1 + x) e^{-x}$. For $\mu, x > 0$, we have
\formula[eq:gest3]{
 G_\mu(x) & < \frac{0.226 u(x)}{\mu^2 x^2} , &
 -G_\mu'(x) & < \frac{0.458 u(x/2)}{\mu^2 x^3} \, , &
 G_\mu''(x) & < \frac{1.589 u(x/2)}{\mu^2 x^4} \, .
}
\end{proposition}

\begin{proof}
By~\eqref{eq:gamma} and estimates $\mu / (1 + \mu^2)^{1/2} \le 1$, $(r^2 - 1)^{1/2} \le r$, $\mu^2 + r^2 \ge \mu^2$, we have
\formula{
 \gamma_\mu(r) & \le \frac{\sqrt{2}}{2 \pi} \, \frac{r}{\mu^2} \, \ind_{(1, \infty)}(r) .
}
By integration, with the notation $u(x) = (1 + x) e^{-x}$,
\formula{
 G_\mu(x) & = \frac{\sqrt{2}}{2 \pi \mu^2} \int_1^\infty e^{-x r} r dr \le \frac{\sqrt{2}}{2 \pi} \, \frac{(1 + x) e^{-x}}{\mu^2 x^2} = \frac{\sqrt{2}}{2 \pi} \, \frac{u(x)}{\mu^2 x^2} .
}
In a similar manner, using $(1 + x + x^2/2) e^{-x} \le 2 (\sqrt{3} - 1) e^{-(\sqrt{3} - 1)/2} u(x/2) < 1.016 u(x/2)$ (which is easily proved by elementary calculus),
\formula{
 -G_\mu'(x) & = \frac{\sqrt{2}}{2 \pi \mu^2} \int_1^\infty e^{-x r} r^2 dr \le \frac{\sqrt{2}}{\pi} \, \frac{(1 + x + x^2/2) e^{-x}}{\mu^2 x^3} < \frac{\sqrt{2}}{\pi} \, \frac{1.016 u(x/2)}{\mu^2 x^3} \, ,
}
and by $(1 + x + x^2/2 + x^3/6) e^{-x} \le 4 e^{-\sqrt{3/2}} u(x/2) < 1.176 u(x/2)$ (again easily proved by a short calculation),
\formula{
 G_\mu''(x) & \le \frac{3 \sqrt{2}}{\pi} \, \frac{(1 + x + x^2/2 + x^3/6) e^{-x}}{\mu^2 x^4} < \frac{3 \sqrt{2}}{\pi} \, \frac{1.176 u(x/2)}{\mu^2 x^4} \, .
\qedhere
}
\end{proof}

Notation $u(x) = (1 + x) e^{-x}$ is kept throughout this section. Note that $u(x)$ is a decreasing function of $x \in (0, \infty)$.

\subsection{Estimates for eigenvalues}

We are now in position to apply Lemmas~\ref{lem:approx0} and~\ref{lem:norm0}. Recall that $\lambda_n$ and $E_n = \lambda_n + 1$ are the eigenvalues of $A$ and $H$, respectively, and $0 < \lambda_1 < \lambda_2 \le \lambda_3 \le ... \to \infty$. The corresponding normalized eigenfunctions are denoted by $\Phi_n$. On the other hand, $\tilde{\lambda}_n = (\tilde{\mu}_n^2 + 1)^{1/2} - 1$, $\tilde{E}_n = (\tilde{\mu}_n^2 + 1)^{1/2}$ and $\tilde{\ph}_n / \|\tilde{\ph}_n\|_{L^2(D)}$ (see~\eqref{eq:tildemu},~\eqref{eq:tildelambda} and~\eqref{eq:phitilde}) are approximations to $\lambda_n$, $E_n$ and $\Phi_n$. In the definition of $\tilde{\ph}_n$, we take $b = a/3$ (this choice optimizes some constants below). The following results are stated for $A$, $\lambda_n$ and $\tilde{\lambda}_n$, but they can be trivially reformulated for $H$, $E_n$ and $\tilde{E}_n$.

\begin{lemma}
\label{lem:main}
For every $a > 0$, the eigenvalues $\lambda_n$ are simple. The eigenfunction $\Phi_n$ is an even function when $n$ is odd, and $\Phi_n$ is an odd function when $n$ is even. Furthermore, for all $a > 0$ and $n = 1, 2, ...$,
\formula[eq:dist3]{
 \| A \tilde{\ph}_n - \tilde{\lambda}_n \tilde{\ph}_n \|_{L^2(D)} & < \frac{(1 + a/3) e^{-a/3}}{\tilde{\mu}_n a^{3/2}} \expr{2.497 + \frac{9.356}{\tilde{\mu}_n a} + \frac{10.314}{\tilde{\mu}_n^2 a^2}}^{1/2} . % c12, c12, c12
}
If $a > 0$ and $\tilde{\mu}_n a > 1.251$, then
\formula[eq:dist2]{
 |\lambda_n - \tilde{\lambda}_n| & < \frac{(1 + a/3) e^{-a/3}}{\tilde{\mu}_n a^2} \expr{2.497 + \frac{9.356}{\tilde{\mu}_n a} + \frac{10.314}{\tilde{\mu}_n^2 a^2}}^{1/2} \expr{1 - \frac{1.251}{\tilde{\mu}_n a}}^{-1/2} . % c12, c13
}
\end{lemma}

\begin{proof}
The argument is divided into four steps.

\emph{Step 1.}
We estimate each term on the right-hand side of~\eqref{eq:approx0} and~\eqref{eq:norm0}, so that in the next step we can apply Lemmas~\ref{lem:approx0} and~\ref{lem:norm0}. We use the notation~\eqref{eq:constants}.

Recall that $b = a / 3$. We introduce the function $w(\xi) = (\xi + 1)^{1/2} - 1$, as in Preliminaries. Denote, in this step only, $\kappa = 1/3$, so that $b = \kappa a$. Since $(a - b)/2 \ge b$ and $a - b \ge b$, by~\eqref{eq:gest3} we have
\formula{
 G_{\mu,b}(a) & = G_\mu(a - b) + G_\mu(a + b) < 0.226 \, \frac{(1 - \kappa)^{-2} + (1 + \kappa)^{-2}}{\mu^2 a^2} \, u(b) \le \frac{0.636 u(b)}{\mu^2 a^2} \, , \\ % c1
 -G_{\mu,b}'(a) & = -G_\mu'(a - b) - G_\mu'(a + b) < 0.458 \, \frac{(1 - \kappa)^{-3} + (1 + \kappa)^{-3}}{\mu^2 a^3} \, u(b) \le \frac{1.739 u(b)}{\mu^2 a^3} \, , \\ % c2
 G_{\mu,b}''(a) & = G_\mu''(a - b) + G_\mu''(a + b) < 1.589 \, \frac{(1 - \kappa)^{-4} + (1 + \kappa)^{-4}}{\mu^2 a^4} \, u(b) \le \frac{8.548 u(b)}{\mu^2 a^4} \, . % c3
}
Hence,
\formula{
 G_{\mu,b}(a) - 2 b G_{\mu,b}'(a) + b^2 G_{\mu,b}''(a) & < (0.636 + 2 \cdot 1.739 \kappa + 8.548 \kappa^2) \, \frac{u(b)}{\mu^2 a^2} \le \frac{2.746 u(b)}{\mu^2 a^2} \, . % c1, c2, c3, c4
}
By~\eqref{eq:nu0est}, $\nu_0(b) \le b / \pi$ and $\nu_0(2b) \le 2 b / \pi$. Hence,
\formula{
 & \frac{G_{\mu,b}(a) \nu_0(2 b)}{2 b^2} \le \frac{0.636 u(b)}{\kappa \pi \mu^2 a^3} \le \frac{0.608 u(b)}{\mu^2 a^3} \, , \\ % c1, c5
 & \frac{(G_{\mu,b}(a) - 2 b G_{\mu,b}'(a) + b^2 G_{\mu,b}''(a)) \nu_0(b)}{b^2} \le \frac{2.746 u(b)}{\kappa \pi \mu^2 a^3} \le \frac{2.623 u(b)}{\mu^2 a^3} \, . % c4, c6
}
By~\eqref{eq:nuinftyest} we have $\nu_\infty(b) \le 1 / (\pi b)$, so that
\formula{
 2 G_{\mu,b}(a) \nu_\infty(b) & < \frac{2 \cdot 0.636 u(b)}{\kappa \pi \mu^2 a^3} \le \frac{1.215 u(b)}{\mu^2 a^3} \, . % c1, c7
}
Since $w(\mu^2) = (\mu^2 + 1)^{1/2} - 1 \le \mu$, we also have
\formula{
 \frac{w(\mu^2) G_{\mu,b}(a)}{2} & < \frac{0.636 u(b)}{2 \mu a^2} = \frac{0.318 u(b)}{\mu a^2} % c1, c8
}
By~\eqref{eq:nuest}, $\nu(x) \le u(x) / (\pi x^2)$. This and~\eqref{eq:gest2} give
\formula{
 \nu(b) I_\mu & < \frac{0.217 u(b)}{\kappa^2 \pi \mu a^2} \le \frac{0.622 u(b)}{\mu a^2} \, , & \nu(2b) I_\mu & < \frac{0.217 u(b)}{4 \kappa^2 \pi \mu a^2} < \frac{0.156 u(b)}{\mu a^2} \, , \\ % c, c9, c, c10
 \frac{2 \nu(a)}{\mu} & \le \frac{2 u(a)}{\pi \mu a^2} < \frac{0.637 u(a)}{\mu a^2} \, . % c11
}

\emph{Step 2.}
Recall that $\tilde{\lambda}_n = w(\tilde{\mu}_n^2)$. For brevity, denote $\tilde{\mu} = \tilde{\mu}_n$. By Lemma~\ref{lem:approx0}, we obtain
\formula{
 \| A \tilde{\ph}_n - \tilde{\lambda}_n \tilde{\ph}_n \|_{L^2(D)}^2 & < 2 (a - b) \expr{\frac{0.608 u(b)}{\tilde{\mu}^2 a^3} + \frac{0.156 u(b)}{\tilde{\mu} a^2} + \frac{0.637 u(b)}{\tilde{\mu} a^2}}^2 \\
 & \hspace*{-4em} + 2 b \expr{\frac{2.623 u(b)}{\tilde{\mu}^2 a^3} + \frac{1.215 u(b)}{\tilde{\mu}^2 a^3} + \frac{0.622 u(b)}{\tilde{\mu} a^2} + \frac{0.318 u(b)}{\tilde{\mu} a^2} + \frac{0.637 u(b)}{\tilde{\mu} a^2}}^2 \\ % c5, c10, c11, c6, c7, c9, c8, c11
 & \hspace*{10em} \le \frac{(u(b))^2}{\tilde{\mu}^2 a^3} \expr{2.497 + \frac{9.356}{\tilde{\mu} a} + \frac{10.314}{\tilde{\mu}^2 a^2}} . % c12, c12, c12
}
This proves~\eqref{eq:dist3}. Lemma~\ref{lem:norm0}, $\thet_\mu \le \pi/8$ and~\eqref{eq:gest2} give
\formula[eq:norm1]{
 \|\tilde{\ph}_n\|_{L^2(D)}^2 & \ge a - \frac{\sin \thet_{\tilde{\mu}}}{\tilde{\mu}} - 4 I_{\tilde{\mu}} > a - \frac{0.383}{\tilde{\mu}} - \frac{0.868}{\tilde{\mu}} = a - \frac{1.251}{\tilde{\mu}} \, . % c, c, c13
}
By Proposition~\ref{prop:dist0}, there is an eigenvalue $\lambda_{k(n)}$ (\textit{a priori} $k(n)$ may depend on $a$, though later we show that in fact $k(n) = n$) such that
\formula[eq:dist1]{
 |\lambda_{k(n)} - \tilde{\lambda}_n| & < \frac{u(b)}{\tilde{\mu}_n a^2} \expr{2.497 + \frac{9.356}{\tilde{\mu}_n a} + \frac{10.314}{\tilde{\mu}_n^2 a^2}}^{1/2} \expr{1 - \frac{1.251}{\tilde{\mu}_n a}}^{-1/2} , % c12, c13
}
provided that the right-hand side is well-defined, that is, $\tilde{\mu}_n a > 1.251$. In the next step we show that this is the case when $n \ge 3$. Proposition~\ref{prop:dist0} also states that we may assume that the eigenfunction $\Phi_{k(n)}$ has the same type of parity as $\tilde{\ph}_n$, that is, it is an even function for odd $n$, and an odd function for even $n$.

For $n \ge 3$, we denote the right-hand side of~\eqref{eq:dist1} by $\eps_n$, and define the intervals
\formula{
 I_n & = (\tilde{\lambda}_n - \eps_n, \tilde{\lambda}_n + \eps_n) .
}
We have thus proved that each of the intervals $I_n$ ($n \ge 3$) contains an eigenvalue $\lambda_{k(n)}$. Below we show that these intervals are mutually disjoint.

\emph{Step 3.}
Clearly, $\eps_n$ decreases with $n$. Furthermore, by~\eqref{eq:muest1}, $a \tilde{\mu}_3 \ge 3 \pi / 2 - \pi/8 > 4.319$ and $a \tilde{\mu}_4 \ge 4 \pi/2 - \pi/8 > 5.890$. It follows that the right-hand side of~\eqref{eq:dist1} is well-defined for $n \ge 3$ (as claimed in the previous step), and % c14, c16
\formula{
 \eps_3 & < \frac{0.200 \pi u(b)}{a} \, , & \eps_4 & < \frac{0.128 \pi u(b)}{a} \, . % c15, c17
}
Since $w(\mu^2) = (\mu^2 + 1)^{1/2} - 1$ is a convex function of $\mu$, by the mean value theorem, for $\mu_1 < \mu_2$ we have
\formula[eq:convex]{
 w(\mu_2^2) - w(\mu_1^2) & \ge 2 \mu_1 w'(\mu_1^2) (\mu_2 - \mu_1) = \frac{\mu_1 (\mu_2 - \mu_1)}{\sqrt{1 + \mu_1^2}} \, .
}
Hence, using also~\eqref{eq:muest1} and $\tilde{\mu}_n \ge 4.319 / a$ ($n \ge 3$), we obtain that for $n \ge 3$, % c14
\formula{
 \expr{w(\tilde{\mu}_{n+1}^2) - \eps_{n+1}} - \expr{w(\tilde{\mu}_n^2) + \eps_n} & > \frac{\tilde{\mu}_n (\tilde{\mu}_{n+1} - \tilde{\mu}_n)}{\sqrt{1 + \tilde{\mu}_n^2}} - \frac{(0.200 + 0.128) \pi u(b)}{a} \\ % c15, c17
 & \hspace{-10em} \ge \frac{4.319}{\sqrt{a^2 + 4.319^2}} \, \frac{3 \pi}{8 a} - \frac{3 \pi u(b)}{8 a} \ge \frac{3 \pi}{8 a} \expr{\frac{4.319}{\sqrt{a^2 + 4.319^2}} - u(a/3)} > 0 ; % c14, c14, k
}
the last inequality is proved as follows:
\formula{
 u(s/3) \, \frac{\sqrt{4.319^2 + s^2}}{4.319} & = \expr{1 + \frac{s}{3}} e^{-s/3} \sqrt{1 + \frac{s^2}{4.319^2}} \le \expr{1 + \frac{s}{3}} e^{-s/3} \sqrt{1 + \frac{s^2}{18}} \\
 & \le \expr{1 + \frac{s}{3}} e^{-s/3} \expr{1 + \frac{s^2}{36}} = \expr{1 + \frac{s}{3} + \frac{s^2}{36} + \frac{s^3}{108}} e^{-s/3} ,
}
and the right-hand is equal to $1$ for $s = 0$, and is strictly decreasing in $s > 0$:
\formula{
 \frac{d}{d s} \expr{\expr{1 + \frac{s}{3} + \frac{s^2}{36} + \frac{s^3}{108}} e^{-s/3}} & = -\frac{s e^{-s/3}}{324} \expr{s^2 - 6s + 18} < 0 .
}
It follows that the intervals $I_n$ ($n \ge 3$) are mutually disjoint. (In fact we have proved that the larger intervals $(w((n \pi / 2 - \pi/8)^2 / a^2) - \eps_n, w((n \pi / 2)^2 / a^2) + \eps_n)$ are mutually disjoint, see Figure~\ref{fig:intervals}.) Since $\lambda_{k(n)} \in I_n$, all numbers $k(n)$ ($n \ge 3$) are distinct. Furthermore, by an argument similar to the one used above, % c14, k, k, c14, k
\formula{
 \expr{w(\tilde{\mu}_3^2) - \eps_3} - w(\pi^2 / a^2) & > \frac{\pi (\tilde{\mu}_3 - \pi / a)}{a \sqrt{1 + \pi^2 / a^2}} - \frac{0.200 \pi u(b)}{a} \\
 & \ge \frac{\pi}{a} \expr{\frac{3}{8} \, \frac{\pi}{\sqrt{a^2 + \pi^2}} - 0.200 u(a/3)} > 0 ; % c15, c15, k
}
indeed, for the last inequality we have
\formula{
 u(s/3) \, \frac{\sqrt{\pi^2 + s^2}}{\pi} & \le \expr{1 + \frac{s}{3}} e^{-s/3} \sqrt{1 + \frac{s^2}{9}} \le \expr{1 + \frac{s}{3}}^2 e^{-s/3} \le \frac{4}{e} < \frac{3/8}{0.200} \, .
}
Hence, $\lambda_{k(n)} > w(\pi^2 / a^2)$ for $n \ge 3$. On the other hand, by~\eqref{eq:cs}, $\lambda_j \le w((j \pi)^2 / (2 a)^2)$. It follows that $k(n) \ge 3$ for $n \ge 3$. To show simplicity of eigenvalues, it remains to prove that there are no other eigenvalues than $\lambda_1$ (which is known to be simple), $\lambda_2$ (which lies in the interval $(\lambda_1, w(\pi^2 / a^2))$) and $\lambda_{k(n)}$ ($n \ge 3$; all distinct and greater than $w(\pi^2 / a^2)$). This is done in the next step.

\begin{figure}
\includegraphics[width=0.7\textwidth]{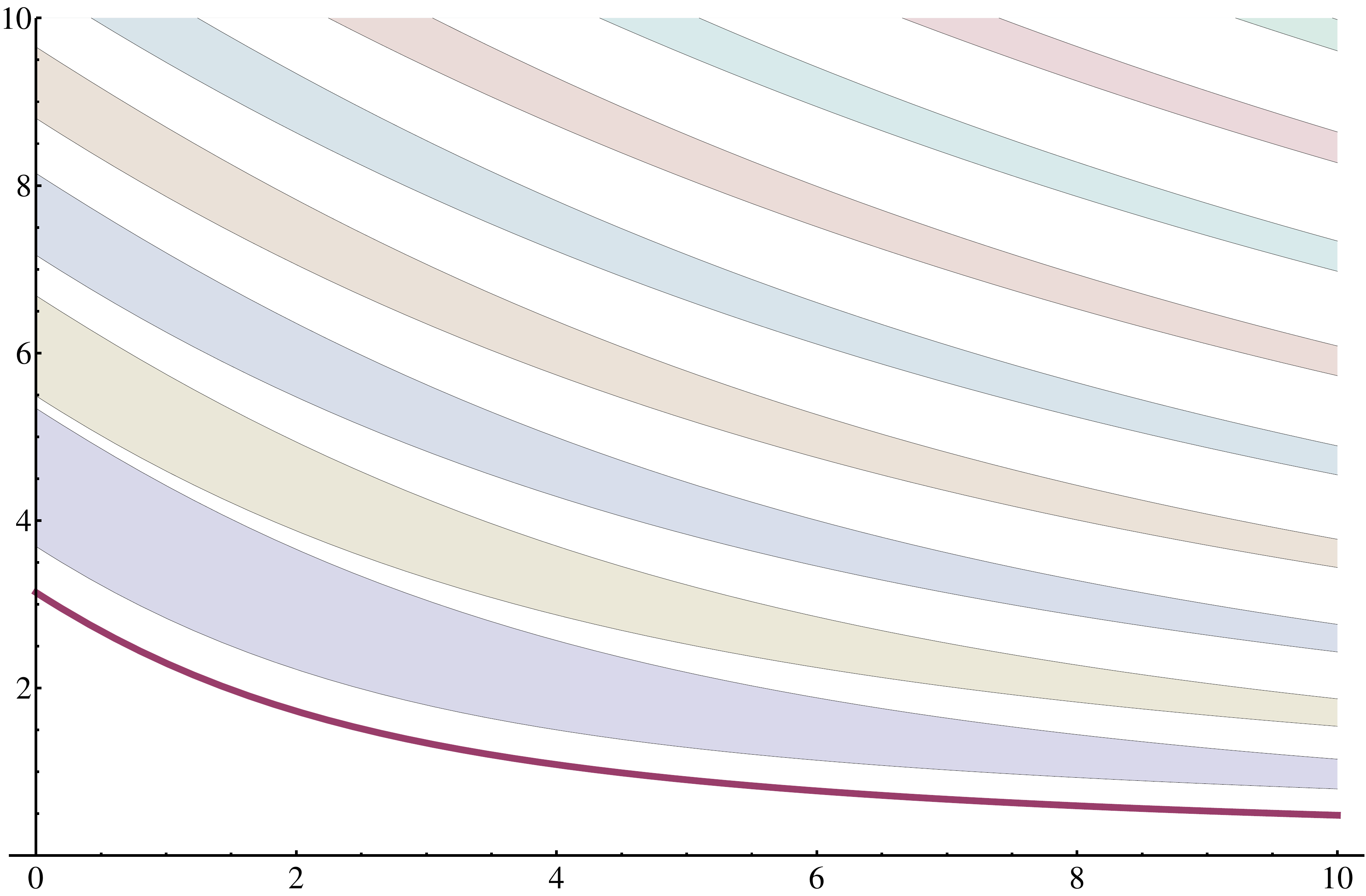}
\caption{Plot of the intervals $(a \, w((n \pi/2 - \pi/8)^2 / a^2) - a \eps_n, a \, w((n \pi/2)^2 / a^2) + a \eps_n)$ for $n = 3, 4, 5, ...$ (each of them contains $a \lambda_n$, an eigenvalue multiplied by $a$), and the function $a \, w(\pi^2 / a^2)$ (the upper bound for $a \lambda_2$, the second eigenvalue multiplied by $a$); here $w(\xi) = (\xi + 1)^{1/2} - 1$.}
\label{fig:intervals}
\end{figure}

\emph{Step 4.}
Recall that $T(t; x, y)$ ($x, y \in \R$, $t > 0$) and $T_0(t; x - y)$ ($x, y \in D$, $t > 0$) are kernel functions of the (sub-)Markov operators $\exp(-t A)$ and $\exp(-t A_0)$. The Fourier transform of $T_0(t; x)$ is $\exp(-t w(\xi^2))$, and $0 \le T(t; x, y) \le T_0(t; x - y)$ for all $x, y \in D = (-a, a)$, $t > 0$. For $t > 0$, we have the following trace estimate (cf.~\cite{bib:bk09, bib:k98}; \cite[Section~9]{bib:kkms10}; \cite[Section~5]{bib:k11})
\formula{
 \sum_{j = 1}^\infty e^{-\lambda_j t} & = \int_D \expr{\sum_{j = 1}^\infty e^{-\lambda_j t} (\Phi_j(x))^2} dx = \int_D T(t; x, x) dx \\
 & \le \int_D T_0(t; 0) dx = \frac{2a}{\pi} \int_0^\infty e^{-t w(s^2)} ds .
}
In the last step, Fourier inversion formula was used. Below this estimate is compared with the lower bound for the sum of $\exp(-\lambda_{k(n)} t)$.

For $a > 0$ and $n \ge 3$, let
\formula{
 \delta_n & = \frac{\sqrt{1 + \tilde{\mu}_n^2}}{\tilde{\mu}_n} \, \eps_n
}
Clearly, $\delta_n$ is decreasing and $\delta_n \to 0$ as $n \to \infty$. By~\eqref{eq:convex},
\formula{
 w((\tilde{\mu}_n + \delta_n)^2) - w(\tilde{\mu}_n^2) & \ge 2 \tilde{\mu}_n w'(\tilde{\mu}_n^2) \delta_n = \frac{\tilde{\mu}_n}{\sqrt{1 + \tilde{\mu}_n^2}} \, \delta_n = \eps_n .
}
Hence,
\formula{
 \lambda_{k(n)} & \le w(\tilde{\mu}_n^2) + \eps_n \le w((\tilde{\mu}_n + \delta_n)^2) .
}
Furthermore, by the definition~\eqref{eq:tildemu} of $\tilde{\mu}_n$ and monotonicity of $\thet_\mu$,
\formula{
 \tilde{\mu}_{n+1} - \tilde{\mu}_n & = \frac{\pi}{2 a} - \frac{1}{a} \, (\thet_{\tilde{\mu}_{n+1}} - \thet_{\tilde{\mu}_n}) \le \frac{\pi}{2 a} \, .
}
Using these two inequalities and monotonicity of $\delta_n$, we obtain that for any $N \ge 3$,
\formula{
 \frac{\pi}{2 a} \sum_{n = N}^\infty e^{-\lambda_{k(n)} t} & \ge \frac{\pi}{2 a} \sum_{n = N}^\infty \exp\bigl( -t w((\tilde{\mu}_n + \delta_N)^2) \bigr) \\
 & \ge  \sum_{n = N}^\infty (\tilde{\mu}_{n+1} - \tilde{\mu}_n) \exp\bigl( -t w((\tilde{\mu}_n + \delta_N)^2) \bigr) \ge \int_{\tilde{\mu}_N + \delta_N}^\infty e^{-t w(s^2)} ds .
}
Let $J_N$ be the set of those $j \ge 1$ for which $j \ne k(n)$ for all $n \ge N$. It follows that
\formula{
 \sum_{j \in J_N} e^{-\lambda_j t} & = \sum_{j = 1}^\infty e^{-\lambda_j t} - \sum_{n = N}^\infty e^{-\lambda_{k(n)} t} \\
 & \le \frac{2 a}{\pi} \int_0^{\tilde{\mu}_N + \delta_N} e^{-t w(s^2)} ds \le \frac{2 a}{\pi} \, (\tilde{\mu}_N + \delta_N) .
}
Taking the limit as $t \to 0^+$, and using the definition~\eqref{eq:tildemu} of $\tilde{\mu}_N$, we obtain that
\formula{
 \# J_N & \le \frac{2a}{\pi} \, (\tilde{\mu}_N + \delta_N) = N - \frac{2}{\pi} \, (\thet_{\tilde{\mu}_N} - a \delta_N) .
}
Note that as $N \to \infty$, we have $\delta_N \to 0$, and $\thet_{\tilde{\mu}_N} \to \pi/8$. Take $N$ sufficiently large, so that $a \delta_N < \thet_{\tilde{\mu}_N}$ ($a > 0$ is fixed). For this $N$, we have $\# J_N < N$. On the other hand, $J_N$ must contain $\lambda_1$, $\lambda_2$ and $\lambda_{k(n)}$ for $n = 3, 4, ..., N-1$. It follows that $\# J_N = N - 1$, and therefore there are no other eigenvalues than $\lambda_1$, $\lambda_2$ and $\lambda_{k(n)}$ ($n \ge 3$), as claimed in the previous step. In particular, $k(n) = n$ for all $n \ge 3$, and all eigenvalues are simple.

Formula~\eqref{eq:dist2} follows now from~\eqref{eq:dist1}. Furthermore, $\Phi_n = \Phi_{k(n)}$ is an even function for odd $n \ge 3$, and an odd function for even $n \ge 3$ (by Proposition~\ref{prop:dist0}; see Step~2). Clearly, $\Phi_1$ is an even function (e.g. because it is positive in $D$). Finally, the fact that $\Phi_2$ is an odd function follows by a minor modification of~\cite[proof of Theorem~3.4]{bib:cs05}: by restricting attention to either even or odd functions, one obtains that (see Proposition~\ref{prop:dist0} for the notation)
\formula[eq:chen-song]{
\begin{array}{c}
 \text{the $n$-th smallest eigenvalue of $A^{\even}$ does not exceed $w((n \pi - \pi/2)^2 / a^2)$;} \\
 \text{the $n$-th smallest eigenvalue of $A^{\odd}$ does not exceed $w((n \pi)^2 / a^2)$.}
\end{array}
}
In particular, the smallest eigenvalue of $A^{\odd}$ is less then $w(\pi^2 / a^2)$, so the only candidates are $\lambda_1$ and $\lambda_2$ (as shown in the previous step). Since $\Phi_1$ is an even function, we conclude that $\Phi_2$ is odd. The lemma is proved.
\end{proof}

\begin{remark}
Alternatively, one could prove that $\Phi_2$ is odd as follows. Clearly, $a \tilde{\mu}_2 \ge \pi - \pi/8 > 2.748$. Hence, if $\eps_2$ denotes the right-hand side of~\eqref{eq:dist1} for $n = 2$, then $\eps_2 < 0.424 \pi u(b) / a$, and by Proposition~\ref{prop:dist0}, there is an eigenvalue $\lambda_{k(2)}$ of $A^{\odd}$ in the interval $I_2 = (\tilde{\lambda}_2 - \eps_2, \tilde{\lambda}_2 + \eps_2)$. By using the results of Step~3, one easily proves that
\formula{
 \tilde{\lambda}_2 + \eps_2 < \tilde{\lambda}_4 - \eps_4 ,
}
and so $k(2) < 4$. Since $\Phi_1$ and $\Phi_3$ are even functions, we must have $k(2) = 2$, and so $\Phi_2$ is an odd function.
\end{remark}

\begin{remark}
Clearly, a higher-dimensional analogue of~\eqref{eq:chen-song} holds true: Let $D = B(0, a)$ be a ball in $\R^d$ and $V(x) = 0$ for $x \in D$, $V(x) = \infty$ otherwise. Fix a $d$-dimensional spherical harmonic $Y(u)$ ($|u| = 1$) of order $l \ge 0$. Suppose that $w(\xi) = (\xi + 1)^{1/2} - 1$ (or, more generally, that $w(\xi)$ is an operator monotone function), and let $A_0 = w(-\Delta)$, where $\Delta$ is the Laplace operator in $\R^d$. Let $(\mu_n^Y)^2$ and $\lambda_n^Y$ denote the $n$-th smallest eigenvalue of $-\Delta + V(x)$ and $A_0 + V(x)$, respectively, among the eigenvalues corresponding to the eigenfunctions of the form $f(|x|) Y(x / |x|)$ (therefore, $a \mu_n^Y$ is the $n$-th zero of the Bessel function $J_{d/2+l-1}$). Then $\lambda_n^Y \le w((\mu_n^Y)^2)$. The proof again is a minor modification of the argument used in~\cite[proof of Theorem~3.4]{bib:cs05}.
\end{remark}

\begin{proof}[Proof of Theorem~\ref{th2}]
Recall that by a scaling argument, we only need to prove the statement of the theorem for $\hbar = c = m = 1$; that is, we prove Theorem~\ref{th2}'. The proof relies on Lemma~\ref{lem:main}. By~\eqref{eq:muest1}, we have $a \tilde{\mu}_n \ge 7 \pi/8 > 2.748$ for $n \ge 2$, and by~\eqref{eq:muest2} and~\eqref{eq:thetaest}, $a \tilde{\mu}_1 \ge \pi/2 - \thet_{\pi/(2 a)} \ge \pi/2 - 1/a > 1.520$ when $n = 1$ and $a \ge 20$. By~\eqref{eq:dist2}, in both cases, % c18, c19
\formula{
 |\lambda_n - \tilde{\lambda}_n| & < 8.610 \, \frac{(1 + a/3) e^{-a/3}}{\tilde{\mu}_n a^2} \le 8.610 \, \frac{(1 + a/3) e^{-a/3}}{(n \pi/2 - \pi/8) a} \\ % c20, c20
 & \le 8.610 \, \frac{8 (1 + a/3) e^{-a/3}}{3 n \pi a} \le 7.309 \, \frac{(1 + a/3) e^{-a/3}}{n a} \, . % c20, c21
}
Formulae~\eqref{eq:asymp2} and~\eqref{eq:asymp2prime} for $n \ge 2$ or $a \ge 20$ are proved. In the remaining case $n = 1$ and $0 < a < 20$, we use a more direct approach. By~\eqref{eq:cs} and the definition of $\tilde{\lambda}_n$ (see~\eqref{eq:tildemu} and~\eqref{eq:tildelambda}), both $\lambda_1$ and $\tilde{\lambda}_1$ are bounded above by $w(\pi^2 / (4 a^2)) = (\pi^2 / (4 a^2) + 1)^{1/2} - 1$. Hence,
\formula{
 |\lambda_1 - \tilde{\lambda}_1| & \le \sqrt{\frac{\pi^2}{4 a^2} + 1} - 1 = \frac{1}{2a} \, \frac{\pi^2}{\sqrt{\pi^2 + 4 a^2} + 2 a} \le \min\expr{\frac{\pi}{2 a} , \frac{\pi^2}{8 a^2}} .
}
Observe that $e^{a/3} / (6 + 2 a)$ is increasing in $a > 0$. Hence, for $a \in (0, 6]$,
\formula{
 \frac{a}{(1 + a/3) e^{-a/3}} \, \frac{\pi}{2 a} & = \frac{3 \pi e^{a/3}}{6 + 2 a} \le \frac{3 \pi e^2}{6 + 12} < 3.869 . % c22
}
In a similar manner, $e^{a/3} / (a (3 + a))$ is increasing for $a \ge 6$, and so for $a \in [6, 20]$,
\formula{
 \frac{a}{(1 + a/3) e^{-a/3}} \, \frac{\pi^2}{8 a^2} & = \frac{3 \pi^2 e^{a/3}}{8 a (3 + a)} \le \frac{3 \pi^2 e^{20/3}}{8 \cdot 20 \cdot 23} < 6.323 . % c23
}
It follows that
\formula[eq:asymp2a]{
 |\lambda_n - \tilde{\lambda}_n| & < 7.309 \, \frac{(1 + a/3) e^{-a/3}}{n a} % c21
}
for all $n = 1, 2, ...$ and $a > 0$. Since $E_n = \lambda_n + 1$ and $\tilde{\lambda}_n = (\tilde{\mu}_n + 1)^{1/2} - 1$, formula~\eqref{eq:asymp2a} implies~\eqref{eq:asymp2prime}.
\end{proof}

\begin{proof}[Proof of Theorem~\ref{th1}]
Recall that by~\eqref{eq:thetaest}, $\thet_\mu = \pi/8 + O(1/\mu)$ as $\mu \to \infty$. Furthermore, $\tilde{\mu}_n \ge n \pi / (2 a) - \pi / (8 a)$, so that $\thet_{\tilde{\mu}_n} = \pi/8 + O(1/n)$ as $n \to \infty$. Hence,
\formula{
 \tilde{\mu}_n & = \frac{n \pi}{2 a} - \frac{\thet_{\tilde{\mu}_n}}{a} = \frac{n \pi}{2 a} - \frac{\pi}{8 a} + O\expr{\frac{1}{n}} ;
}
$O(1/n)$ here and below corresponds to asymptotic estimates as $n \to \infty$. It follows that
\formula{
 \tilde{\lambda}_n & = \sqrt{\tilde{\mu}_n^2 + 1} - 1 = \tilde{\mu}_n - 1 + O(\tilde{\mu}_n^{-1}) = \frac{n \pi}{2 a} - \frac{\pi}{8 a} - 1 + O\expr{\frac{1}{n}} .
}
By~\eqref{eq:asymp2a}, we obtain that 
\formula{
 \lambda_n & = \frac{n \pi}{2 a} - \frac{\pi}{8 a} - 1 + O\expr{\frac{1}{n}} ,
}
as desired (recall that $E_n = \lambda_n + 1$). Simplicity of eigenvalues is a part of Lemma~\ref{lem:main}.
\end{proof}

\begin{proof}[Verification of Remark~\ref{rem:limits}]
In this part the general case is studied, without the scaling assumption $c = \hbar = m = 1$.
\begin{enumerate}
\item
We fix $\hbar$, $c$, $a$ and $n$. When $m \to 0^+$, the right-hand side of~\eqref{eq:asymp2} converges to $8 \hbar c / (a n)$. On the other hand, by~\eqref{eq:tildemu0} and the estimates~\eqref{eq:thetaest} of $\thet_\mu$, we have (as in the proof of Theorem~\ref{th1}) $\tilde{\mu}_n \ge (n \pi/2 - \pi/8) \hbar / (m c a)$, $\thet_{\tilde{\mu}_n} = \pi/8 + O(m)$, and
\formula{
 \tilde{\mu}_n & = \frac{\hbar}{m c a} \expr{\frac{n \pi}{2} - \thet_{\tilde{\mu}_n}} = \frac{\hbar}{m c a} \expr{\frac{n \pi}{2} - \frac{\pi}{8} + O(m)} ,
}
as $m \to 0^+$. It follows that
\formula{
 m c^2 \sqrt{\tilde{\mu}_n^2 + 1} & = m c^2 \expr{\tilde{\mu}_n + O(\tilde{\mu}_n^{-1})} = \frac{\hbar c}{a} \expr{\frac{n \pi}{2} - \frac{\pi}{8} + O(m)} .
}
Therefore, passing to a limit $m \to 0^+$ in~\eqref{eq:asymp2}, we conclude that
\formula{
 \abs{E_n - \frac{\hbar c}{a} \expr{\frac{n \pi}{2} - \frac{\pi}{8}}} & \le 8 \, \frac{\hbar c}{a n} \, .
}
\item
Now $\hbar$, $m$, $a$ and $n$ are fixed, and we let $c \to \infty$. The right-hand side of~\eqref{eq:asymp2} converges to $0$ exponentially fast. Furthermore, $\thet_\mu = O(\mu)$ as $\mu \to 0^+$, and by~\eqref{eq:tildemu0}, $\tilde{\mu}_n = O(1/c)$ as $c \to \infty$. Hence, $\thet_{\tilde{\mu}_n} = O(1/c)$, and so
\formula{
 \tilde{\mu}_n & = \frac{\hbar}{m c a} \expr{\frac{n \pi}{2} - \thet_{\tilde{\mu}_n}} = \frac{\hbar}{m c a} \expr{\frac{n \pi}{2} + O\expr{\frac{1}{c}}}
}
as $c \to \infty$. This gives
\formula{
 m c^2 \sqrt{\tilde{\mu}_n^2 + 1} & = m c^2 \expr{1 + \tilde{\mu}_n^2/2 + O(\tilde{\mu}_n^4)} \\
 & = m c^2 + \frac{\hbar^2}{2 m a^2} \expr{\expr{\frac{n \pi}{2}}^2 + O\expr{\frac{1}{c}}} ,
}
and~\eqref{eq:asymp1alt} follows by~\eqref{eq:asymp2}. Higher order (with respect to $1/c$) expansions similar to~\eqref{eq:asymp1alt} can be easily obtained from~\eqref{eq:asymp2} in a similar manner.
\item
Denote, for brevity, $\bar{a} = \hbar^{-1} m c a$. By~\eqref{eq:tildemu0}, \eqref{eq:dthetaest} and again~\eqref{eq:tildemu0},
\formula{
 \abs{\tilde{\mu}_n - \frac{1}{\bar{a}} \expr{\frac{n \pi}{2} - \thet\expr{\frac{n \pi}{2 \bar{a}}}}} & = \frac{1}{\bar{a}} \abs{\thet\expr{\frac{n \pi}{2 \bar{a}}} - \thet_{\tilde{\mu}_n}} \\
 & \le \frac{2}{\pi \bar{a}} \, \frac{1}{1 + \tilde{\mu}_n^2} \expr{\frac{n \pi}{2 \bar{a}} - \tilde{\mu}_n} = \frac{2}{\pi \bar{a}^2} \, \frac{\theta(\tilde{\mu}_n)}{1 + \tilde{\mu}_n^2} .
}
By~\eqref{eq:thetaest}, $\thet_{\tilde{\mu}_n} \le \thet_{n \pi / (2 \bar{a})} \le \min(n / \bar{a}, \pi/8) \le (\pi/2) n / (\bar{a} + n)$. Furthermore, $\tilde{\mu}_n \ge n \pi / (2 \bar{a}) - \pi / (8 \bar{a}) \ge 3 n \pi / (8 \bar{a}) \ge n / \bar{a}$. It follows that
\formula{
 \abs{\tilde{\mu}_n - \frac{1}{\bar{a}} \expr{\frac{n \pi}{2} - \thet\expr{\frac{n \pi}{2 \bar{a}}}}} & \le \frac{n}{\bar{a} + n} \, \frac{1}{\bar{a}^2 + n^2} \le \frac{2 n}{(\bar{a} + n)^3} \, .
}
Finally, since $w(\mu^2) = (\mu^2 + 1)^{1/2} - 1$ is convex, by the mean value theorem,
\formula{
 \hspace*{3em} & \hspace*{-3em} \abs{w(\tilde{\mu}_n^2) - w\expr{\frac{1}{\bar{a}^2} \expr{\frac{n \pi}{2} - \thet\expr{\frac{n \pi}{2 \bar{a}}}}^2}} \\
 & \le 2 \, \frac{n \pi}{2 \bar{a}} \, w'\expr{\expr{\frac{n \pi}{2 \bar{a}}^2}} \abs{\tilde{\mu}_n - \frac{1}{\bar{a}} \expr{\frac{n \pi}{2} - \thet\expr{\frac{n \pi}{2 \bar{a}}}}} \\
 & = \expr{\expr{\frac{2 \bar{a}}{n \pi}}^2 + 1}^{-1/2} \abs{\tilde{\mu}_n - \frac{1}{\bar{a}} \expr{\frac{n \pi}{2} - \thet\expr{\frac{n \pi}{2 \bar{a}}}}} .
}
Using $(2 \bar{a} / (n \pi))^2 + 1 \ge (1/2) (2 \bar{a} / (n \pi) + 1)^2 \ge (2 / \pi^2) (\bar{a} / n + 1)^2$, we conclude that
\formula{
 \hspace*{3em} & \hspace*{-3em} \abs{w(\tilde{\mu}_n^2) - w\expr{\frac{1}{\bar{a}^2} \expr{\frac{n \pi}{2} - \thet\expr{\frac{n \pi}{2 \bar{a}}}}^2}} \\
 & \le \frac{\pi}{\sqrt{2}} \, \frac{n}{\bar{a} + n} \abs{\tilde{\mu}_n - \frac{1}{\bar{a}} \expr{\frac{n \pi}{2} - \thet\expr{\frac{n \pi}{2 \bar{a}}}}} \le \frac{2 \pi}{\sqrt{2}} \, \frac{n^2}{(\bar{a} + n)^4} \, .
}
This and~\eqref{eq:asymp2a} proves that
\formula[eq:asymp2alt2]{
\begin{aligned}
 \hspace*{3em} & \hspace*{-3em} \abs{\lambda_n - m c^2 w\expr{\frac{1}{\bar{a}^2} \expr{\frac{n \pi}{2} - \thet\expr{\frac{n \pi}{2 \bar{a}}}^2}}} \\
 & < \frac{m c^2}{n \bar{a}} \expr{7.309 (1 + \bar{a}/3) e^{-\bar{a}/3} + 4.443 \, \frac{n^3 \bar{a}}{(n + \bar{a})^4}} . % c21, c24
\end{aligned}
}
Formula~\eqref{eq:asymp2alt} follows.
\qedhere
\end{enumerate}
\end{proof}

\begin{proof}[Proof of Proposition~\ref{prop:simplebounds}]

It suffices to consider the case $c = \hbar = m = 1$. The upper bound in~\eqref{eq:cs2} for $n \ge 1$ is just the upper bound in~\eqref{eq:cs}. The lower bound for $n = 1$ is trivial. Let, as usual, $w(\xi) = (\xi + 1)^{1/2} - 1$. Since $w(\xi^2) / 2 > w(\xi^2 / 4)$, the lower bound for $n = 2$ also follows from the corresponding lower bound of~\eqref{eq:cs}. Finally, let $n \ge 3$. As in Step~3 of the proof of Lemma~\ref{lem:main}, using~\eqref{eq:convex} we find that
\formula{
 \lambda_n - w\expr{\expr{\frac{(n - 1) \pi}{2 a}}^2} & > (w(\tilde{\mu}_n^2) - \eps_n) - w\expr{\expr{\frac{(n - 1) \pi}{2 a}}^2} \\
 & \hspace*{-4em} > \expr{\tilde{\mu}_n - \frac{(n - 1) \pi}{2 a}} \frac{(n - 1) \pi / (2 a)}{\sqrt{1 + ((n - 1) \pi)^2 / (2 a)^2}} - \frac{0.200 \pi u(b)}{a} \\ % c15
 & \hspace*{-4em} > \frac{\pi}{a} \expr{\frac{3}{8} \, \frac{\pi / a}{\sqrt{1 + \pi^2 / a^2}} - 0.200 u(a/3)} > 0 , % c15
}
and the lower bound in~\eqref{eq:cs2} for $n \ge 3$ follows.
\end{proof}

\subsection{Properties of eigenfunctions}

We already know that approximations $\tilde{\lambda}_n$ and $\tilde{E}_n$ are indeed close to true eigenvalues $\lambda_n$ and $E_n$. Define $\tilde{\Phi}_n(x) = \tilde{\ph}_n(x) / \| \tilde{\ph}_n \|_{L^2(D)}$. In this subsection we show that also $\tilde{\Phi}_n(x)$ are good approximations to true eigenfunctions $\Phi_n(x)$. Recall that $\tilde{\ph}_n(x)$ is defined by~\eqref{eq:phitilde}, with $b = a / 3$, as in the previous subsection.

In order to prove Propositions~\ref{prop:l2} and~\ref{prop:loo}, we need the following result.

\begin{proposition}
\label{prop:l2approx}
The sign of $\Phi_n$ can be chosen so that
\formula{
 \norm{\tilde{\Phi}_n - \Phi_n}_{L^2(D)} & < 1.455 \, \frac{(1 + a/3) \sqrt{4 a^2 + \pi^2} \, e^{-a/3}}{n \norm{\tilde{\ph}_n}_{L^2(D)}} \, \sqrt{a} \, . % c27
}
\end{proposition}

\begin{proof}
Let $\alpha_{n,j} = \scalar{\tilde{\ph}_n, \Phi_j}_{L^2(D)}$, so that $\tilde{\ph}_n = \sum_{j = 1}^\infty \alpha_{n,j} \Phi_j$ in $L^2(D)$. Note that $\alpha_{n,j} = 0$ when $n$ and $j$ are of different parity. Since we can replace $\Phi_n$ by $-\Phi_n$ if necessary, with no loss of generality we assume that $\alpha_{n,n} \ge 0$. Furthermore, we let $\beta_n = \| \tilde{\ph}_n \|_{L^2(D)}$. We have
\formula[eq:l2a]{
\begin{aligned}
 \norm{\tilde{\ph}_n - \beta_n \Phi_n}_{L^2(D)} & \le \norm{\tilde{\ph}_n - \alpha_{n,n} \Phi_n}_{L^2(D)} + \abs{\alpha_{n,n} - \beta_n} \norm{\Phi_n}_{L^2(D)} \\
 & = \norm{\tilde{\ph}_n - \alpha_{n,n} \Phi_n}_{L^2(D)} + \abs{\norm{\alpha_{n,n} \Phi_n}_{L^2(D)} - \norm{\tilde{\ph}_n}_{L^2(D)}} \\
 & \le 2 \norm{\tilde{\ph}_n - \alpha_{n,n} \Phi_n}_{L^2(D)} .
\end{aligned}
}
Observe that by~\eqref{eq:cs2}, \eqref{eq:muest1} and~\eqref{eq:convex}, with $w(\xi) = (\xi + 1)^{1/2} - 1$,
\formula{
 \lambda_{n+2} - \tilde{\lambda}_n & \ge w\expr{\expr{\frac{(n + 1) \pi}{2 a}}^2} - w\expr{\expr{\frac{n \pi}{2 a}}^2} \\
 & \ge \frac{\pi}{2 a} \, \frac{n \pi / (2 a)}{\sqrt{1 + (n \pi)^2 / (2 a)^2}} \ge \frac{\pi^2}{4 a^2 \sqrt{1 + \pi^2 / (2 a)^2}} \, ,
}
and for $n \ge 3$, in a similar manner, 
\formula{
 \tilde{\lambda}_n - \lambda_{n - 2} \ge \frac{\pi}{2 a} \, \frac{(n - 2) \pi / (2 a)}{\sqrt{1 + ((n - 2) \pi)^2 / (2 a)^2}} \ge \frac{\pi^2}{4 a^2 \sqrt{1 + \pi^2 / (2 a)^2}} \, .
}
Since $\alpha_{n, n+1} = \alpha_{n, n - 1} = 0$, it follows that
\formula[eq:l2b]{
\begin{aligned}
 \norm{\tilde{\ph}_n - \alpha_{n,n} \Phi_n}_{L^2(D)}^2 = \sum_{|j - n| \ge 2} \abs{\alpha_{n,j}}^2 & \le \expr{\frac{4 a^2 \sqrt{1 + \pi^2 / (2 a)^2}}{\pi^2}}^2 \sum_{|j - n| \ge 2} (\lambda_j - \tilde{\lambda}_n)^2 \abs{\alpha_{n,j}}^2 \\
 & \le \expr{\frac{2 a \sqrt{4 a^2 + \pi^2}}{\pi^2}}^2 \norm{A \tilde{\ph}_n - \tilde{\lambda}_n \tilde{\ph}_n}_{L^2(D)}^2 .
\end{aligned}
}
Finally, since $\tilde{\mu}_n a \ge 3 n \pi/8 > 1.178 n \ge 1.178$, formula~\eqref{eq:dist3} gives % c25, c25
\formula[eq:l2c]{
 \norm{A \tilde{\ph}_n - \tilde{\lambda}_n \tilde{\ph}_n}_{L^2(D)} < 4.228 \, \frac{(1 + a/3) e^{-a/3}}{\tilde{\mu}_n a^{3/2}} < 4.228 \, \frac{(1 + a/3) e^{-a/3}}{1.178 n \sqrt{a}} . % c26, c26, c25
}
Formulae~\eqref{eq:l2a}, \eqref{eq:l2b} and~\eqref{eq:l2c} combined together yield
\formula{
 \norm{\tilde{\ph}_n - \beta_n \Phi_n}_{L^2(D)} & < \frac{4 a \sqrt{4 a^2 + \pi^2}}{\pi^2} \cdot 4.228 \, \frac{(1 + a/3) e^{-a/3}}{1.178 n \sqrt{a}} \\ % c26, c25
 & < 1.455 \, \frac{(1 + a/3) \sqrt{4 a^2 + \pi^2} \, e^{-a/3}}{n} \, \sqrt{a} \, , % c27
}
as desired.
\end{proof}

\begin{proof}[Proof of Proposition~\ref{prop:l2}]
The last statement of the proposition was already proved in Lemma~\ref{lem:main}. Hence, we only need to prove~\eqref{eq:phl2est}.

By a scaling argument, it suffices to consider only the case $c = \hbar = m = 1$. As in the proof of Proposition~\ref{prop:l2approx}, we let $\beta_n = \| \tilde{\ph}_n \|_{L^2(D)}$. We have
\formula{
 \sqrt{a} \, \norm{\Phi_n - f_n}_{L^2(D)} & \le \norm{(\sqrt{a} - \beta_n) \Phi_n}_{L^2(D)} + \norm{\beta_n \Phi_n - \tilde{\ph}_n}_{L^2(D)} + \norm{\tilde{\ph}_n - \sqrt{a} \, f_n}_{L^2(D)} \, .
}
By Lemma~\ref{lem:norm0}, as in~\eqref{eq:norm1},
\formula[eq:norm2]{
\begin{aligned}
 \abs{\beta_n^2 - a} & = \abs{\norm{\tilde{\ph}_n}_{L^2(D)}^2 - a} \le \frac{\sin \thet_{\tilde{\mu}_n}}{\tilde{\mu}_n} + 4 (1 + \sin \thet_{\tilde{\mu}_n}) I_{\tilde{\mu}_n} \\
 & < \frac{0.383}{\tilde{\mu}_n} + \frac{4 \cdot 1.383 \cdot 0.217}{\tilde{\mu}_n} < \frac{1.584}{\tilde{\mu}_n} \, . % c, c, c, c28
\end{aligned}
}
In a similar manner, by~\eqref{eq:thetaest} and~\eqref{eq:gest2},
\formula{
 \abs{\beta_n^2 - a} & \le \frac{\sin \thet_{\tilde{\mu}_n}}{\tilde{\mu}_n} + 4 (1 + \sin \thet_{\tilde{\mu}_n}) I_{\tilde{\mu}_n} < \frac{2}{\pi} + \frac{4 \cdot 1.383 \tilde{\mu}_n}{8} < 0.637 + 0.692 \tilde{\mu}_n . % c, c29
}
Hence,
\formula{
 \norm{(\sqrt{a} - \beta_n) \Phi_n}_{L^2(D)} & = \abs{\beta_n - a} \le \frac{\abs{\beta_n^2 - a}}{\sqrt{a}} < \frac{1}{\sqrt{a}} \, \min \expr{\frac{1.584}{\tilde{\mu}_n} , 0.637 + 0.692 \tilde{\mu}_n} . % c28, c29
}
By a simple calculation, the minimum is always less than $1.414$. Hence, % c31
\formula{
 \norm{(\sqrt{a} - \beta_n) \Phi_n}_{L^2(D)} & \le \frac{1}{\sqrt{a}} \, \min \expr{\frac{1.584}{\tilde{\mu}_n} , 1.414} < \min \expr{\frac{1.345 \sqrt{a}}{n} , \frac{1.414}{\sqrt{a}}} \, ; % c28, c31, c32, c32
}
the last inequality follows from $\tilde{\mu}_n \ge 3 n \pi / (8 a)$. By the definition of $\tilde{\ph}_n$ and $f_n$ (see~\eqref{eq:sincos} and~\eqref{eq:tildephapprox}),
\formula{
 \abs{\tilde{\ph}_n(x) - \sqrt{a} \, f_n(x)} & \le G_{\tilde{\mu}_n}(a + x) + G_{\tilde{\mu}_n}(a - x) ,
}
so that by~\eqref{eq:gest1} and~\eqref{eq:gest2} (and $3 n \pi / (8 a) \le \tilde{\mu}_n \le n \pi / (2 a)$ in the last inequality),
\formula{
 \norm{\tilde{\ph}_n - \sqrt{a} \, f_n}_{L^2(D)} & \le 2 \norm{G_{\tilde{\mu}_n}}_{L^2((0, \infty))} \le 2 \expr{G_{\tilde{\mu}_n}(0+) \norm{G_{\tilde{\mu}_n}}_{L^1((0, \infty))}}^{1/2} \\
 & \le 2 \expr{\min\expr{\frac{0.383 \cdot 0.217}{\tilde{\mu}_n}, \frac{\tilde{\mu}_n^2}{4 \pi}}}^{1/2} \le \min\expr{\sqrt{\frac{0.333}{\tilde{\mu}_n}}, \frac{\tilde{\mu}_n}{\sqrt{\pi}}} \\ % c, c, c33
 & < \min\expr{\frac{0.532 \sqrt{a}}{\sqrt{n}}, \frac{0.887 n}{a}} . % c34, c35
}
Finally, by Proposition~\ref{prop:l2approx}, we have
\formula{
 \norm{\tilde{\ph}_n - \beta_n \Phi_n}_{L^2(D)} & < 1.455 \, \frac{(1 + a/3) \sqrt{4 a^2 + \pi^2} \, e^{-a/3}}{n} \, \sqrt{a} \, . % c27
}
By an explicit calculation, the numerator on the right-hand side does not exceed $5.318$, and a careful estimate gives that it is also less than $35 / a$. % c, c
% All of the following: $x \exp(-x / 8.1)$, $(1 + x/3) \exp(-x / (8.1 + 3))$ and $\sqrt{4x^2 + \pi^2} \exp(-x / (8.1 + \pi^2 / (4 \cdot 8.1)))$ attain maximum (among positive $x$) for $x = 8.1$. The product of these three functions is not less than $x (1 + x/3) \sqrt{4 x^2 + \pi^2} \exp(-0.33253 x)$, which is still greater than the function in the numerator. The value of the product for $x = 8.1$ is less than $33.455$.
Hence,
\formula{
 \norm{\tilde{\ph}_n - \beta_n \Phi_n}_{L^2(D)} & < \min\expr{\frac{7.738 \sqrt{a}}{n} , \frac{50.925}{n \sqrt{a}}} . % c, c
}
By collecting all above estimates, we conclude that
\formula{
 \sqrt{a} \norm{\Phi_n - f_n}_{L^2(D)} & < \min \expr{\frac{1.345 \sqrt{a}}{n} , \frac{1.414}{\sqrt{a}}} + \min\expr{\frac{7.738 \sqrt{a}}{n} , \frac{50.925}{n \sqrt{a}}} \\ % c32, c31, c, c
 & \hspace*{13em} + \min\expr{\frac{0.532 \sqrt{a}}{\sqrt{n}}, \frac{0.887 n}{a}} \\ % c34, c35
 & \le \min\expr{\frac{9.615 \sqrt{a}}{\sqrt{n}} , \frac{1.414}{\sqrt{a}} + \frac{50.925}{n \sqrt{a}} + \frac{0.887 n}{a}} . % c36, c37
}
When $a \le 0.204 n$, then the first argument of the minimum is smaller than $0.887 n / a$, which is less than the other argument of the minimum. For $a > 0.204 n$ (and $n \ge 1$), % c, c
\formula{
 \frac{1.414}{\sqrt{a}} + \frac{50.925}{n \sqrt{a}} + \frac{0.887 n}{a} & < \frac{1.414 \sqrt{n}}{\sqrt{a}} + \frac{50.925 \sqrt{n}}{\sqrt{a}} + \frac{0.887 \sqrt{n}}{\sqrt{0.204 a}} < \frac{54.303 \sqrt{n}}{\sqrt{a}} \, , % c37, c37, c38
}
and the proof is complete.
\end{proof}

\begin{proof}[Proof of Proposition~\ref{prop:loo}]
Again we only consider the case $c = \hbar = m = 1$. Recall that $T(t; x, y)$ and $T_0(t; x - y)$ are the kernel functions of the operators $T(t) = \exp(-t A)$ and $T_0(t) = \exp(-t A_0)$. By definition, $T(t) \Phi_n = e^{-\lambda_n t} \Phi_n$. By the formula for the Fourier transform of $T_0(t; x)$, we have
\formula[eq:heatl2]{
\begin{aligned}
 \norm{T(t; x, \cdot)}_{L^2(D)}^2 & = \int_{-a}^a (T(t; x, y))^2 dy \le \int_{-\infty}^\infty (T_0(t; x - y))^2 dy \\
 & = \frac{1}{2 \pi} \int_{-\infty}^\infty (e^{-t (\sqrt{s^2 + 1} - 1)})^2 ds = \frac{1}{\pi} \int_0^\infty e^{-2 t (\sqrt{s^2 + 1} - 1)} ds \\
 & \le \frac{1}{\pi} \int_0^1 ds + \frac{1}{\pi} \int_1^\infty e^{-2 t (s - 1)} ds = \frac{2 t + 1}{2 \pi t} \, .
\end{aligned}
}
Let $t = 1 / (2 \lambda_n)$ and $\beta_n = \| \tilde{\ph}_n \|_{L^2(D)}$. We have
\formula[eq:uniformest]{
\begin{aligned}
 e^{-1/2} \norm{\Phi_n}_{L^\infty(D)} & = \norm{T(t) \Phi_n}_{L^\infty(D)} \\
 & \le \beta_n^{-1} \norm{T(t) (\beta_n \Phi_n - \tilde{\ph}_n)}_{L^\infty(D)} + \beta_n^{-1} \norm{T(t) \tilde{\ph}_n}_{L^\infty(D)}
\end{aligned}
}
Since $T(t)$ is a contraction on $L^\infty(D)$, we have (by~\eqref{eq:phitilde} and~\eqref{eq:gest1})
\formula{
 \norm{T(t) \tilde{\ph}_n}_{L^\infty(D)} & \le \norm{\tilde{\ph}_n}_{L^\infty(D)} \le \sup_{x \in D} \expr{q(-x) |F_{\tilde{\mu}_n}(a + x)| + q(x) |F_{\tilde{\mu}_n}(a - x)|} \\
 & \le \norm{F_{\tilde{\mu}}}_{L^\infty(\R)} \le 1 + G_{\tilde{\mu}_n}(0+) \le 1.383 ;
}
for the third inequality, note that $q(-x) + q(x) = 1$ and $q(x), q(-x) \ge 0$. Furthermore,
\formula{
 \norm{T(t) (\beta_n \Phi_n - \tilde{\ph}_n)}_{L^\infty(D)} & = \sup_{x \in D} \abs{\int_D T(t; x, y) (\beta_n \Phi_n(y) - \tilde{\ph}_n(y)) dy} \\
 & \le \sup_{x \in D} \expr{\norm{T(t; x, \cdot)}_{L^2(D)} \norm{\beta_n \Phi_n - \tilde{\ph}_n}_{L^2(D)}} .
}
By Proposition~\ref{prop:l2approx}, \eqref{eq:heatl2} and the identity $(2 t + 1) / (2 t) = 1 + \lambda_n$,
\formula{
 \norm{T(t) (\beta_n \Phi_n - \tilde{\ph}_n)}_{L^\infty(D)} & \le \expr{\frac{2 t + 1}{2 \pi t}}^{1/2} \cdot 1.455 \, \frac{(1 + a/3) \sqrt{4 a^2 + \pi^2} \, e^{-a/3}}{n} \, \sqrt{a} \\
 & \hspace*{-8em} < 0.821 \, \expr{\frac{(1 + \lambda_n) a}{n}}^{1/2} \, \frac{(1 + a/3) \sqrt{4 a^2 + \pi^2} \, e^{-a/3}}{\sqrt{n}} \, .
}
By~\eqref{eq:cs},
\formula{
 \frac{(1 + \lambda_n) a}{n} & \le \frac{a + n \pi / 2}{n} \le a + \frac{\pi}{2} \, ,
}
and therefore
\formula{
 \norm{T(t) (\beta_n \Phi_n - \tilde{\ph}_n)}_{L^\infty(D)} & < 0.821 \, \frac{\sqrt{a + \pi/2} \, (1 + a/3) \sqrt{4 a^2 + \pi^2} \, e^{-a/3}}{\sqrt{n}} \, .
}
Clearly, the numerator in the right-hand side is a bounded function of $a > 0$. By a careful estimate, we find that it is less than $14$.
% All of the following: $\sqrt{x + \pi/2} \exp(-x / (6 \cdot 2 + \pi))$, $(1 + x/3) \exp(-x / (6 + 3))$ and $\sqrt{4x^2 + \pi^2} \exp(-x / (6 + \pi^2 / (4 \cdot 6)))$ attain maximum (among positive $x$) for $x = 6$. The product of these three functions is not less than $\sqrt{x + \pi/2} (1 + x/3) \sqrt{4 x^2 + \pi^2} \exp(-0.33314 x)$, which is still greater than the function in the numerator. The value of the product for $x = 6$ is less than $13.875$.
We conclude that
\formula{
 \norm{T(t) (\beta_n \Phi_n - \tilde{\ph}_n)}_{L^\infty(D)} & < \frac{11.494}{\sqrt{n}} \, .
}
Applying the above estimates to~\eqref{eq:uniformest} gives
\formula{
 e^{-1/2} \norm{\Phi_n}_{L^\infty(D)} & \le \frac{1}{\beta_n} \, \expr{\frac{11.494}{\sqrt{e n}} + 1.383} .
}
Finally, by~\eqref{eq:norm2}, $\abs{\beta_n^2 - a} < 1.584 / \tilde{\mu}_n$. Together with $\tilde{\mu}_n \ge 3 \pi n / (8 a) > 1.178 n / a$, for $n \ge 2$ this gives
\formula{
 \beta_n & > \sqrt{a - \frac{1.584 a}{1.178 n}} > \sqrt{1 - \frac{1.345}{n}} \, \sqrt{a} \, ,
}
and so (for $n \ge 2$)
\formula[eq:loo1]{
 \norm{\Phi_n}_{L^\infty(D)} & \le \frac{\sqrt{e}}{\sqrt{1 - 1.345 / n}} \expr{\frac{11.494}{\sqrt{n}} + 1.383} \frac{1}{\sqrt{a}} \, .
}
This already proves that $\tilde{\Phi}_n(x)$ are bounded uniformly in $x \in D$ and $n = 1, 2, ...$. To prove~\eqref{eq:phsupest}, we need to combine~\eqref{eq:loo1} with a similar estimate for small $n$.

First, we need a different version of~\eqref{eq:heatl2}:
\formula{
 \norm{T(t; x, \cdot)}_{L^2(D)}^2 & \le \frac{1}{\pi} \int_0^\infty e^{-2 t (\sqrt{s^2 + 1} - 1)} ds \le \frac{1}{\pi} \int_0^1 e^{-t s^2 / 2} ds + \frac{1}{\pi} \int_1^\infty e^{-2 t (s - 1)} ds \\
 & \le \frac{1}{\sqrt{2 \pi t}} + \frac{1}{2 \pi t} \, .
}
For $t = 1 / (2 \lambda_n)$, we obtain
\formula{
 \norm{T(t; x, \cdot)}_{L^2(D)}^2 & \le \sqrt{\frac{\lambda_n}{\pi}} + \frac{\lambda_n}{\pi} \le \sqrt{\frac{\lambda_n}{\pi}} \, (1 + \sqrt{\lambda_n}) \le \sqrt{\frac{2 \lambda_n (1 + \lambda_n)}{\pi}} \, .
}
Let $w(\xi) = (\xi + 1)^{1/2} - 1$. Since $w(\xi^2) (1 + w(\xi^2)) = \xi^2 + 1 - \sqrt{1 + \xi^2} \le \xi^2$, the upper bound of~\eqref{eq:cs} gives
\formula{
 \norm{T(t; x, \cdot)}_{L^2(D)}^2 & \le \expr{\frac{2}{\pi} \, w\expr{\expr{\frac{n \pi}{2 a}}^2} \expr{1 + w\expr{\expr{\frac{n \pi}{2 a}}^2}}}^{1/2} \le \sqrt{\frac{2}{\pi}} \, \frac{n \pi}{2 a} = \sqrt{\frac{\pi}{2}} \, \frac{n}{a} \, .
}
Arguing as in the previous case, we obtain
\formula{
 e^{-1/2} \norm{\Phi_n}_{L^\infty(D)} & = \norm{T(t) \Phi_n}_{L^\infty(D)} = \sup_{x \in D} \abs{\int_D T(t; x, y) \Phi_n(y) dy} \\
 & \le \sup_{x \in D} \expr{\norm{T(t; x, \cdot)}_{L^2(D)} \norm{\Phi_n}_{L^2(D)}} \le \expr{\sqrt{\frac{\pi}{2}} \, \frac{n}{a}}^{1/2} .
}
Hence,
\formula[eq:loo2]{
 \norm{\Phi_n}_{L^\infty(D)} & \le \sqrt{e} \, \expr{\frac{\pi}{2}}^{1/4} \sqrt{\frac{n}{a}} < 1.846 \, \sqrt{\frac{n}{a}} \, .
}
A combination of~\eqref{eq:loo1} (for $n \ge 16$) and~\eqref{eq:loo2} (for $n \le 15$) gives $\| \Phi_n \|_{L^\infty(D)} \le 7.333 a^{-1/2}$ for all $n \ge 1$.
\end{proof}

%
%                            ---------- o ----------
%

%
%                            ---------- o ----------
%

\end{document}